\documentclass[pra, letterpaper, onecolumn, 10pt, superscriptaddress]{revtex4}
\pdfoutput=1

	\usepackage{graphicx}

\usepackage{amsmath} 		
\usepackage{amssymb}		
\usepackage{amsthm}   		

\usepackage[mathscr]{eucal}	
\usepackage{mathrsfs}		
\DeclareMathAlphabet{\mathpzc}{OT1}{pzc}{m}{it}  

\usepackage{bm}			

\usepackage{epigraph}    		

\renewcommand{\epigraphflush}{center}
\renewcommand{\textflush}{flushleft}
\renewcommand{\sourceflush}{flushright}

\providecommand{\textquote}[1]{\textsl{#1}}      
\providecommand{\textperson}[1]{\textsc{#1}}   
\providecommand{\textsource}[1]{\textit{#1}}     

\def\be{\begin{equation}}      
\def\ee{\end{equation}}

\def\beu{\begin{equation*}}   
\def\eeu{\end{equation*}}

\def\bsub{\begin{subequations}}  
\def\esub{\end{subequations}}

\def\ie{i.e.}  

\providecommand{\stext}[1]{\text{\tiny{#1}}}  
\def\realsymbol{\mathbb{R}}

\DeclareMathOperator{\realpart}{Re} 
\DeclareMathOperator{\impart}{Im} 

\providecommand{\abs}[1]{\left\lvert#1\right\rvert}   

\def\half{\frac{1}{2}}
\def\smallhalf{\tfrac{1}{2}}

\providecommand{\bv}[1]{\boldsymbol{#1}}        
\providecommand{\vv}[1]{\bv{#1}}  
\providecommand{\unitvec}[1]{\hat{\boldsymbol{#1}}} 

\DeclareMathOperator{\trace}{Tr}      
\providecommand{\opdet}[1]{\abs}  

\providecommand{\cc}{^{\ast}}		
\providecommand{\hc}{^{\dag}}		
\providecommand{\inv}{^{-1}}			

\providecommand{\ket}[1]{\left|#1\right\rangle}
\providecommand{\bra}[1]{\left\langle#1\right|}

\providecommand{\mean}[1]{\left\langle#1\right\rangle}
\providecommand{\innerp}[2]{\left\langle#1\left\lvert\vphantom{#1#2}\right.\!#2\right\rangle}  

\providecommand{\innerpa}[2]{\left(#1\, , \, #2\right)}  

\providecommand{\outerp}[2]{\left\lvert#1\left\rangle\vphantom{#1#2}\right\langle#2\right\rvert}  
\providecommand{\qamp}[3]{\left\langle#1\left\lvert\vphantom{#1}#2\vphantom{#3}\right\rvert#3\right\rangle}   

\providecommand{\comm}[2]{\left[ #1, #2 \right]}  		

\providecommand{\var}[1]{\mbox{Var}\left[#1\right]}

\providecommand{\grad}{\vv{\nabla}}
\providecommand{\divergence}{\grad\cdot}
\providecommand{\curl}{\grad\times}

\providecommand{\del}{\partial}

\DeclareMathOperator{\Si}{Si}			

\def\pskip{\mbox{ }\vspace{10pt}\\}   

\begin{document}


\title{Quantum Mechanical Treatment of Transit-Time Optical Stochastic Cooling of Muons}
\author{A.~E.~Charman}%
\affiliation{Department of Physics, U.C. Berkeley, Berkeley, CA 94720}%
\email{acharman@physics.berkeley.edu}
\author{J.~S.~Wurtele}
\affiliation{Department of Physics, U.C. Berkeley, Berkeley, CA 94720}
\affiliation{Center for Beam Physics, Lawrence Berkeley National Laboratory, Berkeley, CA 94720}
\date{April 9, 2009}

\begin{abstract}
\noindent
Ultra-fast stochastic cooling (i.e., on microsecond time-scales) would be desirable in certain applications, for example, in order to boost final luminosity in a muon collider or neutrino factory, where even with relativistic dilation, the short particle lifetimes severely limit the total time available to reduce beam phase space.  But fast cooling requires very high-bandwidth amplifiers so as to limit the incoherent heating effects from neighboring particles.  A method of transit-time optical stochastic cooling has been proposed which would employ high-gain, high-bandwidth, solid-state lasers to amplify the spontaneous radiation from the charged particle bunch in a strong-field magnetic wiggler.  This amplified light is then fed back onto the same bunch inside a second wiggler, with appropriate phase delay to effect cooling. But before amplification, the usable signal from any one particle is quite small, on average much less than one photon for each pass through the wiggler, suggesting that the radiation should be treated quantum mechanically, and raising doubts as to whether this weak signal even contains sufficient phase information necessary for cooling, and whether it can be reliably amplified to provide the expected cooling on each pass.  A careful examination of the dynamics, where the radiation and amplification processes are treated quantum mechanically, indicates that fast cooling is in principle possible, with cooling rates which essentially agree with classical calculations, provided that the effects of the unavoidable amplifier noise arising from quantum mechanical uncertainty are included.  Thus, in this context quantum mechanical uncertainties do not present any insurmountable obstacles to optical cooling, nor do they lead to any significant differences than in the purely classical regime, but do establish a lower limit on cooling rates and achievable emittances.

\end{abstract}

\maketitle


\begin{epigraphs}
\qitem{\textquote{A.~A violent order is disorder; and\\B.~A great disorder is order.  These\\\hspace{12pt}Two things are one.}}{ \textperson{Wallace Stevens}\\ \textsource{``Connoisseur of Chaos''}  }
\vspace{1.8 ex}
\qitem{\textquote{In all chaos there is a cosmos, in all disorder a secret order.}}{\textperson{Carl Jung}}
\vspace{1.8 ex}
\qitem{\textquote{No phenomenon is a phenomenon until it is an observed phenomenon.}}{\textperson{John Archibald Wheeler}}
\end{epigraphs}

\section{Introduction and Overview}

Developed by Simon van der Meer and collaborators (see \cite{vandermeer:1985, mohl:1977, marriner:2004} for reviews), conventional stochastic cooling using radio-frequency (RF) signals has achieved great success in increasing phase space density of particle bunches in storage rings, for heavier particles such as protons or anti-protons where synchrotron radiation damping is inefficient.  Cooling time-scales typically range from minutes to hours.
Ultra-fast stochastic cooling (\ie, on microsecond time-scales) would be
desirable in certain applications\cite{zolotorev:1993}, for example, to boost final luminosity
in the proposed muon collider, where the short particle lifetime severely
limits the time available to reduce beam phase space.  But fast cooling
requires very high-bandwidth amplifiers so as to limit the incoherent
heating effects from neighboring particles. A method of transit-time
optical stochastic cooling (OSC) has been proposed\cite{zolotorev:1994, zholents:2001} which would employ high-gain, high-bandwidth, solid-state lasers to amplify the spontaneous radiation
emitted from the charged particle bunch in a strong-field magnetic wiggler. This
amplified light is then fed back onto the same bunch inside a second
wiggler, with appropriate phase delay to effect cooling. But before
amplification, the usable signal from any one particle is quite small, on
average much less than one photon for each pass through the wiggler,
suggesting that the radiation must be treated quantum mechanically, and
raising doubts as to whether this weak signal even contains sufficient
phase information required for cooling, and whether it can be reliably
amplified to provide the needed cooling on each pass.  A careful treatment of the
dynamics, where the radiation and amplification processes are treated
quantum mechanically, indicates that fast cooling is in principle
possible, with cooling rates that essentially agree with a simple
classical calculation, provided that the effects of the unavoidable
amplifier noise arising from quantum mechanical uncertainty are
included. Thus, quantum mechanical uncertainties do not appear to present any
insurmountable obstacles to optical cooling, but do establish a lower
limit on cooling rates and achievable emittances, so the effectiveness of such schemes will probably be limited by more prosaic classical concerns over the required laser power and phase control. 
These sources of noise, whether classical or quantum mechanical in origin, might be expected to act, more or less,  much like some effective number of extra classical particles in the sample, affecting the cooling rate and equilibrium emittances in degree but not in kind, although they would nevertheless impose a lower bound on the achievable noise floor and ultimately limit the marginal effectiveness of efforts to dilute the beam density before cooling.

\section{Stochastic Cooling: General Features and Considerations}

\begin{figure}[tb!]
\begin{centering}
  \includegraphics[width=0.85\textwidth]{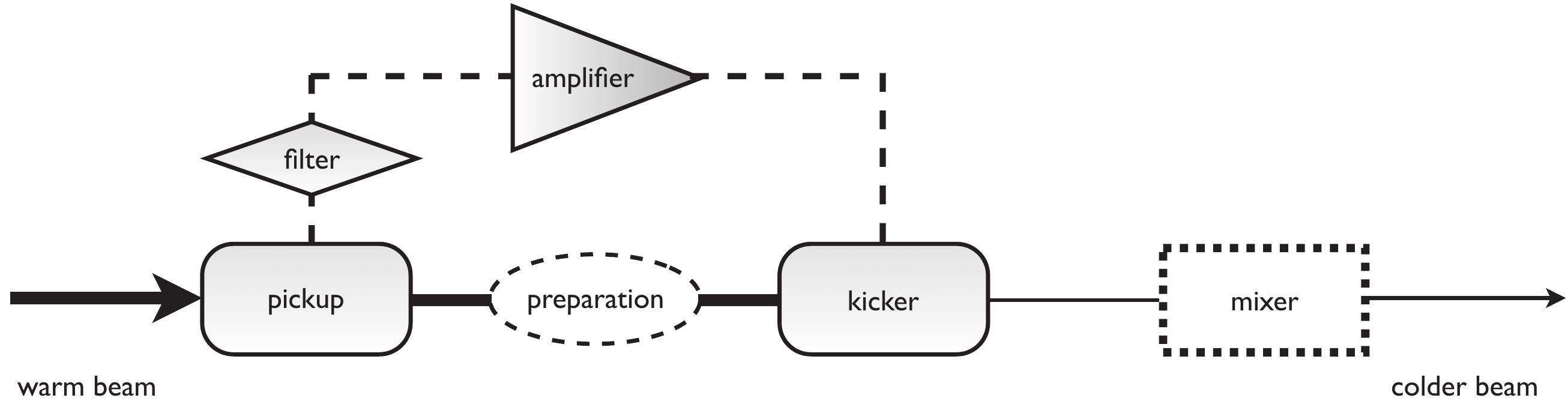}
  \caption{Schematic of an electromagnetically-based stochastic cooling scheme.  A signal containing information about particle phase space deviations from the reference orbit are produced in the kicker, possibly filtered, then amplified, and fed back onto the beam in the pickup, possibly after suitable beam-optical preparations induced on the latter.  If necessary, a mixer section is included between cooling passes to ensure that the incoherent heating effects contribute diffusively.}    
		\label{fig:osc_schematic}
\end{centering}
\end{figure}

In stochastic cooling \cite{vandermeer:1985, mohl:1977, marriner:2004},  an interaction of a charged particle beam with a ``pickup'' measuring device generates a weak electromagnetic cooling signal containing partial information about the phase-space structure of the particles bunch, ideally in a non-perturbative limit, without any appreciable distortion of or back-action on the beam itself.  A general schematic is shown in Fig.~\ref{fig:osc_schematic}.  If, while the beam is properly diverted and prepared, this signal is then suitably amplified, manipulated, and then fed back and made to interact with the same beam, in some ``kicker'' section or device, then the overall dynamics of the reduced ($6D$) beam phase space are non-Liouvillian, and both the longitudinal and transverse beam emittance can be reduced.  Stochastic cooling is non-evaporative, in the sense that it does not decrease available phase space at the expense of removing the outlying particles; beam brightness can be increased as emittance is reduced.

In essentially all stochastic cooling schemes so far proposed or implemented, in which the perturbations to certain particle phase
space coordinates are approximately linearly proportional to some type of electromagnetic kicker fields, the damping time-scale $\tau_{c}$
for cooling of a particular degree of freedom, or equivalently the cooling rate $\tau_{c}^{-1},$  at a particular time in
the cooling process, can be at least roughly approximated by:
\begin{equation}\label{cooltime}
\tau^{-1}_{c}(t) \approx f_{c}\left[B(t) \sqrt{G(t)}- \frac{1}{2}A(t) G(t)\right],
\end{equation}  where $f_{c}$ is the frequency of passages through the cooling sections; $G(t)$ is the net power gain of the
amplifier acting on the pick-up signal;
$B(t)$ is a dimensionless positive parameter, depending on both the nature of the cooling scheme and on the current particle
phase space distribution, and which represents the cooling effects arising from interaction with the feedback signal in the kicker;
and 
\be
A(t) = A_{0} + A_{1}\left(N_{s} + N_{n}\right) + \dotsb,
\ee
represents the heating effects due to each particle's interaction with its own and neighboring particles' signals, where 
$A_{0}$ arises from self-fields and $A_{1}$ and higher-order terms are due to fields from neighboring particles;
$N_{n}$ is a measure of extra noise introduced in the kicker signal by the amplifier, expressed as an equivalent number of extra
particles; and $N_{s}$ is known as the effective sample size, or number of samples, and represents the effective number of
particles with whose kicker signal a given particle also interacts in addition to its own amplified self-field.  Because the
pick-up and amplifier system has finite bandwidth and therefore finite time-response, the signals of some number of
neighboring particles always corrupt a given particle's signal, decreasing the effectiveness of cooling.  In fact, the first term
of (\ref{cooltime}) is the drift, or coherent term, and arises solely from the interaction of each particle with its own kicker
signal, which can be the only source of actual cooling; while the second term is the diffusive, or incoherent, contribution
resulting primarily from amplifier noise and the interaction of a given particle with signals from the roughly $N_{s}$ other
particles present on average in its sample, and contributes only to heating the beam, as is apparent from its negative sign.  (The incoherent
term typically also includes a contribution from the self-field, in order to account for the tendency of particles far out in the tails of the distribution to receive too large of a correcting kick and over-shoot the target trajectory.)

Because the incoherent heating contribution in (\ref{cooltime}) grows faster with amplifier gain than the coherent term, at
any time there is then some locally optimal value of the gain, 
\begin{equation}\label{optimal_gain}
G(t) \approx \left\{\frac{B(t)}{A(t)} \right\}^2
\end{equation}
which maximizes the instantaneous cooling rate then given by 
\be
\tau^{-1}_{c}(t) \approx \smallhalf f_{c}  \frac{B(t)^2}{A(t)}.
\ee
Typically, this locally optimal gain will typically start relatively high
when the beam is noisy and the coherent corrections are large, and then tend to decrease as the beam cools and approaches
an asymptotic distribution of finite emittance, in which the cooling and heating terms just balance.  Note, however,
because the range of possible subsequent cooling rates depend on the current particle phase space distribution, which
in turn depends on the history of past cooling, locally optimal cooling rates do not necessarily result in global optima, \ie, in
the  fastest possible overall cooling achievable in a prescribed time interval.  The globally optimal cooling history is an interesting but non-trivial problem in nonlinear control
theory, left for future research.

 \section{Requirements on and Uses for Fast Stochastic Cooling}

Both the  achievable cooling rates and asymptotic equilibrium emittances therefore depend both on the absolute power delivered by the
feedback signal, and on the relative power in the usable ``coherent'' pickup signal from any single particle, which contains the
phase space information necessary for cooling, as compared to the corrupting ``incoherent'' signal arising from nearby particles
or from noise in the amplifier, which actually contributes to heating during feed-back. 
 
In order to increase the cooling rate and typically simultaneously decrease the equilibrium emittance achievable, the incoherent term
represented by $A(t)$ must therefore be made smaller.  From the form of $A(t)$ we can see that fast cooling times 
therefore require that the amplifier noise and effective sample size both be made as small as possible. 
The sample size will scale like
\begin{equation}\label{sample}N_{s}
\sim \rho_{b} \min\left[\pi\sigma^2_{b\perp}, S_{c}\right] v_{0}\Delta \omega^{-1},
\end{equation} where $\rho_b = \frac{N_{b}}{\pi\sigma^2_{b\perp} L_{b}}$ is the spatial number density of
particles in the bunch, $\sigma_{b\perp}$ is the transverse beam radius, $L_{b}$ is the longitudinal beam length,
$v_{0} = c\beta_0$ is the mean longitudinal beam velocity, $S_{c}$ is a measure of the transverse coherence area of the kicker fields, and $\Delta \omega$ is the limiting bandwidth of the pick-up/amplifier/kicker system.

Rapid cooling will therefore require relatively low beam densities and high-bandwidth, high-peak but variable-gain, low-noise amplifiers. 
Existing stochastic cooling schemes relying on radio-frequency or microwave technology are limited by the
$O(\mbox{GHz})$ bandwidths available for high-gain amplifiers at these frequencies, and typical cooling time-scales range from minutes to hours.

Yet much faster cooling might be necessary or desirable in some situations -- for example, in order to achieve ultra-high luminosity proton beams, because phase space reduction can be offset by particle losses due to collisions, diffusion, etc, in the cooling ring at long time-scales.  Typically time-scales for current radio-frequency (RF) stochastic cooling are $\tau_{\stext{RF-SC}} \sim O(10^2\;\mbox{s})$ or $O(10^3 \;\mbox{s})$, while with realistic technology OSC might achieve $\tau_{\stext{OSC}} \sim O(1\;\mbox{s})$.
For electron beams, OSC can also work at lower energies where synchrotron damping is Inefficient.   Rapid cooling for any proposed muon collider will be essential, because of the finite lifetime of the muons; all stages of particle beam production, collection, collimation, acceleration, cooling, and experimentation must take place in only a few lab-frame decay times, which for muons at $O(10^2 \mbox{ Gev})$ energies is only $O(1\; \mbox{ms})$.  

Such ultra-fast stochastic cooling, on microsecond time-scales, would require beam densities much lower than those typically
achieved for particle beams of useful current with bunch sizes chosen for acceleration in RF structures, and would require 
amplifiers which can achieve high gain over very broad bandwidths with minimal added noise.  Some beam stretching and
subsequent compression can be used to suitably dilute the beam density during cooling and restore bunch sizes after
cooling, and can be achieved in a stable, essentially reversible manner using conventional beam optics,  but sufficiently
broad gain bandwidths cannot be achieved with existing RF technology.
Barring any unforeseen breakthroughs, fast cooling will require moving to optical wavelengths, where
solid-state lasers amplifiers (such as those using Ti:Sapphire crystals) have achieved high gain over
$O(\mbox{THz})$ bandwidths centered around $O(1 \mbox{ }\mu\mbox{m})$ wavelengths.  Because of the high
gain and high bandwidth achievable, optical stochastic cooling shows great promise, but also poses significant technological
challenges.  At such extremely fast time-scales, the pick-up signal cannot be manipulated electronically, but must be suitably
collected, controlled, amplified, and re-directed into the kicker for feedback entirely through optical means; in order to reduce longitudinal emittance, transverse optical fields must be made to effect longitudinal momentum kicks, requiring very high gain; and particle beam optics must control particle positions within a fraction of an optical wavelength, presumably through some active monitoring and feedback.   These pose important questions and difficult challenges, but none in themselves are feared to invalidate the possibility of OSC in principle.  Here we focus primarily on a fundamental question of principle that has been raised, namely whether the wiggler signal from the beam in like regimes of operation contains adequate information to cool quickly or even cool at all, or instead whether it may be hopelessly corrupted by quantum ``noise'' or ``fluctuations.''

\section{Why Consider a Muon Beam?}

The possibility of a muon collider has received significant attention in the past decade or so.  The muon is a fundamental particle that has been little studied at high energies in controlled experiments.  In a muon-muon collider, lepton physics similar to that studied in linear electron-positron colliders might be pursued, but because with a  rest mass of  $m_{\mu} = 105.7 \mbox{ MeV}/c^2$, the muon is $\sim 207$ times heaver than the electron, so synchrotron radiation is comparatively suppressed, and muons can be accelerated and stored in circular rings at high energies ($\sim O( 10^2 \mbox{ GeV})$), as opposed to electrons or positrons which require linear accelerators.  The higher mass also suppresses the so-called ``beamstrahlung'' effects which can lead to energy loss and/or energy spread, so larger bunch sizes and higher luminosities are in principle possible, while radiative corrections are smaller.  In terms of their potential for particle creation, collisions between leptons that seem t behave as point particles are intrinsically more efficient than collision between baryonic particles like protons with significant internal structure.  The larger mass of the muon also translates into larger cross sections for certain interactions, especially for Higgs production, and precise measurements of the muon lifetime or $g-2$ factor, or searches for a muon electric dipole moment (EDM) or for lepton-flavor-violating decays offer promising avenues upon which to search for signatures of SUSY or other physics beyond the standard model.

The catch of course is that muons are unstable.  They must be created through pion capture, so intense sources are expensive and produce initial beams with poor collimation (high transverse emittance) and large energy spread (large longitudinal emittance).   Even with time dilation in the lab frame corresponding to relativistic factors of $\gamma \sim O(10^3)$, a proper lifetime of only  of $\tau_{\mu} \approx 2.2 \mbox{ }\mu\mbox{s}$ leaves only a matter of milliseconds to collimate, manipulate, cool, accelerate, and collide a muon bunch.  This might in part be turned to advantage by optimizing the ring not for muon collisions as such but instead for the production of a highly-collimated, high-flux, terrestrially-based source of neutrinos via spontaneous decay of the muons.

In either mode of operation, as a muon-collider or a neutrino factory, the ring would pose many severe technological challenges, including the need for ultra-fast and intensive cooling, with a characteristic damping time $\tau_{\stext{damp}} \sim O(10^{-6} s)$ or less.
Ionization-based rapid transverse cooling,  as well as longitudinal cooling through shaped absorbers and/or emittance exchange, will be necessary, but transit-time optical stochastic cooling has been proposed as a possible means, after the beam is already collimated and highly relativistic,  to boost final luminosity beyond that which can be achieved by ionization cooling alone, which is limited by multiple scattering and trade-offs between transverse and longitudinal emittance effects.

\section{Transit-Time Optical Cooling}

\begin{figure}[tb!]
\begin{centering}
  \includegraphics[scale=0.85]{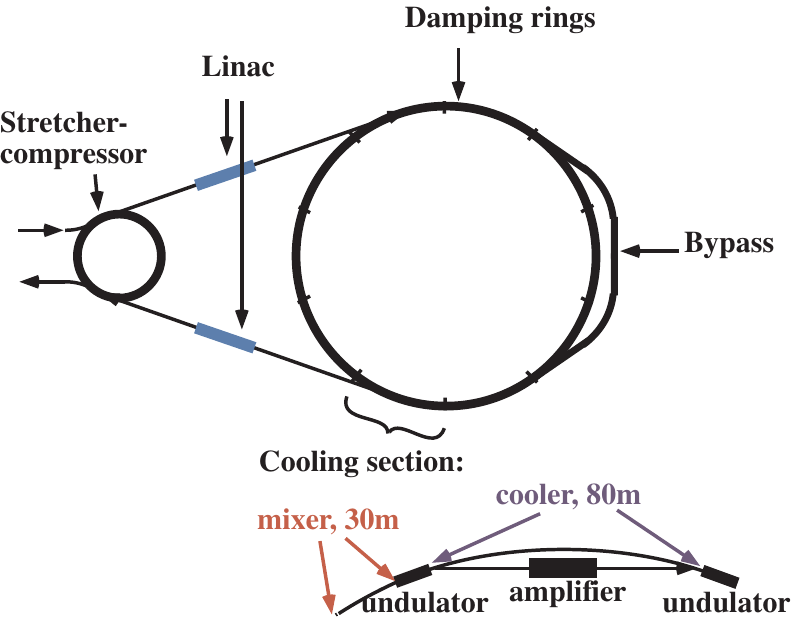}
  \caption{Schematic of a system for fast transit-time optical stochastic cooling. The stretcher/compressor ring, together with the linacs,  expands the beam to lower the density before cooling, then reversibly compresses it after cooling.  The cooling ring may contain one or more cooling sections.} 
	\label{fig:osc_ring}
\end{centering}
\end{figure}

Zolotorev, \textit{et al.}\cite{zolotorev:1994, zholents:2001} have proposed and explored a possible method of ultra-fast transit-time optical stochastic cooling, in which both the pick-up and kicker consist of large-field magnetic wigglers; as shown schematically in  Fig.~\ref{fig:osc_ring}.   In the pick-up magnetic wiggler, Lorentz forces will produce transverse quiver motion of the charged beam particles, which in turn generates a small amount of spontaneous synchrotron
radiation.  This wiggler radiation is then collected and greatly amplified in a low-noise, solid state optical amplifier system and directed into the second wiggler.  While the light is being amplified, the particles are directed through a bypass lattice,
whose beam optics are designed so that each particle receives a time-of-flight delay proportional to the deviation of its
longitudinal and/or transverse phase-space coordinates from the desired reference orbit.  
Particles then rejoin the amplified light in the kicker wiggler, where they again undergo transverse quivering with nearly the same polarization and at nearly the same frequency as the electric field of the optical radiation, resonantly exchanging some amount of energy, with the magnitude and sign of the net energy kick depending on the
relative phase between particle quiver and field carrier oscillation.

Actual cooling can be effected only by the interaction of each
particle with its own amplified field, and, if the transit-time in the bypass lattice is adjusted so that the delay relative to the
self-field is proportional to the longitudinal momentum deviation, then the interaction in the kicker can produce a restoring
force leading to reduced momentum spread.  If particle time-of-flight is manipulated so as to also depend on a transverse
betatron coordinate, then strong dispersion in the lattice in the region of the kicker wiggler can also lead to transverse emittance reduction in that direction, and alternating the polarization of the wigglers or
rotating the transverse phase space of the particle beam can then result in cooling the full transverse phase space.

Cooling is therefore critically sensitive to particle phase relative to the optical signal.  Between the pickup and the kicker, particle positions must be carefully controlled, within a fraction of the micron-scale optical wavelength, presumably through active feedback on the particle beam optics.  As cooling proceeds and deviations from the ideal orbit decrease, continued efficient cooling demands that the amplifier gain be decreased, and that the particle beam optics be adjusted in the bypass, if not continuously then at least intermittently, so that the probable range of particle deviations continues to be mapped into approximately one-half of an optical period in time-of-flight variation.

Between one cooling pass and the next, efficient cooling demands that the lattice must be designed to provide good mixing, or effective ``randomization'' of particles within each bunch, so that the ``incoherent'' signal arising from the sample particles neighboring any given particle is really incoherent, becoming effectively randomized for each cooling kick.  Because of the intrinsically low signal-to-noise ratios (SNRs), stochastic cooling can really only work at all because the coherent drift term acts during every pass as a restoring force on average, tending to kick each particle towards the desired reference orbit on every pass, while the heating terms, although always present, ideally are 
more or less random from pass to pass, and therefore contribute in a diffusive fashion, with no bias in a particular direction, and with a net standard deviation accumulating only in proportion to the square root of the number of kicks.

However, if the net effect from all the sample particles is not independent from kick to kick, so a particle can experience a substantially similar ``incoherent'' signal for several passes, or even even partially correlated heating kicks in the same direction over many passes, then the resulting heating can become super-diffusive, and cooling can be greatly slowed or even suppressed altogether.  

Any given particle will be subject to the heating effects due to signals primarily from
its neighboring sample of particles, so the size of the heating signal will scale inversely with the bandwidth of the cooling system.  In most traditional
stochastic cooling schemes, good mixing essentially requires that the identity of particles in a given particle's sample be randomized between each pass, because the heating signal may depend strongly on particle degrees-of-freedom that do not change significantly from one pass to the next.  To shuffle the make-up of each sample between each cooling
pass, each particle should be shifted in position by a distance comparable to a sample length or more, between exiting the kicker and traveling to the next passage through the pickup, in some manner that does not lead to persistent correlations between longitudinal particle position and the DOFs undergoing cooling.

For the high-bandwidth, transit-time OSC method, this sample length is typically very short compared to the beam dimensions, about $L_s \sim N_u \lambda_0$ \footnote{This is what we will shortly recognize as the \textit{coherence length} of the radiation, not the much longer wiggler length $L_u = N_u \lambda_u$.} so this kind of mixing should be even easier to achieve than in RF schemes.

However, this amount of mixing is  really more than is needed here.
With transit-time optical stochastic cooling applied to highly relativistic beams of already moderately low emittance and energy spread, each particle will produce wiggler radiation of roughly the same envelope shape and spectrum, just with a different overall phase.
The heating term for any given particle consists essentially of a non-stationary shot noise, the superposition of about $N_{s}$,  nearly-identical wave-packets of duration $N_u\lambda_0,$ but with essentially random phases.
Since the phase-space information useful for cooling is also encoded in the phases of the signals, good mixing in this context does not really require that particles move between samples between kicker and pickup, but can be achieved merely by assuring that longitudinal particle positions within a sample are effectively randomized, or equivalently that the relative distances between particles shift on the order of one optical wavelength $\lambda_0$, or perhaps a little more, in a suitably ``random'' fashion, so as to randomize the phases of  the approximately $N_{s}$ wave-packets making up the incoherent kick signal in a given particle's sample.
By ``random,'' we here mean that, ideally, the shifts should be largely uncorrelated with relative longitudinal positions before the kick, but can actually depend on other degrees-of-freedom, even ones that might be actively cooled, such as transverse betatron coordinates or even longitudinal momentum.  Because $\lambda_0$ is so short compared to the stretches of beam line between cooling sections available for mixing (typically
$O(10 \mbox{ m})$), this level of sample mixing should not be difficult to achieve, and in fact the greater challenge will be to
suppress unwanted mixing on these optical length-scales between the pickup and corresponding kicker.   (Actually, if the longitudinal positions are initially uncorrelated with beam energy or betatron deviations, then the time-of-flight delays purposefully introduced in the bypass lattice between pickup and kicker during a single pass themselves can provide much of the needed mixing for the next pass.)

In order to sufficiently reduce the incoherent heating effects from neighboring particles, beam density is reversibly lowered before cooling, and then restored following cooling.  Even with high-bandwidth optical amplifiers and therefore relatively short sample lengths, typical densities for
particle beams of interest are still much too large for microsecond-scale cooling with conceivable amplifier powers.  RF acceleration requires reasonably short bunch lengths ($O(10 \mbox{ cm})$ or  so) for proper phasing, and  high luminosity
requires large bunch charge, $N_{b} \sim O(10^{9})$ or even higher if possible, so each bunch will enter the cooling section with rather high linear charge density, leading to unacceptably large sample sizes, on the order of  $N_{s} \sim O(10^{4})$ or
greater.  So before cooling, each bunch emerging from the acceleration sections must be greatly stretched, from the order of tens of centimeters to a few hundred meters, so that so that cooling can take place at lower particle density and smaller sample sizes, say $N_{s} \sim
O(10^2)$  After cooling, the bunch is de-compressed in order to restore luminosity.  The reversible compression and stretching is effected using a linear accelerator (LINAC) and a specially designed lattice (see Fig.~ \ref{fig:osc_ring}).  
Because the beam is highly relativistic, all beam particles travel at almost the same nearly-luminal velocity, but with a distribution of relativistic momenta, so drift in free space will not efficiently expand the beam, but instead the beam may be first stretched by transport through a ring with very high momentum
compaction factor, such that particles with greater longitudinal momentum will travel along longer orbits and lag behind those with
less momentum.   Assuming such stretching separately conserves longitudinal action (clearly something of an approximation, since the compaction ring must correlate transverse coordinates with longitudinal momentum), this corresponds to a simple symplectic rotation in the energy-duration plane of beam phase space, so as it increases the beam length, it also introduces a head-to-tail energy chirp, \ie,  a correlation between longitudinal intra-bunch longitudinal particle position and energy, and as an added bonus, decreases the effective range of relative energy spread, having effectively transformed part of the random energy variation into an ordered energy correlation.  The chirp would tend to impede particle mixing between cooling passes, and so is removed by applying a suitably ramped current in an induction LINAC
before the beam passes into the cooling ring proper.   The beam stretching and energy compression is essentially completely Hamiltonian,  analogous to the adiabatic expansion of a gas,  and can be reversed after cooling, when a complementary energy chirp is introduced in the beam by another LINAC, and the beam is de-compressed to its original length (ideally in the same compaction ring), but now hopefully with reduced emittance and higher luminosity.  For the muon cooling scenario, the stretching and compression phases would consume approximately half of the allotted cooling time, \ie, typically a few to several microseconds, but the benefits of beam-stretching more than make up for this extra time by increasing the achievable cooling rate via reduction in sample size and reduction in the energy variation that must be addressed by the cooling sections.

\section{Spontaneous Wiggler Radiation}

The detailed dynamics of a transit-time optical cooling system will depend on the physics and form of the spontaneous wiggler radiation emitted in the pickups.  The central wavelength $\lambda_{0}$ of the wiggler radiation is downshifted from the wiggler period
$\lambda_{u}$ itself by relativistic effects; for a planar wiggler,
\begin{equation} \lambda_{0} \approx
\frac{\lambda_{u}}{2\gamma_{0}^{2}}\left( 1 + \frac{a_{u}^{2}}{2}\right),
\end{equation}
where  $a_{u} = \frac{|q| B_{u}}{k_{u}mc^{2}}$ is the wiggler parameter, $k_{u} \equiv
2\pi/\lambda_{u},$ is the lab-frame wiggler wavenumber,  $B_{u}$ is the peak wiggler magnetic field, $\gamma_{0}
mc^{2}$ is the average energy of a beam particle and $q$ is its charge.   This can be viewed as a resonance condition, such that the radiation slips ahead of the sources by one optical wavelength in the time the charges advance by one wiggler wavelength.

Given the average beam energy, the wiggler
parameter and wiggler period are chosen to approximately match the resonant radiation wavelength $\lambda_0$ \footnote{It might also be possible to work at an harmonic of this fundamental, but this case will not be considered here.} to the center of the gain bandwidth for the solid-state laser amplifiers, which is typically $O(1\;\mu\mbox{m})$. The homogeneous ``coherent'' bandwidth\cite{kim:1989} of such wiggler radiation is given by
\begin{equation}
\Delta\omega \approx \frac{1}{2N_{u}}\omega_{0},
\end{equation}
where $\omega_{0} = ck_{0} = 2\pi c/\lambda_{0}$ is the central radiation frequency and $N_{u}$ is the
number of undulator periods in the pickup wiggler.   This just follows from the Fourier-Heisenberg uncertainty product, since each particle will obviously radiate almost exactly $N_u$ periods of radiation. 

In the so-called undulator limit, where the angular deflection of particles due to their induced quiver motion is small compared to the characteristic opening angle for synchrotron radiation, \ie $a_u < \gamma_0\inv$, the so-called  coherent component or coherent mode \footnote{Coherent here refers to radiation that is is approximately diffraction-limited transversely and approximately Fourier-limited longitudinally or temporally.  The radiation from a single particle into this bandwidth and in this transverse mode is coherent in this sense, but the radiation from the bunch as a whole is incoherent in general, consisting of approximately $\tfrac{L_b}{N_u\lambda_0}$ longitudinal modes, essentially randomly-phased with respect to each other, where $N_u \lambda_0$ is the so-called coherence length, proportional to the inverse bandwidth.} of the radiation
corresponds to a nearly diffraction-limited beam with angular spread 
\begin{equation}
\delta\theta \approx \tfrac{ \sqrt{1 + \smallhalf a_u^2} }{2 \sqrt{N_{u}} \gamma_{0}},
\end{equation}
smaller than characteristic synchrotron radiation opening angle by a factor of $O(N_u^{-1/2})$, and with spot size (\ie, optical beam waist)
\begin{equation}
w_0 \approx \frac{\lambda_{0}}{4\pi}\frac{1}{\delta\theta}
=  \tfrac{\sqrt{1 + \smallhalf a_u^2} }{4\pi } \frac{\sqrt{N_u}\lambda_u}{\gamma_0},  \end{equation}
satisfying the optical uncertainty principle for transverse degrees-of-freedom (DOFs).

Assuming that the light fields in the kicker remain spatially coherent over the transverse extent of the particle beam, and the amplifier bandwidth $\Delta \omega_{A}$ centered on $\omega_{A}$ is matched to the coherent bandwidth
$\Delta\omega$ of spontaneous radiation, centered on $\omega_{0},$ the effective sample size $N_{s}$, meaning the average number of particles contributing appreciably to the incoherent heating signal experienced by any other particle, will scale like:
\begin{equation}
N_{s} \sim \rho_b\sigma^{2}_{b \perp} L_{s},\end{equation}
where $n_b =  \rho_b\sigma^{2}_{b \perp}$ is the average number line density of particles in the bunch, and $L_{s} \sim  N_{u} \lambda_{0}$ is the sample length length \footnote{Actually, a particle \textit{trailing} another particle
by any distance will see almost none of the other particle's radiation, since the radiation is largely confined to a small forward angle by relativistic effects, and the faster radiation slips forward relative to the particles, while a particle \textit{leading} another particle by about $N_u \lambda_0$ will interact with only a small fraction of that particle's radiation that slips sufficiently far ahead over the finite length of the kicker wiggler, so the effective sample length $L_{s} $ will be somewhat shorter than $N_u \lambda_0$, as will be seen below.} over which particles will affect their neighbors.  For fast
cooling, the beam must be sufficiently stretched so that $N_{s}$ is quite small compared to conventional stochastic
cooling regimes,  \ie., $N_{s} \lesssim O(10^{2})$ or so.

In addition, the undulator magnets themselves will be optimized quite differently than for conventional light-source applications.  Most applications of wiggler radiation benefit from high coherence (narrow-bandwidth)
spectra, and therefore rely on moderate (in the context of particle physics) energy beams traveling through wigglers of
moderate wavelength but with many periods.  In optical stochastic cooling, it is desirable to produce very broad bandwidth
optical radiation from extremely relativistic beams, so the wigglers will consist of a relatively small number ($N_u \sim O(10)$) of long-period ($\lambda_u  \sim O(50\mbox{ cm})$) magnets with field strengths essentially as high as is practical ($B_{u} \sim O(10 \mbox{ T})$).

From classical radiation theory and Planck's law, the average number of photons  emitted per particle into the coherent
component of undulator radiation produced in a planar wiggler may be roughly approximated as
\begin{equation}\label{power1}
\EuScript{N}_{\stext{ph}} = \alpha \,\tfrac{\pi}{2} \tfrac{1 + a_{u}^{2}}{1 + \half a_{u}^{2}}\left[
J_{1}\!\left(\tfrac{1 + a_{u}^{2}}{4\left(1 + \half a_{u}^{2}\right)}\right) - 
J_{0}\!\left(\tfrac{1 + a_{u}^{2}}{4\left(1 + \half a_{u}^{2}\right)}\right)\right]^{2} \sim O(\alpha),
\end{equation}
where $J_{\ell}(x)$ is the $\ell$th-order ordinary Bessel function, and $\alpha  = \tfrac{e^2}{\hbar c} \approx \frac{1}{137}$ is the fine structure constant.

While the total energy radiated per particle as determined by the Larmor formula will of course be proportional to the number $N_{u}$ of wiggler periods,  the power radiated into the coherent mode remains constant (or nearly so) as $N_{u}$ varies, because the coherent bandwidth scales inversely with $N_{u}.$  The remaining energy is radiated too far off-axis or at frequencies too far
from the fundamental to be particularly useful for OSC.  Actually, equation (\ref{power1}) is not entirely
accurate for small-period wigglers where $N_{u} < O(10)$, but in any case, for $\lambda_{0} \sim O(1\mbox{ }\mu\mbox{m})$, $N_{u} \sim O(10)$, and achievable magnetic field strengths corresponding to $a_{u} \sim O(1),$ the number of photons emitted per particle into the coherent mode should still be $O(\alpha)$.  

\section{``Naive'' Quantum Mechanical Considerations}
 
So optical stochastic cooling poses serious technological challenges, at least for the very fast cooling required for muons, but in pondering the possibilities for such fast cooling based on wiggler radiation, serious concern arose over possible fundamental rather than merely practical limitations of this scheme. The apparent problem is that the actual cooling
arises only through the interaction of each particle with its own wiggler radiation, which as we have just seen, is extremely weak before amplification.  

In any one pass through a pickup, each particle only radiates on average $O(\alpha) \sim
10^{-2}$ photons that can be collected, amplified, and fed back to actually effect cooling, so the 
optical cooling signal from each particle will very weak and presumably may be subject to significant quantum mechanical effects. Yet naive quantum mechanical considerations then raise fundamental doubts as to whether the pickup signal even contains the phase information needed for transit-time cooling, whether this information can be reliably amplified and extracted, and whether quantum fluctuations in the incoherent signal from neighboring
particles or arising in the optical amplifier itself lead to more significant heating than is accounted for classically.

\subsection{Quantum Cooling Catastrophes?}

With so few photons on average in the relevant cooling component of the pickup signal for any given particle, simple-minded quantum mechanical thinking suggests possible quantum catastrophes for cooling.  These lines of argument, while ultimately flawed,  truly raised concerns and engendered debate, and are not merely straw men erected only to be demolished by more careful analysis.

\subsubsection{Do individual particles radiate at random in ``quantum jumps''?}

If, as in the standard treatment by Sands\cite{sands:1970} of synchrotron radiation damping in electron storage rings, particles are assumed to radiate independently and at random, in a series of discrete, stochastic ``quantum jumps'' corresponding to Poissonian emission of a whole number of photons, at some average rate but at
random times, and if the amplifier is imagined to faithfully multiply whatever photons are emitted,
perhaps with the addition of some extra randomly-phased noise photons due to spontaneous emission or
thermal noise \footnote{Note that the \textit{independent} Poissonian photon emission model, in which all emissions are assumed uncorrelated,  can predict  any average emitted photon number, while the variance in photon number is always proportional to the mean.  At least for the amplifier noise, we would imagine something closer to a so-called chaotic or thermal state, where due to interference effects the statistics are not those of shot noise, but rather the standard deviation is proportional to the mean.  This is the first hint that something might be wrong with the reasoning advanced here....}, then on average any one particle emits a photon only once in every $O(\alpha^{-1})$ passes through the pickup, while the neighboring particles in its
sample emit on average a total of $O(N_{s} \alpha)$ photons, give or take $O\left(\sqrt{N_{s}\alpha}\right)$.  So for $N_{s} \gtrsim O(10^2)$, one or more photons within a coherence length, randomly phased with respect to the particle in question will likely be present in the pickup signal and be amplified on every pass.  It might seem that each particle will be subject to an appreciable heating kick on each pass, but usually experience no cooling kick whatsoever on most passes, but then every $O(\alpha^{-1})$ turns or so, suddenly receive a large cooling kick, comparable in magnitude to the typical heating kick per pass.  Such stochastic discreteness effects might be expected to drastically lower the cooling rate compared to that calculated classically, or possibly even lead to unstable feedback preventing or suppressing cooling altogether.

\subsubsection{When individual particles do radiate, is the phase even well-defined?}

Because the sample size $N_{s}$ is quite small compared to that in most conventional stochastic cooling schemes, and the  coherent signal intrinsically small, it also seems possible that the fluctuations in the pickup signal may no longer be dominated by the classical shot noise associated with random particle positions within the beam, but instead or additionally include quantum fluctuations of
some sort.  In particular, if the radiation emission from each particle consists of the occasional random emission of a photon in a ``quantum
jump'' at some random time while the particle undulates in the pickup wiggler, then the phase of these photons would be
expected to be very poorly determined.  Even if the photon is perhaps more
realistically considered to be emitted over some finite formation length, presumably the wiggler length itself in the lab frame, in the relevant
regime of small emission probability per pass, this might seemingly make the phase uncertainty worse, not better.  But in transit-time cooling schemes, the particle phase-space information used for cooling is encoded almost exclusively in the
phase of the signal, so that even if the self-field is present, it seems that it might not carry the phase information necessary for transit-time cooling to work.  Because the average photon number emitted by any one particle is small, the variation or
uncertainty in this number is also small in absolute terms, and the ``number-phase'' Heisenberg uncertainty principle,
\be
\Delta \EuScript{N} \Delta \phi \ge \frac{1}{2}
\ee
suggests that the phase \footnote{At this heuristic level, fortunately we can
simply ignore the well-known technical difficulties associated with defining a self-adjoint phase operator in quantum mechanics conjugate to the usual bosonic number operator.} of any emitted photons is very poorly determined. 
 If photon emission is assumed to be at least approximately Poissonian, then $\Delta \EuScript{N} \approx \EuScript{\bar{N}}^{1/2},$ where
$\EuScript{\bar{N}}$ is the average number of photons; but if $\EuScript{\bar{N}} \sim O(\alpha),$ the phase uncertainty must satisfy
\be
\Delta \phi \ge \EuScript{\bar{N}}^{-1/2}  \sim  O\bigl(\alpha^{-\frac{1}{2}}\bigr) \gtrsim  2\pi.
\ee 
But if essentially all of the particle phase-space information useful for cooling is contained in the optical phase of the pickup radiation, and this is unavailable, cooling cannot occur.

\subsubsection{What of spontaneous emission or thermal noise in the amplifier?}

In addition to the discreteness effects in photon-emission and photon-phase noise, the amplifier is expected to add the equivalent of one or more photons to the pre-amplified signal due to unavoidable spontaneous emission within the active gain medium, and perhaps  as well some thermal or other additional noise photons.  These photons will be randomly-phased, so this source of noise acts more or less like having some extra number of particles $N_{n}$  in the sample, in addition to, and for our parameter regime comparable or greater to,  the actual number $N_{s}$.  Therefore, if the cooling does work despite our concerns addressed above , this may thereby affect the cooling rate in degree, but should not quell cooling altogether.

\subsubsection{Stochastic Cooling in the Quantum-Jump Model}

Fortunately, these fears of quantum effects catastrophically slowing or suppressing cooling will turn out to be misguided; the only source of quantum noise actually present is of the final sort mentioned: additive amplifier noise.  But it will be informative to incorporate these various considerations into very approximate quantitative estimates for the longitudinal (or rather, energy-spread) cooling rates with various naive quantum effects incrementally included for comparison, in order to make this
intuitive reasoning more precise and to better understand later where it goes wrong. 

The skeptical reader may again suspect that these arguments are introduced disingenuously, but they accurately represent the very concerns that motivated this investigation, and with some basis.  As mentioned previously, a very similar discrete model of radiation by photon emission was used, and by all appearances quite successfully, by Sands\cite{sands:1970} to treat radiation
damping in a synchrotron ring, where quantum fluctuations offset damping, leading to finite equilibrium emittances, and is also
commonly used to analyze laser cooling of particle beams by Thomson scattering\cite{esarey:2000}.

The energy kick given to the $jth$ particle on a
single pass is
\begin{equation}
mc^{2}\Delta \gamma_{j} = q\!\!\!\!\!\int\limits_{t_{j}}^{t_{j} + \Delta t_{j}} \!\!\!\!\!dt\, \bv{E}\bigl(\bv{x}_{j}(t),t\bigr) \cdot \bv{v}_{j}(t),
\end{equation}
where $\bv{E}$ is the transverse optical electric field seen by the particle in the kicker,  $\bv{x}_j(t)$ is
the particle's quivering spatial trajectory, governed predominately by the large magnetic
field of the plane-polarized kicker wiggler, $\bv{v}_(t) = c\bv{\beta}_{j}(t) = \frac{d}{dt}\bv{x}(t)$ is its velocity, $t_{j}$ is its arrival time
at the front of the kicker wiggler, and $\Delta t_{j}$ is the time spent inside. 

As we are interested in a rough comparative scaling, we will use a simplified ``back-of-the-envelope'' model of longitudinal cooling where we assume all fields and trajectories oscillate sinusoidally,  neglect end effects in the wigglers, diffractive and other transverse variation in the fields, dispersion and nonuniform gain in the amplifier, and various other details, consider beam particles to be highly relativistic (\ie, $\gamma_j \gg 1$) but already relatively cold, (\ie, $\abs{\gamma_{j} - \gamma_{0}} \ll \gamma_{0}$), and assume the static wiggler fields are large compared to radiation fields both before and after amplification.  Keeping only the lowest order contributions in $\gamma\inv$, the (normalized) energy kick per pass can then be estimated as
\begin{equation}\label{energy_kick}
\Delta \gamma_{j} \approx \frac{q a_{u} N_{u}\lambda_{u}}{2 mc^{2}\beta_{0}\gamma_{0}}
\sqrt{G}\left[E_{j}\sin(\phi_{jj}) + \sum_{k \neq j}E_{k}\sin(\phi_{jk}) + \eta_j \right],
\end{equation}
where $\beta_{0}c$  and $mc^{2}\gamma_{0}$ are the average beam velocity and energy, respectively;
$E_{j}$ is the amplitude of the wiggler field produced by particle $j$ before amplification, assumed to be a sinusoidal plane wave with a rectangular envelope corresponding to exactly $N_{u}$ periods in the pickup wiggler; $\phi_{ij}$ is the relative
phase delay or advance between the transverse quiver velocity of particle $i$ and the amplified pickup radiation from
particle $j$ at the entrance to the kicker, as determined by their separation in the pickup and the bypass beam optics; $G$ is the power gain of the amplifier, assumed constant over the relevant bandwidth for wiggler radiation, and
$\eta_j$ is a stochastic variable with vanishing mean, representing the additional amplifier noise arising from thermal effects
and/or spontaneous emission. 
We have supposed that the spatial trajectory of each particle is classical, determined by the
wiggler fields and initial conditions upon entering the cooling system, \ie, neither space charge nor other collective effects,
nor recoil or multiple scattering effects from the radiation fields, will appreciably perturb the spatial trajectories of the particles on any one pass, although by design the
amplified radiation fields will perturbs the energy of the particle while in the pickup.  The sum in (\ref{energy_kick}) is taken over some effective number $N_{s}$
particles in particle $j$'s sample, where to compensate for neglecting the assumed finite
temporal extent of each particle's pickup radiation, such that any one particle is subjected to only part of the field arising from a neighboring particle while quivering in the kicker wiggler, this effective sample size will be somewhat smaller (\ie,
by some numerical factor of $O(1)$) than the actual number of particles within a coherence length $N_{u} \lambda_{0}$ of
the given particle.  We have also assumed that after amplification the optical field can be treated classically, although it can retain the (amplified) stochastic fluctuations arising from its quantum origins.

Excluding phase noise in the radiated fields, ideally the phase-delay $\phi_{jj}$ is arranged through appropriate choice of bypass optics to be a function, ideally a nearly linear function, of the deviation $\delta \gamma_j$ in particle energy:
\be
\phi_{jj} = \mu_1\, \delta\gamma_j + \mu_2 \,\delta\gamma_j^2 + \dotsc
\ee
where $\delta\gamma_j = \gamma_j - \gamma_0$ is the (normalized) energy deviation of the $j$th particle from the
reference orbit,  and the sign of $\mu_1$ is chosen so that the coherent signal on average provides a \textit{restoring} force, nudging the $j$th particle toward the reference energy.  (If transverse phase space is also to be cooled, then the phase delay will have contributions proportional to the betatron errors as well.  This is neglected here.)

The longitudinal cooling rate, or inverse cooling time-scale $\tau_{c}\inv$, may be defined as
the instantaneous rate at which RMS energy deviations in the beam are damped:
\begin{equation}
\tau^{-1}_{c} = -\smallhalf \tfrac{d}{dt}\log\mean{\delta \gamma_{j}^{2}} \approx
-\smallhalf  f_{c}  \left[\frac{\mean{\Delta\gamma_{j}^{2}} +
2\mean{\delta\gamma_{j}\Delta\gamma_{j}}}{\mean{\delta\gamma_{j}^{2}}}  \right], 
\end{equation}
where $f_{c}$ is the frequency of passes through the cooling system(s),  $\Delta\gamma_{j}$ is the cooling kick as given by (\ref{energy_kick}), and averages are performed over the current phase space distribution of the particles in the beam, and over any stochasticity in their radiation emission as well as  over any noise from the amplifier.

Assuming purely classical emission, the self-field amplitudes 
$E_{j}$ and phases $\phi_{j j}$ are deterministic functions of the single particle energy, (itself known only in a statistical or ensemble sense), while the phase differences $\phi_{i j}$ for $i \neq j$ are determined by shot noise describing the random particle positions within the beam.  Further supposing the time-of-flight delays and amplifier gain are chosen to
be locally optimal at all times during the cooling (not necessarily leading to the globally optimal minimum cooling time), and for simplicity taking particle energy deviations to be Gaussian in the lab frame, and also assuming $(N_s + N_n) \ge 1,$ the longitudinal cooling rate becomes approximately
\begin{equation}\label{cooling_rate1}
\tau_{c}^{-1} \approx f_{c} \frac{   \tfrac{1}{e} - \tfrac{2}{e^5} \frac{1}{(N_{s} + N_{n})^2}   }{1 - \tfrac{1}{e^2} + (N_{s}+N_{n})  + \tfrac{4}{e^4} \frac{1}{(N_{s} + N_{n})}  },
\end{equation}
where $N_{n}$ is proportional to the noise power $\langle \eta^{2}\rangle$ in the amplifier, expressed in terms of an
equivalent number of extra sample particles, and we have neglected corrections of order $O\left(\frac{1}{(N_{s} + N_{n})^3} \right)$ or higher.  Classically, $N_{n}$ cannot vanish, but it could in principle be made arbitrarily small by sufficiently strong pumping and simultaneously cooling of the amplifier.

Instead, if we incorporate our naive quantum mechanical intuitions about emission fluctuations by imagining that the particles emit discrete
photons in a Poissonian fashion, but continue for the moment to ignore Heisenberg phase noise, then $\phi_{j j}$ is still regarded as deterministic (in the sense explained above) but $E_{j}^{2}$ may be taken to be proportional to a Poissonian random variable with some probability $p_{\stext{rad}} \sim \alpha$ of photon emission per pass.  Since $p_{\stext{rad}}  \ll 1$ in our regime,  we can approximate each Poissonian random variable by a binomial variable, effectively neglecting the very rare emission of two or more photons, and we find after some algebra that in this case the optimized cooling rate becomes
\begin{equation}\label{cooling_rate2}
\tau_{c}^{-1} \approx p_{\stext{rad}}\, f_{c} \frac{ \tfrac{1}{e} - \tfrac{2}{e^5}\frac{1}{(N_{s} + N_{n})^2}}{1 - \tfrac{1}{e^2} + (N_{s}+N_{n})  + \tfrac{4}{e^4} \frac{1}{(N_{s} + N_{n})}},
\end{equation}
which as expected is indeed slower by a factor of $p_{\stext{rad}} \sim  O\left(\alpha^{-1}\right)$ than the fully classical prediction.  This drastic slow-down occurs, of course,  because the coherent cooling signal is by assumption only present on average in a fraction $O( p_{\stext{rad}}) \sim O(\alpha)$ of passes through the cooling section.

While throughout our simple analysis we have assumed linearity of the cooling kicks in the fields and particle deviations, in which case stochastic cooling can work at some rate regardless of the signal-to-noise-ratio (SNR), nonlinear effects might lead to instabilities beyond some finite range of SNRs, so it is also possible that this large multiplicative noise might not just drastically slow down cooling but frustrate cooling altogether.

Next, to incorporate the phase noise in addition to the Poissonian emission into our model, it seems natural to also treat the $\phi_{j j }$
as random variables with conditional means determined as above, but subject to random fluctuations, ostensibly with RMS deviations
\be
\langle \delta \phi_{j j}^{2} \rangle^{1/2} \sim \tfrac{1}{2 \sqrt{N_{\stext{ph} }}} \sim O\left(\alpha^{-1/2}\right),
\ee
which are comparable to $2\pi$.  Taking these assumed phase fluctuations to be Gaussian for simplicity, the cooling rate becomes approximately
\begin{equation}\label{cooling_rate3}
\tau_{c}^{-1} \approx p_{\stext{rad}}  f_{c}  e^{-\langle \delta \phi_{j j}^{2} \rangle}\frac{1/e}{1 -e^{-2(\langle \delta \phi_{j j}^{2}
\rangle +1)} + (N_{s}+N_{n})},
\end{equation}
to leading order in the small quantities $e^{-\langle \delta \phi_{j j}^{2} \rangle}$ and $\frac{1}{N_{s} + N_{n}}$.
With such phase uncertainty, the cooing rate would be suppressed by more than a factor of 
$e^{-\langle \delta \phi_{j j}^{2} \rangle} \sim O(10^{-3})$, because the cooling information carried
by the phase of the self-fields has been corrupted with intrinsically quantum noise.

Finally, quantum mechanics will enforce a lower bound for the amplifier noise  $N_{n} \approx \tfrac{1}{2}\langle \eta(t)^2\rangle$ due to unavoidable spontaneous emission in the pumped medium responsible for the gain.  Various quantum mechanical and semi-classical arguments (see for example, \cite{yariv:1989} for a survey) suggest a minimum amount of added noise equivalent in its  final effects to one-half photon per mode entering the amplifier along with any actual signal present and amplified along with it.  With our simple model of windowed plane waves,  and an effective interaction length in the kicker (accounting for slippage between the particles and fields)  equal to the coherence length $N_{u}\lambda_{0}$ of the radiation, each particle effectively interacts with a single mode, so that one
would predict that at best $N_{n} \sim \half \frac{1}{\EuScript{N}_{ph}} \sim O(\alpha^{-1}).$  While not appearing as a
multiplicative slow-down in the cooling rate and therefore not as devastating as the other possible quantum effects described
above, such spontaneous emission noise does indicate that for small sample sizes quantum fluctuations may be as important a contribution to the incoherent heating as classical shot noise from actual particles, and that these fluctuations will ultimately limit the cooling gains achievable by diluting or stretching the beam.  So indeed we can we can roughly account for
the spontaneous emission by increasing the effective sample size from $N_{s}$ to $N_{s} + N_{n}$ for some effective noise number $N_{n} \sim N_{\stext{amp}}/\EuScript{N}_{\stext{ph}},$ where $N_{\stext{amp}}$ is a measure of the  effective added noise by the amplifier
expressed as an equivalent number of photons at the amplifier front-end (\ie, prior to amplification. )

\section{Towards a More Careful Treatment of Quantum Effects}

Fortunately, a more careful quantum mechanical analysis will reveal that most of these naive intuitions are incorrect, and that in effect only the additive amplifier noise is present, so that cooling rates are accurately estimated by classical continuous
emission results, provided allowance is made for the unavoidable amplifier noise, which is ultimately quantum
mechanical in origin, and cannot be made arbitrarily small without violating the Heisenberg Uncertainty Principle or the unitary nature of quantum dynamics.  That is, in a more 
careful quantum mechanical analysis neither the multiplicative emission-noise leading to the $O(\alpha)$ slow-down in the
naive Poissonian-emission model (\ref{cooling_rate2}), nor the catastrophic slowdown appearing in the naive phase-noise model
(\ref{cooling_rate3}) actually occurs; in effect, only additive noise arising from spontaneous
emission or thermal effects in the amplifier appears, leading to a cooling rate of the approximate form (\ref{cooling_rate1}) for some effective noise number $N_{n},$ whose minimum value is essentially constrained by the uncertainty principle.

A more rigorous treatment of OSC will force us to examine carefully each stage of the cooling dynamics: the particle
motion in the pickup wiggler, the resulting radiation emitted, the quantum mechanics of the optical amplification process,  the
particle-radiation interaction in the kicker wiggler, and the resulting changes in the beam phase-space distribution.  Because we
are here interested in a proof-of-principle question rather than a detailed design assessment, we will make a number of
simplifying assumptions, yet still incorporate the essential classical, quantum, and statistical physics of the processes so as to
address the fundamental question of whether optical stochastic cooling based on amplification of small
pickup signals is intrinsically flawed due to the effects of quantum noise, and ultimately arrive at a less pessimistic answer than what we arrived at above.

A fully self-consistent, quantum mechanical (or worse, QED-based) treatment of beam particles, radiation fields, amplifiers and other optical elements would be prohibitively difficult, but fortunately is not actually necessary either.  Rather than explaining the quantum mechanical features 
of the dynamics of the beam particles, we will in effect carefully explain them away, arguing that particles in the beam can be treated classically in their interaction with the wiggler, radiation, and any external focusing fields.

Then we will verify that such particles do not radiate into photon number states, but rather Glauber coherent states, which are actually the states closest to classical radiation fields allowed by quantum mechanics,  and quite different in their statistics from the states consisting of whole numbers of photons.  Making certain idealizations, The radiation essentially retains this form as it passes through dielectric optic elements. 

Of course, the inverted population of atoms in the active medium of a laser amplifier behaves in a highly non-classical manner, but the precise dynamics of the amplifier need not be evolved explicitly.  Very general considerations of amplifier action will be sufficient to
determine the action of the amplifier on the pickup field in an ``input-output'' formalism where the specifics of the intermediate dynamics can be neglected, and to characterize the best-case limits on the additional noise introduced of the amplification process without resorting to an explicit microscopic model of the amplifying medium.

Once amplified, the radiation behaves entirely like a classical but noisy field, both in its statistics and its interaction with the beam particles in the kicker.

So once these results are established, a simple estimate of the cooling rates can be made essentially along classical lines, just accounting for any extra noise in the field due to amplified spontaneous emission.

\section{Particle Dynamics are Classical}

For beams of electrons, muons, or protons,  at  relevant energies and emittances,  and
with realistic wiggler strengths, the de Broglie wavelengths associated with the particles' longitudinal and transverse motion
are extremely small compared to the radiation wavelength, wiggler period,  beam dimensions, and other relevant scales, so we
will argue that the particles can be treated as classical point particles obeying classical relativistic kinematics.  The subsequent
analysis of the radiation will be vastly simplified by assuming the dynamics of the particles in the pickup wiggler to be classical, and in fact with prescribed classical trajectories determined by external fields only, so the effort needed to carefully establish and justify the accuracy of these assumptions will be subsequently rewarded.

Clearly, in order to adequately describe particles classically, all \textit{statistical and dynamical} manifestations of quantum-mechanical or quantum-electrodynamical effects on the particle DOFs must be negligible.   Often,
statistical and dynamical effects are not clearly distinguished, but it not difficult to find regimes where either the classical
statistical limit or classical dynamical limits may be valid, but not both.  However, it is often the case for a
system consisting of many particles that classical statistical noise can also tend to mask the dynamical manifestations of
quantum noise.  Both limits of course are related in part to the size
of the typical de Broglie wavelength $\lambda_{dB}$ associated with a quantum mechanical particle, but
ignoring quantum statistical effects typically requires that $\lambda_{dB}$ be small compared to average inter-particle spacings, while
ignoring quantum dynamical effects typically requires that $\lambda_{dB}$ remain small compared to all relevant dynamical length
scales.  For example, in a plasma or ionized gas at rest (on average),  if $\lambda_{dB} \sim \frac{h}{\sqrt{m k_{\stext{B}}T}}  \ll
n^{-\frac{1}{3}},$ where $n$ is the particle number density, $T$ is the temperature, and $k_{\stext{B}}$ is Boltzmann's constant,
then quantum statistical effects can be ignored and the particles can be
taken to satisfy Maxwell-Boltzmann statistics,  as if they were classical point particles, whether they are identical fermions or
bosons.  But unless $\lambda_{dB} \ll \ell_{\stext{L}}$ where $\ell_{\stext{L}}
\sim \frac{q^2}{k_{\stext{B}}T}$ is the Landau length, or typical distance of closest approach between two plasma particles, then
the Coulomb scattering between particles should be treated quantum mechanically. 

\subsection{Quantum Statistical Degeneracy}

In the case at hand of a relativistic, charged particle beam of indistinguishable fermions, the safe neglect of quantum statistical effects
requires that the beam be non-degenerate in the average rest frame:
\be
(\rho_b')^{-\frac{1}{3}} \gg \lambda_{dB}',
\ee
where $\lambda_{dB}' \sim \frac{h}{\sqrt{m k_{\stext{B}}T'}}$ is the typical thermal deBoglie wavelength, $T'$ is an effective
temperature, which may be different for the longitudinal and transverse DOFs, and $m$ is the invariant particle rest mass.
We will assume throughout that the average beam energy $mc^2 \gamma$ and beam energy spread $mc^2 \delta\gamma$  
satisfy $\gamma \gg 1$ and $\tfrac{\delta \gamma}{\gamma} \ll 1$, even before stochastic cooling is applied.  Since $a_u \sim O(1),$
this also implies that the longitudinal momentum $\tilde{p}$ inside the wiggler is only negligibly smaller than the average beam momentum $p = mc\gamma \beta$ outside the wiggler.  Some of the longitudinal beam kinetic energy is converted into energy
of  transverse quiver in the wiggler magnetic field, but assuming that transverse canonical momentum is conserved in the
presumed transversely-uniform wiggler fields, and that particles enter the wiggler approximately on axis, one finds 
\be
\tfrac{\tilde{p}}{p} \sim 1 -\tfrac{1}{4}\tfrac{a_u^2}{\beta^2\gamma^2},
\ee
while by assumption $\beta \sim 1$ and $\frac{a_u}{\gamma} \ll 1$, so we will ignore this distinction when convenient.

Since motion in the wiggler field will not significantly effect the beam temperature, average density, or energy, for simplicity
we can analyze the beam while it drifts freely, prior to its entering the pickup wiggler.   In this limit of high energy and
relatively low energy spread, one finds from relativistic velocity addition that the rest-frame longitudinal temperature can be
expressed as 
\be
k_{\stext{B}}T'_{\parallel} \sim mc^2 \beta_{\parallel}'^2
\sim mc^2\frac{1}{\beta^2\gamma^2}\left(\delta\gamma \right)^2,
\ee
where $\beta c$ is the  average beam velocity in the lab frame, with $\beta \sim 1$,  and $\delta\gamma \approx
\frac{\delta p_{\parallel}}{mc}$ is the lab-frame beam energy spread.  Since momenta perpendicular to a Lorentz boost remain
unchanged, the transverse rest-frame temperature can be written as
\be
k_{\stext{B}}T'_{\perp} = \frac{ \mean{p_{\perp}'^2}}{2m} \approx \frac{\delta p_{\perp}^2}{m}.
\ee
where $\delta p_{\perp}$ is the root-mean-squarer transverse momentum spread in the lab frame, related to the conventional
transverse beam emittance $\epsilon_{\perp} \sim \frac{\sigma_{\perp}\delta p_{\perp}}{mc\beta\gamma},$ where
$\sigma_{\perp}$ is the transverse (radial) beam size.  From Lorentz contraction, it follows that the
particle density must transform as $n_b' = n_b/\gamma,$ so the conditions for non-degeneracy, written in terms of
lab-frame quantities, become 
\begin{equation}\label{stat_class_long}
\rho_b \ll \gamma \frac{1}{\lambda_c^3} \left[ \frac{\delta \gamma}{\gamma} \right]^{3} \sim \left( \frac{\epsilon_{\perp}}{\lambda_c}\right)^3 \frac{\gamma}{\sigma_{\perp}^3},
\end{equation}
and
\begin{equation}\label{stat_class_trans}
\rho_b \ll  \gamma \frac{\delta p_{\perp}^{3} }{h^{3}} = \gamma \frac{1}{\lambda_c^{3}}\left[\frac{\delta
p_{\perp}}{mc}\right]^{3}
\end{equation}
where $\lambda_c = \frac{h}{mc}$ the Compton wavelength.  For achievable beam densities and emittances, even after
cooling, these conditions are typically easily achieved by many orders of magnitude; existing particle beams are extremely far from
being degenerate, and quantum statistical effects arising from the Pauli Exclusion Principle can be completely ignored.

\subsection{Pair Creation or other QED Effects}

Although the particles are assumed to be highly relativistic in the lab frame, they are all streaming in the same general
direction, with small relative variations in momentum, so particle-scattering will involve small center-of-momentum energies,
and hence particle/anti-particle pair creation or other exotic QED scattering effects should be negligible.  In the average rest
frame, this requires 
\be
mc^2 \langle \beta'^2 \rangle \ll 2mc^2,
\ee
equivalent in the lab frame (after dropping some factors of two) to the requirement that
\begin{equation}\label{class_noQED}
\frac{\delta \gamma}{\gamma} \ll 1, 
\end{equation}
which has already been assumed.  Of course, muons will spontaneously decay in a random, Poissonian
fashion, which is a completely non-classical process mediated by electroweak interactions, but the resulting electrons will be
lost from the beam as soon as it is bent in magnetic fields calibrated for the heavier muons, so this process can be ignored for
particles which remain in the beam throughout the full cooling process, requiring several turns in the cooling ring.

\subsection{Spin Effects}

Since spin degrees of freedom are intrinsically quantum mechanical, at least for apparently non-composite particles such as muons or electrons \footnote{While the energy of a spin state can be made arbitrarily large by increasing the magnetic field, the \textit{action} associated with any spin states of a single lepton is bounded in magnitude by  $\tfrac{\sqrt{3}}{2} \hbar$ and so cannot participate in any sort of Correspondence Principle limit.  See the discussion in the sequel.}, all spin effects should be negligible in a truly classical treatment.  The energy associated with the spin of an elementary fermion in the wiggler field $B_{u}$ is $U_{spin} \sim \frac{q \hbar}{2mc} B_{u} = \frac{1}{2} a_u \hbar
k_{u} c.$  In order to safely neglect spin effects, this energy
$U_{spin}$  should be smaller than all other relevant energy scales, including: the total particle particle energy $\gamma
mc^2;$ particle quiver kinetic energy, which for $\gamma \gg 1 $ and $a_u \sim O(1)$ is about
$\half \gamma m v_{\perp}^2=\half \gamma m \left[ c\tfrac{a_u}{\gamma}\right]^2 = 
\half m c^2\tfrac{a_u^2}{\gamma};$ the mean photon energy $\hbar\omega_0;$ and the total radiated energy per
particle, $U_{rad},$ which is larger by about $O(N_u)$ than the power radiated into the coherent mode.  Using the
relativistic Larmor formula, the total radiated energy per particle in the pickup wiggler can be estimated as
\begin{equation}\label{radiation_per_particle}
U_{rad} = P_{rad}\Delta t = \frac{2}{3}\frac{q^2}{m^2 c^3 }\frac{d p^{\mu}}{d\tau}\frac{d
p_{\mu}}{d\tau} \frac{N_u\lambda_u}{c}
\approx \frac{q^2}{c} \gamma^4 \dot{\beta}^2 \frac{N_u\lambda_u}{c} \approx  
\alpha N_u a_u^2\hbar\omega_0.
\end{equation}
These requirements imply, respectively, that
\bsub
\begin{align}
\hbar \omega_0 &\ll  mc^2 \frac{ \gamma^3 }{ a_u };\\
\hbar \omega_0 &\ll \gamma a_u m c^2;\\
\gamma^2 &\gg a_u;\\
\gamma^2 &\gg \frac{1}{\alpha N_u a_u}.
\end{align}
\esub
Since $\hbar\omega_0 = \frac{\lambda_c}{\lambda_0}mc^2,$ the first two
conditions can be written equivalently as $\lambda_c \ll \frac{\gamma^3}{a_u} \lambda_0$ and 
$\lambda_c \ll \gamma a_u \lambda_0.$ For optical or near infrared frequencies, a wiggler parameter $a_u \sim
O(1)$, $N_u \sim O(10)$ wiggler periods, and $\gamma \sim O(10^2)$ or $O(10^{3}),$ these conditions are all
easily satisfied. 

To be thorough, we should examine not just energies but forces associated with spin DOFs.  As pointed out by Bohr
and Mott as early as 1929 \cite{mott:1929}, consistency of the Correspondence Principle demands that any non-composite particle whose
spatial DOFs are behaving classically should also exhibit no observable spin effects, because the magnitude of the
spin and associated action is bounded, and there is therefore no way to consider its classical (high-action) limit independently from that of the spatial DOFs.  Evidently, this suppression of spin effects must be enforced by a ``conspiracy'' between the Heisenberg Uncertainty Principle and the Maxwell-Lorentz equations\footnote{Incidentally, this makes the prospects for using optical stochastic cooling to perform beam polarization look rather bleak.  Actually, partially spin-polarized particle beams have been achieved, and would seem to offer a counter-example to the Bohr-Mott Conjecture, except that only collective spin observables on many particles are ever really measured, which can behave classically.}.  In particular, for charged elementary fermions, the magnetic field produced by the spin of one particle and felt by another particle will be swamped by the magnetic field produced by the moving charge associated with the first particle.  At a characteristic distance $r' \sim n'^{-\frac{1}{3}},$ in the average rest frame, the
magnetic field due to one spin is roughly
$B_{\stext{spin}}' \sim \frac{q \hbar}{2mc}\frac{1}{r'^{3}},$ while the magnetic field due to the same particle's motion will
scale like $B_{v}' \sim \frac{q v'}{c}\frac{1}{r'^2}$.  If the uncertainty in position
$\delta r'$ satisfies $\delta r' \ll r',$ consistent with the particle's spatial DOFs behaving classically, then it can
easily be shown that $B_{\stext{spin}}' \ll \delta B_{v'}',$ where $\delta B_{v'}'$ is the uncertainty in the Biot-Savart field
$B_{v'}'$ due to the Heisenberg uncertainty in the velocity $\bv{v}'.$  Therefore, spin-spin effects will be negligible in a
non-degenerate beam of charged leptons.

Although ideally the magnetic field of the wiggler is taken to be transverse, transversely uniform, and plane polarized,  such
spatial variation is not strictly consistent with Maxwell's equations, which demand that the wiggler magnetic fields are both
curl-free and divergence-free in the neighborhood of the beam line, requiring in addition to the principle sinusoidal transverse
component at least some small longitudinal component and some transverse gradients with scale-length
$O(\lambda_0)$ in the neighborhood of the beam axis. With such self-consistent field profiles, it can be directly shown that
the transverse component of the dipole force on the spin,  given by, $\bv{F}_{dp} = \frac{g \hbar}{2
mc}(\bv{s}\cdot\grad)\bv{B}_{u},$ will be small compared to the transverse component of total Lorentz force on the
charge, which is given by  $\bv{F}_{L}
\frac{q}{c}\bv{v}\times\bv{B}_{u},$  if
\bsub
\begin{align}
\delta v_{\perp} &\ll v_{z},\\ 
\hbar k_0 &\ll \gamma^2\beta mc,
\end{align}
\esub
where $\delta v_{\perp} \approx \frac{1}{\gamma m}\delta p_{\perp}$ is the transverse velocity spread, and $k_0 =
\frac{2\pi}{\lambda_0}$ is the central optical wavenumber; while the longitudinal dipole force on the spin will be
negligible compared to the longitudinal component of the Lorentz force, provided 
\bsub
\begin{align}
\hbar\omega_0 &\ll \gamma mc^2,\\
\frac{1}{\gamma} N_u^2a_u\lambda_c &\ll \lambda_0,\\
\hbar k_0 &\ll \frac{1}{N_ua_u}\gamma^3\beta mc,
\end{align}
\esub
all readily satisfied.  Spin degrees-of-freedom are here unimportant to all aspects of the particle dynamics.

\subsection{Transverse Motion}

Next, we argue that quantum mechanical effects in the transverse motion are unimportant.  In order that this be the case, the
spread in a particle's transverse wavefunction must remain small relative to other relevant length-scales,
including the transverse beam size, the amplitude of the transverse quiver motion, the range of transverse variation of the
wiggler field, as well as the optical wavelength.  Also, the transverse velocity fluctuations associated with quantum
mechanical uncertainty demanded by the Heisenberg uncertainty principle must remain small compared to the transverse
quiver velocity and to the transverse velocity spread in the beam.  Neglecting any initial transverse particle velocity and any
transverse spatial variation in the wiggler fields, conservation of canonical momentum implies that the transverse quiver
velocity will be $v_{\perp} \sim c \frac{a_u}{\gamma}$.   Since we assume $\frac{a_u}{\gamma} \ll 1,$ the
transverse quiver motion can be taken to be non-relativistic, provided we replace the rest mass $m$ everywhere with the
``effective'' relativistic mass $\gamma m$ to account for the increase in inertia arising from the high longitudinal velocity in the
lab frame.  Furthermore, we will neglect the transverse forces due to the wiggler fields or external beam-optical fields.  Because
classically these fields will provide focusing (at least in one direction) on average, semi-classical considerations suggest that
they will act mostly to inhibit the spread of the wavefunction, so ignoring them altogether should constitute a conservative
approximation.

To get an idea of the scaling, we suppose each particle is described by a Gaussian wavepacket, initially (at $t = 0$) separable in the transverse ($x$ and $y$) coordinates, and having some minimum initial variance in each component of position and momentum consistent with the Uncertainty Principle, but then freely streaming during the total
interaction time $\Delta t \approx \frac{N_u\lambda_u}{\beta c} \sim  \frac{N_u\lambda_u}{c} $ in the wiggler.
The dynamics for the different transverse directions are identical, so we need follow only one component, say in the $\unitvec{x}$ direction.
The variance in position at a subsequent time $t$ will be given by
\begin{equation}\label{var_x}
\Delta x^2(t) =  \Delta x^2(0) + \tfrac{1}{4} \tfrac{\hbar^2t^2}{\gamma^2m^2 }\Delta x^2(0) = \Delta
x^2(0) + \tfrac{\Delta p_{x}^2(0) t^2}{\gamma^2m^2},
\end{equation}
while the variance in momentum,
\begin{equation}\label{var_px}
\Delta p_{x}^2(t) =  \Delta p_{x}^2(0),
\end{equation}
is constant, since transverse forces have been neglected, and we have further assumed that the initial conditions satisfy
\begin{equation}
\Delta x(0) \Delta p_{x}(0) \approx \frac{\hbar}{2}.
\end{equation}
By differentiating the expression for $\Delta x^2(t)$  with respect to $\Delta x^2(0)$, one finds that the spatial variance is minimized at any subsequent time $t > 0$ by 
\begin{equation}
\Delta x^2(t) =  \smallhalf \tfrac{\hbar}{\gamma m} t.
\end{equation}
Such a wave-packet with minimal spatial spread is the closest thing to a classical point-particle
allowed by quantum mechanics, and evaluating the spatial variances at $t = \Delta t,$  we have 
\begin{equation}
\Delta x^2 (\Delta t) \sim  \tfrac{1}{4\pi}\tfrac{N_u\lambda_c\lambda_u}{\gamma}.
\end{equation}
In order that the particle behavior remain classical during this interaction time, 
this variance should be sufficiently small in several senses: it is necessary that this
transverse spatial spread be small compared to the transverse beam dimensions, \ie, that $\Delta x(\Delta t ) \ll \sigma_{\perp}$, or 
\begin{equation}\label{class_trans}
\lambda_c\lambda_u \ll \tfrac{4\pi}{N_u}\gamma \sigma_{\perp}^2;
\end{equation} that it be small compared to the transverse extent of individual particle orbits, \ie, that
 $\Delta x(\Delta t) \ll  c \tfrac{a_u}{\gamma}\tfrac{\lambda_u}{c}$, or
\begin{equation}\label{class_trans2}
\lambda_c \ll \tfrac{4\pi}{N_u}\tfrac{a_u^2}{\gamma}\lambda_u;
\end{equation}
that it be small compared to the \textit{transverse} range of variation for the wiggler magnetic field, which from
Maxwell's equations (specifically, $\curl\bv{B} = \bv{0}$) will be comparable to the longitudinal length-scale of field
variation, namely $\lambda_u$;  \ie, that  $\Delta x(\Delta t) \ll \lambda_u$, or 
\begin{equation}
\lambda_c \ll \tfrac{4\pi}{N_u} \gamma \lambda_u;
\end{equation}
and finally, in order that the particle's phase in the radiation field be well-defined, we require that the phase uncertainty associated with this transverse uncertainty in the presence of wavefront curvature remain small: $k_0\, \delta\theta \Delta x \ll \pi$, or roughly
\be
\lambda_c \ll \tfrac{\pi}{2} \lambda_0.
\ee
For sufficiently long wigglers ($N_u \gg 10$), some of these conditions could be violated, but in that case a more sophisticated analysis including transverse focusing effects would be needed to assess the classicality of the particle orbits.  In our regime ($N_u \lesssim O(10)$) they are met.

So far we have followed the conventional reasoning, but we actually need to be more careful, because up to now we have only considered spatial spread, but in fact the wave-packet that minimizes transverse spatial variances has arbitrarily large variance
in transverse momenta as $t \to 0$, violating the requirement that quantum mechanical transverse momenta uncertainties
should also be small compared to the transverse momentum spread in the beam and to the wiggler-induced quiver
momentum.   It turns out that with a lengthy calculation for an appropriate Gaussian wavepacket, the uncertainties in both transverse position and momentum can be made sufficiently small simultaneously provided the above conditions remain satisfied, and in addition 
\bsub
\begin{align}
\tfrac{\lambda_c}{\lambda_0} &\ll 4\pi \tfrac{\delta p_{\perp}}{mc},\\
\tfrac{\lambda_0}{\lambda_c} &\ll 4\pi a_u,
\end{align}
\esub
which are readily satisfied.

Including the effects of the wiggler fields should not change these conclusions, provided the wiggler field itself acts classically.   In the average rest frame, the wiggler fields  can be treated via the Weizs{\"a}cker-Williams approximation as a traveling electromagnetic plane wave consisting of virtual photons, which then scatter off the particles to produce the real wiggler radiation.  Over the length of the wiggler and the transverse area of the coherent radiation, the number of virtual wiggler photons is very large:
\be
\alpha\inv N_u^2 \tfrac{a_u^2}{\gamma^2} \tfrac{\lambda_u^2}{\lambda_c^2} \gg 1.
\ee
Actually, even over the much smaller volume defined by the wiggler length, the transverse extent of the particle's quiver, and the classical muon radius $r_{\mu}$,  the number of virtual wiggler photons is still large:
\be
N_u \tfrac{a_u^3}{\gamma}  \tfrac{\lambda_u}{\lambda_c} \gg 1.
\ee           
Additionally, by definition the wiggler field is very coherent, with ``quantum'' uncertainties in the virtual photon number density very small in a relative sense, and uncertainties in phase very small compared to                                    $2\pi$.  One could not demand more from what is regarded as a classical field in what is really a quantum mechanical world.

\subsection{Longitudinal Dynamics}

In a similar manner, we can show that the longitudinal dynamics can be taken to be classical, although unlike the transverse
motion, the longitudinal motion is highly relativistic.  For the moment, we will assume that the particles are freely streaming
longitudinally, unperturbed by wiggler, space-charge, radiation, or other fields.  Because spin, ZBW, and other relativistic
quantum effects can be neglected, we can fortunately forego use of the Dirac or even Klein-Gordon equations, and simply use a
one-dimensional Schrodinger equation with the positive-energy branch of the relativistic dispersion relation:
\begin{equation}\label{rel_omega}
\hbar\omega(k) = \sqrt{m^2c^{4} + c^2\hbar^2k^2} - mc^2.
\end{equation} We assume the initial state is a minimum-uncertainty Gaussian wavepacket:
\be
\psi(z,\,t = 0) = \tfrac{1}{\sqrt{\sqrt{2\pi}\Delta z(0)}}e^{i\bar{k}z}e^{-\frac{z^2}{4\Delta z^2(0)}},
\ee
where
$\Delta z(t)^2$ is the longitudinal spatial variance at time $t,$ and $\hbar\bar{k} = mc\gamma\beta,$ is the average
longitudinal beam momentum.  Taking a Fourier transform, this can be written as
\be
\psi(z,\,  t=0) = \frac{1}{\sqrt{2\pi}}\!\int \!dk\, \psi(k) e^{i k z},
\ee
where
\be
\psi(k) = \left[\tfrac{2 \Delta z (0)}{\sqrt{2\pi}} \right]^{\half}e^{-\left(k - \bar{k}\right)^2\Delta z^2(0)}.
\ee
Using the dispersion relation (\ref{rel_omega}), the freely-propagating solution at later times can then be written as
\begin{equation}\label{rel_wave}
\psi(z,\,t) = \tfrac{1}{\sqrt{2\pi}}\!\int \!dk \, \psi(k) e^{i k z - i\omega(k) t}.
\end{equation}
Since $\frac{\delta \gamma}{\gamma} \ll 1,$ the relativistic dispersion relation (\ref{rel_omega}) can be
expanded in a Taylor series about $\bar{k}:$
\be
\begin{split}
\omega(k) &\approx 
\tfrac{mc^2}{\hbar}\left[ \gamma -1 \right] + c\beta\left(k - \bar{k}\right)  + \smallhalf \tfrac{\hbar}{\gamma^3 m}\left(k
-\bar{k}\right)^2\\
&= \tfrac{mc^2}{\hbar}\left[\tfrac{1}{\gamma} -1\right] + c\beta k  + \smallhalf \tfrac{\hbar}{\gamma^3
m}\left(k -\bar{k}\right)^2 + \ldots.
\end{split}
\ee
After a little algebra, we find the longitudinal variances to be
\begin{equation}
\Delta p_{z}^2(t) = \Delta p_{z}^2(0) = \tfrac{1}{2}\tfrac{\hbar^2}{\Delta z^2(0)},
\end{equation} and
\begin{equation}
\Delta z^2(t) \approx \Delta z^2(0)  + \tfrac{1}{4}\tfrac{\hbar^2t^2}{m^2\gamma^{6} \Delta z^2(0)} = 
\Delta z^2(0)  + \tfrac{\Delta p_{z}^2(0) t^2}{m^2\gamma^{6}}.
\end{equation}
Differentiating, we find that the spatial spread is minimized at any subsequent time $t > 0$ by the choice
\begin{equation}\label{min_z}
\Delta z^2(t) = \smallhalf \tfrac{\hbar t}{m \gamma^3}.
\end{equation}
Over the interaction time $\Delta t \approx \frac{N_u\lambda_u}{c},$ this spread $\Delta z(t)$ must remain small compared to the smallest length-scale associated with variations in longitudinal forces experienced by the particle, namely the optical wavelength
$\lambda_0$, so that the particle can behave like a point particle and can be taken to have  a well-defined phase in the
optical field.

That is, classical behavior requires $\Delta z(\Delta t) \ll \lambda_0$, or
\begin{equation}\label{class_long1}
\tfrac{N_u}{4\pi} \tfrac{\lambda_c\lambda_u}{\gamma^3} \ll \lambda_0^2.
\end{equation}
Since $\lambda_u \sim \gamma^2 \lambda_0$, this condition is usually written in the form  
\begin{equation}\label{class_long2}
\tfrac{N_u}{4\pi} \tfrac{\lambda_c}{\gamma} \ll \lambda_0.
\end{equation}  
In the FEL literature, this condition (\ref{class_long2}) is often the \textit{single requirement} mentioned for allowing a classical treatment of electron dynamics, but again rather more care is needed, because both position and momentum uncertainty must remain sufficiently small for a classical treatment to be accurate.   Consistency demands that the quantum mechanical uncertainty in
particle momentum $\Delta p_{z}(t)$ at least remain negligible compared to the classical longitudinal momentum spread of the
beam, $\delta p \approx \frac{\delta\gamma}{\gamma} p \approx mc \,\delta\gamma$.  For a plane-polarized wiggler field,
another small longitudinal momentum scale exists, namely the extent of longitudinal momentum variation due to the ``figure-eight'' particle orbits in the plane-polarized wiggler fields, which, from energy conservation, can be shown to be approximately $\delta p_{1} \approx \frac{a_u^2}{\gamma^2} \,p_0$ in terms of momentum excursions in the lab frame.  A little algebra reveals that uncertainties in position and momentum can be made simultaneously negligible provided (\ref{class_long2}) is satisfied, as well as the conditions
\bsub
\begin{align}
\tfrac{\lambda_c}{\lambda_0} &\ll 4\pi \,\delta\gamma\, \label{transpread3a}\\
\tfrac{\lambda_c}{\lambda_0} &\ll 4\pi\, \tfrac{a_u^2}{\gamma}. \label{transpread3b}
\end{align}
\esub
We have so far neglected the effects of any longitudinal forces.  The fast harmonic motion along $z$ in a
planar wiggler should only slow the spatial spreading, or at least not significantly increase it, as should any ponderomotive
bunching in the electromagnetic fields.   Coulomb repulsion from the space-charge fields will be de-focusing, and might
lead to additional spreading of the wave-packets, but classically this force is typically negligible compared to the other forces in the present parameter regime, as we will see shortly.

\subsection{Radiation Reaction}

However, in both the longitudinal and transverse dynamics, we have so far ignored a potentially important force, that of
radiation reaction, or direct recoil.  In order to consistently treat the particles classically but the radiation emission quantum mechanically,
the effects of radiation reaction, or particle recoil, must be negligible\footnote{Curiously, there is an apparently consistent
formalism for evolving matter quantum mechanically but radiation classically which includes radiation reaction effects, the
so-called neoclassical radiation theory of E. T. Jaynes, which can capture many effects, such as the photoelectric effect and the Lamb shift, that are purported to be evidence for the quantum nature of \textit{light}.   However, some predictions of quantum optics, such as perfect anti-correlation after a beam-splitter for single photon states, or photon polarization states that violate Bell's inequalities, cannot be described in this model.  Anyway, here we are concerned with the opposite ``hemi-classical'' limit, with particles treated classically and radiation
quantum mechanically.}.  Intuitively, we expect that the total effect on any particle due to the spontaneous wiggler radiation fields, including  self-fields, should be negligible simply because the radiation field strengths are much weaker than those of the external wiggler fields, \ie, $a_{1} \ll a_u$, where
$a_{1}$ is the normalized vector potential for the spontaneous wiggler radiation.  Using (\ref{radiation_per_particle}) as an
estimate for the total power radiated per particle, assuming each particle emits into a cone with synchrotron opening angle $\delta\theta \sim \frac{1}{\gamma}$ (larger than the coherent opening angle by $\sqrt{N_u}$), and assuming the particle positions are governed by shot noise, so the emission from different particles is randomly phased and therefore adds incoherently in intensity, we find that $a_{1} \ll a_u$ provided
\begin{equation}
\alpha^2 N_u
\left[n \sigma_{\perp}^2\lambda_c\right] \tfrac{\lambda_c}{\lambda_u} \sim 
\tfrac{1}{\gamma^2}\alpha^2N_u n \sigma_{\perp}^2\lambda_c \tfrac{\hbar\omega}{mc^2} \ll 1.
\end{equation}
While easily satisfied in the parameter regimes of interest here, this is essentially a far-field condition, while the radiation reaction force is manifestly a near-field effect, so more care is again needed.

In order that we can ignore recoil effects in the
longitudinal motion, the average energy of radiation per particle should be small compared to the average particle energy, \ie, 
\begin{equation}
\alpha N_ua_u^2\hbar \omega_0 \ll \gamma mc^2,
\end{equation}
or equivalently
\begin{equation}
\alpha \,a_u^2\tfrac{N_u\lambda_c}{\gamma} \ll \lambda_0.
\end{equation}
Because the observed power radiated per particle (if it could be isolated) would be subject to large (Poissonian)
fluctuations, we should really demand that the energy of any single radiated photon be small compared to the average particle energy:
\begin{equation}
\hbar \omega_0 \ll \gamma mc^2,
\end{equation} or equivalently
\begin{equation}
\tfrac{1}{\gamma} \lambda_c \ll  \lambda_0.
\end{equation}
Actually, we should impose still stricter conditions.  In the average rest-frame, the energy change due to 
the recoil from a particle scattering one virtual wiggler photon into one real radiation photon ought to be small compared to the typical RMS kinetic energy of a particle.  Because $\delta \gamma' \approx 2 \tfrac{\delta\gamma}{\gamma}$,  back in the lab frame, this requirement entails 
\be
\tfrac{\lambda_c^2}{\lambda_u^2} \gamma^2 \ll \tfrac{\delta \gamma}{\gamma},
\ee
setting an ultimate bound on how cold the particle beam can be or become and retain classical behavior, which is nevertheless easily satisfied for any remotely accessible beam temperatures.
Since the particle cannot actually absorb energy from the static wiggler magnetic field,  perhaps we should even more strongly require that the energy of a single emission or absorption event by a particle (ignoring inconsistencies with momentum conservation) be small compared to the RMS particle kinetic energy, in the average rest frame.  That is, we demand $\hbar \omega' \ll m c^2 \delta \gamma'$, or
\be
\tfrac{\lambda_c}{\lambda_u} \gamma \ll \tfrac{\delta \gamma}{\gamma},
\ee
which apart from neglected factors of $2$ amounts to the same thing as requiring that the lab-frame radiated photon energy is small compared to the lab-frame energy spread, \ie,
\be
\hbar \omega_0 \ll mc^2 \delta \gamma,
\ee
which is reassuring since the Correspondence Principle Limit should be relativistically invariant.  This condition is mathematically equivalent to the previous condition (\ref{transpread3a}) despite being deduced from very different physical considerations.

In order that the particle continue to have a well-defined phase in the optical field after emission, it is also necessary that the total particle recoil over the interaction time, due to one photon emission, remains small compared to the optical wavelength.  For a
photon radiated in the forward direction, the momentum kick is roughly 
\be
\hbar k_0 = \Delta p = mc\Delta\left(\gamma \beta \right) \sim mc \left( \gamma \Delta \beta + \gamma ^{3}\beta^2
\Delta \beta \right) \sim mc \gamma^3\Delta \beta.
\ee
We demand that $c \Delta\beta \frac{N_u \lambda_u}{c} \ll \lambda_0$, or
\begin{equation}
\tfrac{ N_u\lambda_c }{ \gamma } \ll \lambda_0.
\end{equation} This is precisely the same condition found above for the longitudinal quantum mechanical spreading of the wave-packet, without consideration of recoil effects or other longitudinal forces, to remain negligible.

Transversely, off-axis photons are expected to be emitted with a polar angle $\theta$ of at most about $\theta \sim \tfrac{1}{\gamma}$, due to relativistic ``head-lighting'' effects.  The resulting momentum kick will be small compared to the transverse momentum spread if 
$\tfrac{1}{\gamma}\hbar k_0 \ll \delta p_{\perp}$, or 
\begin{equation}
\hbar\omega_0 \ll \gamma^2\tfrac{\delta  p_{\perp}}{p} mc^2,
\end{equation}
where again $p = mc\gamma\beta$ is the average beam momentum. The transverse recoil should also be small
compared to the transverse quiver,
$\frac{1}{\gamma}\hbar k_0 \ll mc \,a_u$, or
\begin{equation}
\hbar\omega_0 \ll a_u \gamma mc^2;
\end{equation}
and finally, the total transverse recoil over the interaction time must be sufficiently small so that the resulting perturbation in the particle's transverse position does not appreciably affect the phase of any \textit{subsequently} emitted radiation.  The
transverse momentum kick is approximately
\be
\tfrac{1}{\gamma}\hbar k_0 \sim \Delta p_{\perp} \approx mc
\gamma \Delta \beta_{\perp}.
\ee
The total perturbation in transverse position due to this recoil is
\be
\Delta z \sim c \Delta \beta_{\perp} \tfrac{ N_u\lambda_u }{c} = \tfrac{1}{\gamma^2} \tfrac{\hbar
\omega_0}{mc^2 N_u\lambda_u}.
\ee
For nearly on-axis particles, the resulting perturbation in the phase of the
emitted radiation, as collected at the end of the wiggler, is somewhere between $\Delta \phi \sim k_0\Delta s$  and 
$\Delta \phi \sim k_0 \frac{ \Delta s^2 }{2 N_u \lambda_u }$, depending on the particle's longitudinal position in the wiggler, so demanding $\Delta \phi \ll 2\pi$, we find 
\begin{equation}
\hbar \omega_0 \ll \gamma^2 N_u \tfrac{\lambda_0}{ \lambda_u } mc^2 \sim N_u mc^2
\end{equation} and
\begin{equation}
\hbar \omega_0 \ll \tfrac{\gamma^2}{ \sqrt{N_u}} \left[ \tfrac{\lambda}{\lambda_u} \right]^{1/2} mc^2
 \sim \tfrac{\gamma  mc^2}{\sqrt{N_u}}.
\end{equation}

We have seen that recoil effects on any one particle are negligible for its subsequent dynamics, but in order that recoil can be fully ignored, its effects on the radiation field must also be small.  From the Compton scattering relation, we see that in the average rest frame of the beam, the shift in frequency $\delta\omega'$ due to particle recoil during the scattering of a virtual
wiggler photon into a real radiation photon in a Weizs{\"a}cker-Williams treatment will be small compared to the central radiation
frequency $\omega_0'$ provided that, in the lab frame: 
\begin{equation}
\tfrac{1}{2\gamma} \tfrac{\hbar \omega_0}{ mc^2} \ll 1,
\end{equation}
or equivalently,
\begin{equation}
\lambda_c \ll 2\gamma \lambda_0.
\end{equation}
Actually, this Compton shift should also be small relative to the coherent bandwidth, requiring
\begin{equation}
\tfrac{1}{\gamma}\tfrac{\hbar \omega_0}{ mc^2 } \ll \tfrac{\hbar \omega_0} {N_u },
\end{equation}
equivalent to the by now very familiar requirement that 
\begin{equation}\label{class10}
\tfrac{N_u}{\gamma}\lambda_c \ll \lambda_0.
\end{equation}
We must also demand that the recoil from emission of a photon will not change the energy of any one particle so much that any \textit{subsequent} emission by that particle will likely lie outside the original coherent bandwidth at the average beam energy, but a little algebra shows that this condition is again (\ref{class10}); we begin to understand why in the literature this is often regarded as \textit{the} condition defining the classical/quantum boundary for wiggler physics.

As pointed out by Benson and Madey \cite{benson:1985} and Brau \cite{brau:1990}, these standard recoil arguments really only indicate that \textit{expectation values} of particle observables can be calculated classically.  We must also verify that
variances or fluctuations about mean values are dominated by classical effects such as shot noise in the beam.  We expect
that the most stringent condition for higher-order moments will arise from considerations of longitudinal motion, so we
concentrate on this behavior here.  We have seen that the average recoil of a particle due to photon emission is sufficiently small for a classical description to be valid, but it should also be the case that additional quantum fluctuations due to the
variance in the photon emission are negligible.

Now, it is known that the photon counting statistics for Compton scattering from plane waves is Poissonian, so for sufficiently
uniform wiggler fields and a sufficiently relativistic beam, such that in the average beam rest frame the wiggler field can be
represented as virtual photons in a plane-wave mode according to the Weizs\"{a}cker-Williams approximation, photon counting statistics
for photons scattered by any one particle should be Poissonian or close to Poissonian.  Of course, it is precisely the correct understanding of the field statistics, before and after amplification, as experienced not by hypothetical photon counters, but by actual individual
particles in the beam which is at issue in our entire analysis, and we are not yet equipped to make a careful calculation. 
But to arrive at a conservative upper bound for the variance or fluctuations due to recoil, we can assume that the photon
emission statistics for each particle in the wiggler are Poissonian, and then demand that the extra quantum mechanical uncertainty in energy due to fluctuations in photon emission is negligible compared to the momentum uncertainty already needed to
satisfy the uncertainty principle, which in the absence of these recoil fluctuations has already been shown to be negligibly
small.  Even though most of the total energy emitted by a given particle will be radiated at frequencies more than one coherent
bandwidth below the central photon energy $\hbar\omega_0$, and/or outside the coherent solid angle $\delta \theta \sim \frac{1}{\sqrt{N_u}\gamma}$, to be conservative we will here assume that all power is emitted close to the forward direction near the frequency
$\omega_0$.  The expected total number of photons radiated per particle (including those outside the coherent mode), in the pickup is then estimated as $\bar{n}_{\stext{rad}} \sim \alpha N_u a_u^2$.  The additional uncertainty in particle energy due to these fluctuations in radiation, assuming Poissonian statistics, is then 
\begin{equation}
\delta \gamma_{\stext{rad}} \sim \tfrac{ \hbar\omega_0 }{ mc^2 }\sqrt{ \alpha N_u a_u^2 } =
\tfrac{ \lambda_c }{ \lambda_0 }\sqrt{ \alpha N_u a_u^2 }.
\end{equation}
For a classical-behaving wave-packet, the uncertainty in longitudinal position at the end of the pickup
interaction should be no more than a few times the minimum uncertainty allowed as given by (\ref{min_z}), which
corresponds to an approximate lower bound on the Heisenberg uncertainty in beam energy:
\begin{equation}
\delta \gamma_{\stext{HUP}} \ge \sqrt{\tfrac{\lambda_c \gamma}{\lambda_0 N_u}}.
\end{equation} 
Demanding that $\delta \gamma_{\stext{rad}} \ll \delta \gamma_{\stext{HUP}}$, we find that 
\begin{equation}
\tfrac{\lambda_c}{\lambda_0}\tfrac{\alpha}{\gamma}N_u^2 a_u^2 \ll 1,
\end{equation}
which will be readily achieved in situations of interest.

\subsection{Summary of Arguments for Classicality of Particle Degrees-of-Freedom}

We have uncovered a plethora of conditions in support of our claim that the particle dynamics in the pickup wiggler can be treated classically.  Perhaps we should review the most important: $\rho_{b} \ll  \bigl(\tfrac{\epsilon_{\perp}}{\lambda_{c}}\bigr)^{3}\tfrac{\gamma}{\sigma_{\perp}^3}$ and $\rho_{b} \ll \tfrac{1}{\lambda_{c}^{3}}\bigl(\tfrac{\delta\gamma}{\gamma}\bigr)^{3},$ so that the beam particles are non-degenerate; $\frac{\delta\gamma}{\gamma} \ll 1,$ so that pair-creation and other exotic QED effects can be ignored;  $\tfrac{a_{u}}{\gamma} \ll 1$, so that certain spin effects can be neglected; and
\be
\tfrac{\hbar\omega_0}{mc^2} \ll \min\!\left[ \delta \gamma, \gamma, a_u, a_u \gamma,  \tfrac{a_u^2}{\gamma} , \tfrac{\gamma^3}{a_u}, N_u, \tfrac{\gamma}{\sqrt{N_u}}, \tfrac{\gamma}{N_u}, \tfrac{\gamma}{N_u^2 a_u}, \tfrac{\gamma^3}{N_u a_u}, \tfrac{\gamma}{\alpha N_u^2 a_u^2},  \tfrac{\delta p_{\perp}}{mc} , \tfrac{\gamma \delta p_{\perp}}{mc} \right],
\ee
so that a variety of other potential spin, transverse, and longitudinal effects may be ignored

Note that the condition $\frac{N_{u}\lambda_{c}}{\gamma} \ll \lambda_{0},$ or $\tfrac{\hbar\omega_0}{mc^2} \ll \tfrac{\gamma}{N_u},$ is often mentioned as the fundamental condition for classical behavior of particles in wigglers, because it  simultaneously ensures at least two fundamental conditions for classicality: that the longitudinal spreading of a single-particle wave-packet over the length of the wiggler in the absence of emission remains small relative to the shortest force length-scale, which is the optical wavelength; and that if a photon is emitted,  then the resulting recoil is sufficiently small so that the resulting shift in particle position over the interaction time in the wiggler is small compared to an optical wavelength.  However, for short wigglers this condition, while important, is not typically the most stringent of the ones we have established.

It should also be acknowledged that, strictly speaking, all the inequalities deduced above, individually or collectively, constitute neither sufficient nor necessary conditions for the observation of exclusively classical behavior on the part of the beam particles.  The failure  of any one condition does not necessarily indicate that quantum behavior will be observed, but only that it will be necessary to perform an actual quantum-mechanical calculation to determine whether the quantum corrections to classical behavior will be appreciable or not.  That is, they are necessary conditions for \textit{avoiding} a detailed quantum treatment.

Conversely, certain of these conditions determine whether sufficiently classical-looking quantum states for the beam exist, but even when such states are available there is no guarantee that the particles will be prepared in or remain in such a state.  However, it remains a poorly understood but widely observed tendency for systems, especially macroscopic systems consisting of either massive
objects or of many interacting constituent parts, to actually behave classically whenever they can behave classically.  The emergence of a classical world from quantum physics remains a  fundamental and only very partially understood problem, although some
recent progress has been made in this direction by Zurek \cite{zurek:1981, zurek:1982, zurek:1993b} and others concerning the analysis of decoherence and so-called environmentally-induced superselection rules. 

In the case of particle beams at least, all empirical evidence suggests this is true.  We believe that the resolution lies in the realization that one does not typically track or make observations on individual particles but rather the particle beam as a whole, or at least some slice or segment of the beam, which always contains many particles and is always subject to uncertainty due to uncontrolled classical noise from the environment and/or our limited initial knowledge and subsequent measurement precision.  Recall that the closest classical analog of a quantum state, even a single-particle state, is not a particle trajectory but rather a phase-space distribution, or statistical ensemble of trajectories.  If there are sufficiently classically-looking quantum states for individual particles, consistent with what is known macroscopically about the beam, typically consisting of only a few parameters such as average dimensions, density, energy, and the longitudinal and transverse emittances, and if furthermore the classical emittances are sufficiently large, then classical statistical uncertainty can simply swamp quantum fluctuations, and the density matrix for the beam as a whole may be essentially indistinguishable from a proper statistical mixture of the most nearly-classical pure states, regardless of whether individual particles were ever actually prepared in or remain in these particular classical-looking states.

\section{Classical Single-Particle Dynamics Are Adequate}

We confidently conclude that particles in the beam can be taken to behave classically throughout the pickup wiggler, and furthermore, assert that they will follow classical trajectories determined by the initial conditions and the external wiggler magnetic fields (and other external focusing fields, if present) and perhaps the mean self-fields, with negligible perturbation from the spontaneously-radiated pickup radiation itself, either through recoil (radiation-reaction), coherent forcing, or incoherent scattering.

Under these conditions, we will see that it is possible to consistently treat the physics in what we will call a ``hemi-classical'' formalism, where the particles are treated classically but the radiation is described quantum mechanically, in contrast to  the ``semi-classical'' regime, a description usually applied either to a system in some sense poised between quantum and classical behavior and approaching a correspondence principle limit or otherwise amenable to a long wavelength asymptotic analysis, or else to interactions between matter and fields in which the matter must be described quantum-mechanically but the radiation can be described classically\footnote{We will also see that within the hemi-classical formalism, the behavior and observable properties of the radiation will actually be almost entirely classical, so the hemi-classical theory is semi-classical in the first sense of the term, but complementary to the second, widely-used meaning.}.

While not strictly necessary for the validity of the hemi-classical description, it will nevertheless be quite convenient to also argue away even the Vlasov (mean) fields, so that in fact the beam can be simply  described by the single-particle physics of classical charges moving in fully external fields.

\subsection{Radiation Scattering Effects Are Small}

We have already seen above that radiation reaction effects may be neglected, so the effects on a given particle of its own wiggler radiation are unimportant,  But this radiation actually slips ahead of the particle, and might interact with other beam particles downstream.  The simplest way to check whether these effects might be important is to compare the density of actual radiated photons inside the wiggler to the effective density of virtual wiggler photons.  In the average rest frame, the latter photon density is  given approximately by 
\be
n'_{u} \sim \pi  \alpha\inv \tfrac{\gamma a_u^2}{\lambda_u \lambda_c^2}.
\ee
Using  (\ref{radiation_per_particle}), and assuming the radiation from each particle is predominately radiated within the synchrotron opening angle $\tfrac{1}{\gamma}$ in the lab frame, and remembering that a given particle will feel the fields from about $N_{s}$ other particles while traveling through the wiggler, the density of radiated photons experienced by a particle in the average rest frame is at most about
\be
n'_{\stext{rad}} \sim 4 \alpha \tfrac{\gamma^3 N_s a_u^2}{N_u^2 \lambda_u^3},
\ee 
and demanding $n'_{\stext{rad}} \ll n'_{u}$, after a little algebra we deduce the requirement
\be
\alpha^2 \gamma^2 N_s \ll \tfrac{\pi}{4} N_u^2 \tfrac{\lambda_u^2}{\lambda_c^2},
\ee
which is expected to be satisfied with many orders-of-magnitude to spare.

With this density of radiation photons, the mean free path $\ell_{C}$ for Thomson/Compton scattering from any one beam particle may be roughly estimated in the lab frame by setting
\be
\ell_{c} \alpha^2 \lambda_c^2  \tfrac{1}{\gamma} n'_{\stext{rad}} \approx 1,
\ee
so we will have $\ell_{c} \ll N_u \lambda_u$ provided
\be
\alpha^2 \tfrac{\lambda_c^2}{\lambda_u^2} \gamma^2 \tfrac{N_s}{N_u} a_u^2 \ll 1,
\ee
which is also readily satisfied, so scattering off the pickup radiation by downstream particles is completely negligible.  Besides, even if subsequent Thomson or Compton scattering from the radiated field does occur, the momentum transfers will be of comparable magnitude to the direct radiation recoil terms during the original emission, already known to be negligible.

\subsection {Mean Space-Charge Forces are Mostly Small}

Assuming a long, nearly azimuthally-symmetric beam with uniform particle density $n_b$, the average transverse Coulomb-Lorentz force due to self-fields on a beam particle at a distance $r$ from beam axis is
\begin{equation} F_{r}(r) = \frac{2\pi q^2 n_b }{\gamma^2} r,
\end{equation}
so that the average electric and magnetic transverse forces due to mean self-fields cancel to
$O\left(\frac{1}{\gamma^2}\right)$ for a symmetric beam.  The residual repulsive force will be negligible compared to the Lorentz force from the wiggler provided $F_{r}(\sigma_{\perp})\frac{N_u\lambda_u}{c} \ll a_u m c$, or
\begin{equation}
\alpha  \frac{1}{a_u\gamma^2} N_u n \sigma_{\perp}\lambda_u\lambda_c \ll 1,
\end{equation}
which is readily satisfied in the stretched beam.  To be completely negligible, this repulsive self-force
should also result in transverse beam spreading which remains small compared to the optical wavelength over the interaction
time, \ie,  $\half \frac{1}{\gamma m}F_{r}(\sigma_{\perp})\left[\frac{N_u\lambda_u}{c}\right]^2 \ll \lambda_0$, or 
\begin{equation}
\alpha \frac{1}{\gamma} N_u^2 n \sigma_{\perp}\lambda_u\lambda_c \ll 1,
\end{equation}
which is also satisfied provided the beam is sufficiently stretched.  For a long uniform beam, the average
longitudinal force on an randomly chosen particle vanishes by symmetry, but the RMS force does not, and particles away from the midpoint of the beam will experience Coulomb repulsion.  The velocity is highly relativistic, so particles are inertially stiff, but on the other hand, the
relevant length-scale for perturbations is only the optical wavelength
$\lambda_0$ over the full interaction time $\Delta t$ in the wiggler.  At about one standard deviation into the tail of a Gaussian beam, the Coulomb repulsion will scale like $F_{z} \sim \pi q^2 n \frac{\sigma_{\perp}^2}{\sigma_{z}}$.  The resulting spread over
the interaction time should be small compared to the optical wavelength, \ie, 
\be
\frac{1}{2\left[\gamma + \beta^2\gamma^3\right] m} F_{z} \left[ \frac{N_u \lambda_u}{c} \right]^2 \ll
\lambda_0,
\ee
which becomes
\begin{equation}
\alpha N_u^2 \frac{1}{\gamma} n \sigma_{perp}^2\lambda_u \frac{\lambda_c}{ \sigma_z } \ll 1,
\end{equation} which is easily met in our parameter regime.

\subsection{Fluctuations and Collective Oscillations are Mostly Small}

In addition to establishing that the \textit{average} space-charge effects are negligibly small, to we must
additionally show that the \textit{fluctuations} and \textit{collective oscillations} about this average behavior, arising from self-fields, are also
negligible.  Collective plasma oscillations will be unimportant if either the relativistic plasma frequency $\omega_{p}'$ in the
average rest frame is sufficiently low so that oscillations simply do not have time to develop during the interaction time, or if any Langmuir waves that might arise are heavily suppressed via Landau damping mechanisms, which will occur if the typical wavelength $\lambda_{p}'$ for plasma waves is small compared to the Debye screening length $\lambda_{\stext{D}}'$.  In the average rest frame, the relativistic plasma frequency is given in Gaussian units by 
\begin{equation}
\omega_{p}'= \left[\frac{4\pi n' q^2}{m}\right]^{\half} = \left[\frac{4\pi n q^2}{\gamma m} \right]^{1/2},
\end{equation}
where from relativistic length contraction $n' = n/\gamma$ and $\Delta t' \approx
\frac{N_u\lambda_u}{\gamma c},$ so demanding $\omega_{p}' \Delta t' \ll 2\pi,$ we find 
\begin{equation}\label{smallomegap}
\left[2 \alpha \frac{1}{\gamma^3}N_u^2 n \lambda_u^2 \lambda_c\right]^{1/2} \ll 2\pi.
\end{equation}
which is expected to be satisfied, at least for sufficiently massive particles like muons or protons (but possibly not for electron beams).

In the average rest frame, the Debye length is
$\lambda_{D}' = \sqrt{\frac{ k_{\stext{B}}T'}{4\pi q^2 n'}}$, where $T'$ is the effective longitudinal rest-frame temperature. 
From relativistic velocity addition, we have seen that 
\be
k_{\stext{B}}T' \approx mc^2 \langle \beta'^2 \rangle \approx mc^2 \frac{\delta \gamma ^2}{\beta^2\gamma^2},
\ee
assuming $\gamma \gg 1$ and $\frac{\gamma}{\gamma} \ll 1$.  Therefore the rest-frame Debye length can be written in
terms of lab-frame quantities as
\begin{equation} 
\lambda_{\stext{D}}' = \left[ \frac{\gamma}{\alpha \lambda_c n} \right]^{\half} \frac{\delta \gamma}{\gamma}.
\end{equation}
The rest-frame wavelength $\lambda_{p}'$ for plasma oscillations is expected to be comparable to that of the
wiggler period in this frame, $\lambda_{p}' \sim \lambda_u' = \lambda_u/\gamma$, so that plasma oscillations will be damped provided $\lambda_{p}' \ll \lambda_{D}'$, or
\begin{equation}
\left[\frac{2\alpha \lambda_u^2\lambda_c n}{\gamma \delta \gamma^2}\right]^{\half} \ll 1.
\end{equation}
Because the wiggler period in the envisioned OSC scheme is so long compared to most other applications, this may be only marginally satisfied or may actually be violated, but because the beam is diffuse and the total number of wiggler periods is so small and the resulting interaction time so short,  waves that are not suppressed by Landau damping will not really have much time to arise, as demonstrated in (\ref{smallomegap}) above.

While we can and will therefore neglect collective oscillations here,  although these conditions could fail for certain beam parameter regimes, and then the effects of collective space-charge oscillations would in principle need to be included in the particle dynamics for a fully consistent treatment.  THis is left for future research.

Collisions, or binary (but possibly shielded) Coulomb interactions between discrete particles not accounted for in the Vlasov (mean) fields, could also cause deviations from the ideal wiggler orbits.  Including Debye screening effects, the effective Coulomb scattering frequency $\nu_{c}'$ in the average rest frame (including the cumulative effects of many small-angle scattering events) is approximately
\be
\nu_{c} \approx \frac{8\pi n' q^{4}}{m^2c^{3}\delta \beta'^3}\ln \left(\Lambda'\right),
\ee
where $\delta \beta'$ is the RMS velocity in the average rest frame, and where
\be
\Lambda' =
\lambda_{D}'^{3}n'^{3} = \left[\frac{\delta \gamma}{\gamma}
\right]^{3}\left[\frac{\gamma}{2\alpha}\right]^{\frac{3}{2}}\left[\frac{1}{\lambda_c^{3} n} \right]^{\half}
\ee
is the rest-frame plasma parameter, and satisfies $\Lambda' \gg 1$, while typically $\ln \left(\Lambda'\right) \le O(10)$ or so.  Comparing to the
plasma frequency, we see that
\be
\frac{\nu_{c}'}{\omega_{p}'} \approx \frac{\ln \left(\Lambda'\right)}{2\pi \Lambda'},
\ee
so that the collision frequency is expected to be much smaller than the plasma frequency, and is expected to remain unimportant.  Specifically,  Coulomb scattering will be negligible if $\nu_{c}'\Delta t' \ll 1$,  or
\begin{equation}
\left[2 \alpha \frac{1}{\gamma^3}N_u^2 n \lambda_u^2 \lambda_c\right]^{\half} \frac{\ln
\left(\Lambda'\right)}{2\pi \Lambda'} \ll 1,
\end{equation} which is readily achieved for beams of interest.

Finally, while we have seen that we can neglect the effects of the space-charge forces on the average density $n,$ or equivalently on the one-particle distribution function, we seek to establish that space-charge effects will not lead to
significant correlations in the two-particle distribution function,  so that the relative positions of beam particles can still be
described by classical shot noise.   Such shot noise is characterized by Poissonian statistics within any given spatial region,
or equivalently by particle positions chosen randomly and independently in proportion to the average density $n$, and so
shot noise in any one Lorentz frame will appear as shot noise in any other frame, just with an appropriately
Lorentz-transformed density.  In the average rest frame, Coulomb repulsion is expected to lead to a certain amount of
anti-correlations in particle positions, which may develop over the full lifetime of the beam, not just the interaction time in
the wiggler.  Roughly on the time-scale $\nu_{c}^{-1}$ corresponding to the Coulomb scattering and dielectric screening times
in a plasma, we expect the beam to have approximately approach a state of thermodynamic equilibrium in its average rest frame, at least in the absence of cooling and when neglecting the slow expansion of the beam, and hence in order to estimate the magnitude of these
particle correlations we can make use of well-known results from equilibrium statistical mechanics, without having to solve
a much more difficult time-dependent problem in kinetic theory.

In the average rest frame, consider a sub-volume $\delta V'$ small compared to the beam as a whole but large enough to contain
many particles on average.  (Strictly speaking, we actually need not demand that $\delta N' \gg 1,$ but only that $\delta
N' \ll N_{b}$.)  Treating this region as a thermodynamic system in thermal and diffusive equilibrium with the rest of the beam
temporarily regarded as a heat and particle reservoir, it is easy to see that in the absence of particle interactions, the particle
number $\delta N'$ in this volume will be subject to Poissonian fluctuations.  In the non-degenerate (low-$n$ and/or high-$T$)
limit of indistinguishable free particles, the partition function for the region of interest can be taken to be 
\begin{equation}
Z'(T',\delta N', \delta V') =
\frac{\left[n_{Q}'\delta V'\right]^{\delta N'}}{\delta N' !},
\end{equation}
where
\be
n_{Q} = n_{Q}'(T') = \left[m
\frac{k_{\stext{B}}T'}{2\pi\hbar^2}\right]^{\frac{3}{2}} \sim \frac{1}{\lambda_{dB}'^{3}}
\ee
is the quantum concentration, corresponding to one particle in a volume whose linear dimensions scale like the thermal de Broglie wavelength $\lambda_{dB}'$, and where
for simplicity we assume the transverse and longitudinal temperatures are equal.  Using the usual Stirling approximation for the factorial term, the Helmholtz free energy is given by
\be
F'(T', \delta N', \delta V') = -k_{\stext{B}}T'\ln Z' = \delta N' k_{\stext{B}}T'\left[
\ln\left(\frac{\delta N'/\delta V'}{n_{Q}'}\right) - 1\right],
\ee
so the chemical potential is 
\be
\mu'(T', \delta N', \delta V') = \frac{\del}{\del \delta N'}F'(T', \delta N', \delta V') = k_{\stext{B}}T' \ln\left(\frac{\delta N'/\delta
V'}{n_{Q}'}\right).
\ee
The grand partition function can then be written as
\be
\mathcal{Z}'(T', \mu', \delta V') = \sum_{\delta N' = 0}^{\infty}\lambda'^{\delta N'}
\frac{\left[n_{Q}'\delta V'\right]^{\delta N'}}{\delta N' !} = e^{\lambda' n_{Q}'\delta V'},
\ee
where $\lambda' = e^{\frac{\mu'}{k_{\stext{B}}T'}}$ is the absolute activity.  The average particle number within $\delta V'$ is then
\be
\bar{\delta N}'(T', \mu', \delta V') = \mean{\delta N'(T', \mu', \delta V')} = \lambda'\frac{\del}{\del \lambda'}\ln
\mathcal{Z}'(T', \lambda', \delta V')  = \lambda' n_{Q}'\delta V',
\ee
and the probability distribution for particle number can be
written as
\begin{equation}
P'(\delta N') =\frac{1}{\mathcal{Z}'}\frac{\left[ \lambda' n_{Q}'\delta V'\right]^{\delta N'}}{\delta N' !}
=\frac{\mean{\delta N'}^{\delta N'}e^{-\mean{\delta N'}}}{\delta N' !},
\end{equation}
which is a Poisson distribution, as expected, so that particle fluctuations satisfy
\be
\mean{\left(\delta N' - \mean{\delta N'}\right)^2} = \mean{\delta N'} .
\ee

When the effects of Coulomb interactions are included, the general problem becomes somewhat intractable, but a perturbation-type expansion for the partition function or free energy can be found in the typical ideal plasma regime, corresponding  to large
Debye shielding, $\Lambda \equiv \lambda_{D}'^{3} n'  \gg 1$, or equivalently weak coupling:
\begin{equation}\label{pertparam}
\frac{q^2n^{1/3} }{k_{\stext{B}}T'} \ll 1,
\end{equation}
so that that the average Coulomb potential energy between neighboring particles is small compared to their mean thermal kinetic energy.  For a single charged species, the Helmholtz free energy becomes
\begin{equation}\label{FC}
F'(T',\delta N', \delta V') = \delta N' k_{\stext{B}}T'\left[ \ln\left(\frac{\delta N'/\delta V'}{n_{Q}'} \right) - 1\right] - \frac{2}{3}\sqrt{\pi}q^{3}\delta N' \left[\frac{\delta N'}{\delta V'}\right]^{\tfrac{1}{2}}+ \dotsb. 
\end{equation}

Because of the long-range nature of Coulomb force, successive terms in the expansion include various
fractional powers and even a logarithm of the local density $\frac{\delta N'}{\delta V'},$ so this is not a simple virial expansion in the density, and because of the appearance of $q^{3}$, neither is it an analytic perturbation expansion in the natural coupling parameter $q^2$.  But assuming the perturbation parameter (\ref{pertparam}) is very small, we need include only the leading correction, the so-called Debye term, and need not worry here about non-analyticity or convergence
of higher-order terms.  Typically, this free energy for a Coulomb gas is derived using textbook Debye-shielding arguments (see for example \cite{kunkel:1966}) leading to an effective screened potential energy of Yukawa form, but such derivations assume net neutrality of the plasma.  However, the expansion (\ref{FC}) remains valid even for non-neutral plasmas, and can be derived from quantum or
classical statistical mechanics by re-summation techniques, or in a kinetic theory formalism by truncating the BBGKY hierarchy beyond the two-particle distribution function, a valid approximation when the inter-particle correlations in fact do remain small, which can be self-consistently checked at the end of the calculation.

At least in the plasma or Debye-shielded regime,  the long-range nature of the Coulomb force is both blessing and curse; it is typically impossible to treat exactly, but the statistical correlations which can develop between a given particle and any one other randomly chosen particle in the
plasma will on average be small because of the screening effects of all the other intervening particles, and the long-range correlations which do persist result from the self-averaged effects of many particles and can be incorporated through a self-consistent mean-field (in our case the beam space-charge field), leading to the well-known collective effects, as distinct from true many-body correlations that can only b described via a joint probability density function.

While it is somewhat troublesome to find a convenient expansion for the full probability distribution for the fluctuation $\delta N'$, it is straightforward to find the so-called Fano factor, or ratio of variance to average, sufficient for our purposes.  Using the
Helmholtz free energy (\ref{FC}), the chemical potential for the Coulomb gas becomes
\begin{equation}
\mu(T', \delta N', \delta V') = \frac{\del}{\del \delta N'} F'(T', \delta N', \delta V') = k_{\stext{B}}T'  \ln\left(\frac{\delta N'/\delta
V'}{n_{Q}'} \right) - \sqrt{\pi}q^{3}\left[\frac{1}{k_{\stext{B}}T'} \frac{\delta N'}{\delta V'} \right]^{\smallhalf}.
\end{equation}
Shifting from the canonical to the grand canonical ensemble, the chemical potential $\mu(T', \overline{\delta N}', \delta V')$
evaluated at the average particle $\overline{\delta N'}$ is formally invertible, and can be taken to
implicitly define the average particle number  $\overline{\delta N'}(T', \mu', \delta V')  =
\mean{\delta N'(T', \mu', \delta V')}$ as a function of the chemical potential
$\mu'$.  The variance in particle number can then be expressed as
\begin{equation}
\mean{\bigl(\delta N' - \mean{\delta N'}\bigr)^2} =
k_{\stext{B}}T'\frac{\del}{\del \mu'}\mean{\delta N'(T', \mu', \delta V')} = \frac{1}{\frac{\del}{\del \overline{\delta N'} } \mu'(T',
\overline{\delta N'},\delta V') },
\end{equation}
which after a little algebra can be written as
\begin{equation}
\begin{split}
\mean{ \bigl(\delta N' - \mean{\delta N'} \bigr)^2 } &= 
\frac{ \overline{\delta N'} }{ 1 - \half \sqrt{\pi} q^3 \left( \frac{1}{ k_{\stext{B}}T'} \right)^{\tfrac{3}{2}} \left(\frac{ \overline{ \delta
N'}}{\delta V'}  \right)^{\smallhalf}}\\
&= \frac{ n' \delta V' }{ 1 - \half \sqrt{\pi} \left(  \frac{ q^2 n^{1/3}   } { k_{\stext{B}}T' } \right)^{3/2} },
\end{split}
\end{equation}
so the actual Fano factor may be written simply as
\be
\tfrac{\mean{ \bigl(\delta N' - \mean{\delta N'} \bigr)^2 } }{\mean{\delta N'}} = \frac{1}{1 - \tfrac{1}{16\pi} \frac{1}{\Lambda'}}.
\ee
Provided the weak-coupling condition (\ref{pertparam}) holds, we see that any deviations from Poissonian fluctuations are indeed small, at least at the level of first and second-order moments.

In summary, to a very good approximation, we may therefore assume that that particles in the beam behave classically in the pickup wiggler, even as the particles radiate, and that their relative positions are well approximated by classical shot noise, and
that their trajectories are determined almost exclusively by the external electromagnetic
fields  and their individual initial conditions at the entrance to the wiggler, and are in most cases largely unaffected by collisions or space-charge forces or by the radiation field itself, either through coherent or incoherent scattering or through radiation reaction\footnote{As we have seen, in some regimes, collective Langmuir oscillations might be marginally appreciable, but as we are interested in a fundamental proofs-of-principle regarding the effects of radiation noise, these collective effects will be subsequently ignored.}.  

With this classical model of prescribed particle trajectories, the subsequent analysis can be
tremendously simplified.  To make matters even simpler, off-axis effects and all betatron or other transverse motion of particles are also ignored, accept for the idealized $1$D quiver motion in the planar wiggler field, as determined by canonical momentum conservation.  Higher harmonics of particle motion in the planar wiggler are ignored, so the longitudinal motion is taken to be uniform.

\section{Quantum Mechanical Description of Wiggler Radiation}

An exact analytic description of the quantum mechanics of emission would be prohibitively difficult, requiring a complete treatment of the dynamics of the coupled particle-field system, and in general leading to highly entangled states for the matter and radiation.  But in the regime established above, where recoil effects are negligible and particles follow classical orbits that can be determined independently of the actual radiation fields, the problem is simplified tremendously.

However, before we can deduce the quantum state of the radiation, we must digress to summarize a quantum optics formalism that can capture the collimation, collection, transport and amplification of the wiggle radiation.

\subsection{Quantum Optics for Dielectric-Guided Paraxial Beams}

After emission in the pickup wiggler, travels essentially paraxially down a mostly open beam line, but which does contain various passive dielectric lenses used collect, and transport the radiation, as well as one or more apertures to select transverse mode
characteristics, in addition to  the active gain medium of the amplifier or amplifiers.  If the field is quantized in the vacuum, then the effects of dielectric elements must be included as interaction terms in the Hamiltonian coupling a large number of vacuum modes, complicating the dynamics.

Instead, we can quantize directly the ``macroscopic fields'' in the presence of an idealized  linear medium described by susceptibility tensors.  Numerous approaches to this problem may be found in the literature, for example those in \cite{glauber:1991, huttner:1992, milonni:1995, gruner:1996, juzeliunas:1996, dalton:1996, dalton:1997, dung:1998, tip:1998, crenshaw:2000, crenshaw:2003, garrison:2004, suttorp:2004, suttorp:2004b} and even earlier antecedents.  Most are restricted to special cases where the medium is homogeneous and isotropic, or the dielectric tensor is frequency-independent, or only certain kinds of frequency response are allowed.  The more general problem, involving anisotropy, slow spatio-temporal variations in  the medium response, and arbitrary frequency dispersion consistent with the Kramers-Kronig relations, remains an area of current research\footnote{It seems unlikely than any fully general, Hamiltonian theory of \textit{macroscopic} quantum optics can be established, because the energy cannot always be written as a well-defined function of the macroscopic fields and free sources at one instant of time.}, and certain fundamental debate persists \footnote{Disagreement continues over whether the electric field $\bv{E}(\bv{x},t)$ or the displacement field $\bv{D}(\bv{x},t)$ should be canonically conjugate to the vector potential $\bv{A}(\bv{x},t)$.  Also, in the microscopic theory it is well known that with the addition of sources, the commutation relations of the field observables are unchanged, and commute with all observables for the material degrees-of-freedom, which does not always seem to be the case with the quantized macroscopic theories.  These questions are not fully resolved, but \cite{cohentannoudji:1989} suggests that some of the disagreement may just be due to different gauge choices, or to different dressed operators going by the same name.  Anyway, this mostly concerns the meaning of ``photons'' and the nature of measurements performed inside the dielectric medium, which are of minor concern here.}.

Here we consider the unmagnetized case where $\mu = 1$ while the dielectric tensor $\epsilon(\bv{x})$ is real (\ie, lossless), positive, isotropic,  dispersionless to a good approximation over the wiggler and amplifier bandwidths, but possibly spatially inhomogeneous, \ie, piecewise smooth in its dependence on position $\bv{x}.$  This model allows for transport of the radiation by dielectric mirrors or by ideal lenses without chromatic aberration,  In practice, some dispersion will be present, so here $\epsilon(\bv{x})$ is chosen to best represent the average effects of the actual lenses or other optical media along the beam path.  Perfectly-conducting boundaries\footnote{We have worked out in detail the case where the quantization region is simply-connected and simply-bounded.  Some subtleties arise if it is not simply-connected, because then the  kernel of the curl operator can contain certain vector fields which are not gradients of a single-valued scalar field, reflecting the non-trivial cohomology of the interior region. However, generalizations of the Helmholtz-Hodge Decomposition Theorem can be used in these cases, and this is a subject of ongoing study on our part.} can also be included to model apertures, beam pipes, or metallic mirrors \footnote{We had worked out this particular quantum theory of dielectrics in some detail, before realizing it was nearly identical to that developed by Glauber and Lewenstein in \cite{glauber:1991} and later by other authors.}.  The resulting eigenmodes in general possess more complicated spatial structure than simple plane waves,  but will retain harmonic time dependence.  The exact form of these modes is not actually needed for our purposes here, but in the vacuum regions can be well approximated by the usual paraxial Gauss-Hermite basis set.

Here we just summarize the main tools and results.  In the presence of the hypothesized medium, the macroscopic fields can still be written (in Gaussian units) as
$\bv{B} = \bv{H} = \curl\bv{A}$ and $\bv{E} = \epsilon\inv \bv{D} = -\grad\phi
-\frac{1}{c}\tfrac{\del}{\del t}\bv{A}.$  We choose to work in the generalized Coulomb, or $\epsilon$-transverse, gauge, where
\be
\divergence \left[ \epsilon \bv{A}\right] = 0.
\ee
For any piecewise-smooth dielectric function satisfying $0 < \epsilon(\bv{x}) < \infty,$ it can be rigorously shown using generalizations of the Helmholtz theorem and existence and uniqueness results for the Laplace equation, that this
is a well-defined gauge choice leading to a unique decomposition of the electric field into an $\epsilon$-transverse component containing all (but not necessarily only) radiation and an irrotational (curl-free) non-radiative component.

From the macroscopic Maxwell's Equations, we deduce that the scalar and vector potentials potentials must satisfy
\bsub
\begin{align}
\divergence\left[\epsilon \grad\phi\right] &= -4\pi \rho,  \label{eqn:phi_eq}\\
-\curl\left(\curl\bv{A}\right) - \tfrac{\epsilon}{c^{2}}\tfrac{\del^{2}}{\del t^{2}}\bv{A} &=
-\frac{4\pi}{c}\bv{J} + \frac{\epsilon}{c}\tfrac{\del}{\del t}\grad \phi \label{eqn:a_eq}
\end{align}
\esub
where $\rho$ and $\bv{J}$ are the ``free'' charge and current flux densities, respectively, for any charges not implicitly accounted for through $\epsilon(\bv{x}),$ namely those in the particle beam.

Given the assumed good behavior of the dielectric function, it can be verified that the generalized Poisson equation (\ref{eqn:phi_eq}) has existence and unique properties essentially identical to the standard Poisson boundary value problem. 
The scalar potential $\phi$ is therefore determined by the instantaneous positions of all the additional charges, as in the usual Coulomb gauge, and satisfies an unretarded elliptic equation,  hence does not embody independent dynamical degrees-of-freedom for the EM fields.

Separation of variables is used to express the vector potential $\bv{A}$ in terms of paraxial eigenmodes $\bv{u}_{j}(\bv{x};\omega)$ of the generalized Helmholtz equation:
\be\label{eqn:eigencurl}
\epsilon(\bv{x})\inv \curl\curl \bv{u}_{j}(\bv{x}; \omega) =
\tfrac{\omega^{2}}{c^{2}}\bv{u}_{j}(\bv{x}; \omega);
\ee
Here the set of integers labeled by $\ell$ is taken to index the transverse polarization and modal structure, assumed to be discrete, as in a paraxial beam or a waveguide, so $\ell$ may have multiple components which are implicitly hidden in the notation for brevity.  The frequency $\omega$ labels \footnote{Following both Caves\cite{caves:1982}, and general conventions for paraxial optics, we labeled the modes by the frequency $\omega$ at which they \textit{would} oscillate \text{if} allowed to freely evolve, because in general they are not characterized by a true fixed wavenumber.  The explicit dependence on $\omega$ is simply a label, and is not meant to indicate that these are in general Fourier transforms of time-domain quantities.  In fact, we remain in the time-domain, where functions labeled by $\omega$ nevertheless evolve in $t$, non-trivially in the presence of free sources.} what is in general a continuum of longitudinal modes.

The generalized Helmholtz operator appearing in (\ref{eqn:eigencurl}) is not Hermitian with respect to the usual $\mathcal{L}_2$ inner product on $\realsymbol^3,$ but it is Hermitian with respect to the $\epsilon$-weighted inner product:
\be
\innerpa{ \bv{u} } {\bv{v} }_{\epsilon} = \! \int \! d^{3}\bv{x} \, \epsilon(\bv{x})\, \bv{u}(\bv{x})\cc \!\cdot\! \bv{v}(\bv{x}), 
\ee
and with suitable boundary conditions can be naturally extended to a self-adjoint operator,
and the eigenmodes may then be chosen to be orthogonal with respect to this inner product:
\be
\innerpa{\bv{u}_{\ell}(\bv{x}; \omega)}{\bv{u}_{\ell'}(\bv{x}; \omega')}_{\epsilon} = \delta_{\ell\, \ell'} \delta(\omega - \omega'),
\ee
where $\delta_{\ell\,\ell'}$ and $\delta(\omega)$ are the Kronecker delta symbol and Dirac delta function, respectively.

These modes will be complete in the sense of spanning the space of $\epsilon$-transverse vector fields in the quantization region, so
the vector potential $\bv{A} = \bv{A}_{\epsilon\perp}$ for the radiation fields can be decomposed as
\be
\bv{A}(\bv{x},t) = \sum\limits_{\ell}\! \int\limits_{0}^{\infty} \!d\omega \, c\sqrt{\tfrac{2\pi\hbar}{\omega}}
\left[a_{\ell}(\omega)\bv{u}_{\ell}(\bv{x}; \omega)e^{-i\omega t} + a_{\ell}(\omega)\hc \bv{u}^{*}_{\ell}(\bv{x};\omega)e^{+i\omega t}
\right],
\ee
where the  macroscopic field operators $a_{\ell}(\omega)$ satisfy canonical commutation
relations
\bsub\label{ccomm2}
\begin{align}
\comm{a_{\ell}(\omega)}{ a_{\ell'}(\omega')\hc} &= \delta_{\ell\, \ell'}\delta(\omega - \omega')\\
\comm{a_{\ell}(\omega)}{ a_{\ell'}(\omega')^{\phantom{\dag}}} &= 0.
\end{align}
\esub
In the presence of the dielectric medium,  the ``quasi-free'' Hamiltonian governing propagation in the absence of free charges becomes
\be
\mathscr{H}_0 =  \tfrac{1}{8\pi} \!\int \!d^{3}\bv{x} \, [\bv{E} \!\cdot\! \bv{D} + \bv{B} \!\cdot\! \bv{H} ] = \sum\limits_{\ell} \int\limits_{0}^{\infty} \! d\omega \, \hbar \omega \, a_{\ell}(\omega)\hc  a_{\ell}(\omega),
\ee
where constants and other terms which do not depend on the field operators have been dropped.  As in the vacuum case, we see that the free-field Hamiltonian remains equivalent to that of a collection of uncoupled harmonic oscillators.  The source/field interaction Hamiltonian remains
\begin{equation}
\label{hamiltonain1a}
\mathscr{H}_{\stext{int}} = -\tfrac{1}{c} \!\int\! d^{3}\bv{x}\, \bv{J} \!\cdot\! \bv{A}.
\end{equation}
Note, however, that the macroscopic operators $a_{j}(\omega)$ are ``dressed,'' in general containing inside regions with nontrivial permittivity, contributions from both the usual creation and annihilation operators associated with the microscopic fields in free space.

Recall that here the background dielectric function must be real and independent of time and/or frequency. Any  effects of dispersion or attenuation must be incorporated explicitly through additional interaction terms.

The gain in the active amplifying medium must also be treated quantum mechanically, but a specific atomic model will not be needed; very general features of quantum mechanics will be sufficient to characterize its best-case noise properties without explicitly evolve the radiation fields through the system.

One could argue that by neglecting dispersion we are throwing away a source of noise for the radiation.  Dispersion at some frequencies implies absorption at other frequencies according to the Kramers-Kronig relations, while fluctuation-dissipation theorems suggest that this absorption must also be associated with noise.   Since quantum noise in the radiation field is of central concern, one might reasonably fear that ignoring dissipation and the attendant fluctuations in the lenses or other dielectric elements is a terrible idea.  However, there are no fundamental lower bounds on the losses and fluctuations introduced by  passive dielectric elements in any given bandwidth, so this perhaps is a useful idealization, 
because the frequencies at which absorption occurs and noise is injected are hopefully well removed by design from the bandwidth of interest.  If not, because we will assume the radiation propagation and amplification are entirely linear, we can actually lump any dispersion, attenuation, gain, and noise into a single generalized amplification process, and need not ascribe them to individual elements.

We will also assume all lenses are suitably coated to suppress reflection losses, which would otherwise lead to additional ``partition noise'' as in a idealized beam-splitter, because any unitary description of such devices would couple into the transmitted mode part of the vacuum fluctuations from the additional mode.  Again, we are seeking a best-case analysis, and any extra additive noise could be easily incorporated at the end of the calculation.

\subsection{Hemi-Classical Wiggler Radiation}

At last we almost equipped to determine the quantum state of the radiation field due to emission in the pickup wiggler, and after it passes through any passive, linear, collection optics.

We have seen that, with high accuracy, the dynamics can be treated in our so-called  ``hemi-classical'' approximation, where the particles and static wiggler field are treated classically, without any radiation reaction forces, while the light is treated quantum mechanically, a regime complementary to the usual semi-classical approximation, in which the matter and its Coulombic interactions are treated quantum mechanically but the radiation fields are assumed classical.

First we briefly review some important properties of the coherent states, first discussed by Schrodinger but introduced and extensively analyzed by Glauber\cite{glauber:1963a, glauber:1963b, glauber:1963c, glauber:1965} in the context of the quantum theory of optical coherence.

\subsubsection{Coherent States}

As eigenstates of the modal annihilation operators for the fields, such coherent states possess many useful mathematical and physical properties, a few of which will be briefly reviewed here.  (See, for example, \cite{barnett:1997} for a thorough modern discussion.)  Given a complete orthonormalized set of field modes, the
(multi-mode) continuum coherent state $\ket{\left\{\alpha_{\ell}(\omega) \right\}}$ is by definition an eigenstate of each modal annihilation operator $a_{\ell}(\omega)$ with complex eigenvalue $\alpha_{\ell}(\omega)$, \ie,
\be
a_{\ell}(\omega) \ket{\left\{\alpha_{\ell}(\omega)\right\} } = \alpha_{\ell}(\omega) \ket{\left\{\alpha_{\ell}(\omega)\right\} }.
\ee
The choice $\alpha_{\ell}(\omega) = 0$ corresponds to the vacuum for that family of modes.  The prescription $\alpha_{\ell}(\omega) = 0$ for all possible $\ell$ and $\omega$ corresponds to the overall vacuum state $\ket{0}$ for the dressed photon field.  In the case of a fully discrete set of modes, where only certain discrete set of natural frequencies $\omega_{\ell k}$ are allowed, this multi-mode Coherent state is manifestly separable, and can be written as tensor product $\bigotimes\limits_{\ell\, k}\ket{\alpha_{\ell\, k}}$.

By convention, phases are chosen so that the coherent state can be written as
\be
\ket{a_{\ell}(\omega)} = D\left[\alpha_{\ell}(\omega)\right] \ket{0},
\ee
where 
\be
D\left[\alpha_{\ell}(\omega)\right] = e^{ \sum\limits_{\ell}\!\int \!d\omega\left[  \alpha_{\ell }(\omega) a_{\ell}(\omega)\hc - \alpha_{\ell}(\omega)\cc a_{\ell}(\omega) \right]}
\ee
is the unitary modal phase-space displacement operator,
which satisfies $D\left[-\alpha_{\ell}(\omega)\right]\inv = D\left[-\alpha_{\ell}(\omega)\right] = D\left[\alpha_{\ell}(\omega)\right]\hc $ and $D\left[\alpha_{\ell}(\omega)\right]\hc a_{\ell}(\omega) \,D\left[\alpha_{\ell}(\omega)\right] = a_{\ell}(\omega) + \alpha_{\ell}(\omega),$  We see that Glauber coherent states just consist of displaced vacuum states, so they possess the same quantum mechanical uncertainties as the vacuum but with a non-zero bias to mode amplitudes.  The states are well-defined and normalizable provided that $\sum_{\ell}\!\int \!d\omega\, \abs{ \alpha_{\ell}(\omega) } ^2 < \infty.$

Using this definition of the coherent state, it is straightforward to show for a discrete mode the coherent state can be written as
\be\label{coherent1}
\ket{ \alpha_{\ell\, k} } \equiv \ket{\alpha_{\ell}(\omega_{\ell k}) } = \sum_{n_{\ell k} = 0}^{\infty}
\frac{\left(\alpha_{\ell\, k} \right)^{n_{\ell k}} e^{-\abs{\alpha_{\ell\, k}}^2}}{\sqrt{n_{\ell k}!}} \ket{n_{\ell k}},
\ee
in which 
\be
\ket{ n_{\ell k} } = \tfrac{1}{ \sqrt{n_{\ell k}!} } \bigl[ a_{\ell\, k}\hc \bigr]^{n_{\ell k}} \ket{0}
\ee
is the usual number eigenstate for this mode.  From (\ref{coherent1}), it is evident that coherent state will have Poissonian photon-counting statistics under ideal measurement conditions.

Glauber coherent states are also minimum uncertainty states in two important senses.  For simplicity , let us consider the case of discrete modes.  For any choice of overall  phase $\phi \in \realsymbol$, the Hermitian quadrature operators, or sine-like and cosine-like components of the field amplitude operator $a_{j}(\omega_i)$
are defined as 
\be
X_{\ell k} = X_{\ell k}\hc = \smallhalf\left[e^{-i\phi} a_{\ell k} + e^{i\phi} a_{\ell k}\hc\right]
\ee
which is said to be in-phase, and
\be
Y_{\ell k} = Y_{\ell k} \hc = \tfrac{1}{2 i}\left[e^{-i\phi} a_{\ell k} - e^{i\phi} a_{\ell k} \hc\right],
\ee
which is said to be in quadrature.  The Hermitian quadrature operators can be shown to satisfy
\bsub
\begin{align}
\comm{X_{\ell k} }{ X_{\ell' k' } }&= 0,\\
\comm{Y_{\ell k}}{ Y_{\ell' k'} } &= 0,\\
\comm{X_{\ell k} } { Y_{\ell' k' } } &= \tfrac{i}{2} \delta_{\ell\, \ell'}\delta_{k k'}.
\end{align}
\esub
Defining $\Delta X_{\ell k} \equiv X_{\ell k} - \mean{X_{\ell k}}$ and $\Delta Y_{\ell k} \equiv Y_{\ell k} - \mean{Y_{\ell k}}$, it follows form
the generalized  Heisenberg-Robertson uncertainty principle that
\be
\mean{\Delta X^2_{\ell k } } \mean{\Delta Y^2_{\ell k} } \ge \tfrac{1}{4}\abs{ \comm{X_{\ell k} }{ Y_{\ell k} }}^2 =
\tfrac{1}{16}.
\ee
It is easily shown that any coherent state symmetrically saturates this uncertainty-principle limit, in the sense that
the variances satisfy $\mean{(\Delta X_{\ell k})^2 } = \mean{(\Delta Y_{\ell k})^2} = \tfrac{1}{4}$.
Also, while it is impossible to define a true Hermitian modal phase observable $\Phi_{\ell k}$ conjugate to the modal number operator $\EuScript{N}_{\ell k} = a_{\ell k}\hc a_{\ell k}$, under reasonable definitions of a quasi-phase operator, such as the Barnett-Pegg phase \cite{barnett:1997}, the coherent states also saturate or nearly saturate the ``number-phase'' uncertainty principle: 
$\mean{(\Delta \EuScript{N}_{\ell k} )^2}\mean{(\Delta \Phi_{\ell k})^2} \ge \tfrac{1}{4}.$

Coherent states also have important overlap and over-completeness properties, but these will be introduced as needed, as well as a beautiful group theoretic structure which we will not need here.

\subsubsection{Optical Correspondence Principle}

Central to our arguments is the so-called \textit{Optical Correspondence Principle}, a theorem first established by Glauber \cite{glauber:1963c} in early work on the theory of coherent states,  and then generalized independently by Glauber\cite{glauber:1966b} and Sudarshan and Mehta\cite{sudarshan:1966b, sudarshan:1967}.  As it turns out, essentially the identical theorem holds within our formalism for ``macroscopic'' radiation propagation in a linear medium described by a real, positive $\epsilon(\bv{x})$ and possibly bounded by perfect conductors.

If the initial state of the radiation field is a multi-mode Glauber coherent state (possibly,  but not necessarily,  the vacuum state), and the free sources all follow prescribed classical trajectories, then the final state of the radiation field is also a multi-mode Glauber coherent state, whose expectation value is given by the corresponding solution to Maxwell's equations with the same sources and propagating through the same medium, with classical initial conditions given by the expectation value of the initial coherent state.

If the sources of interest for the radiation, namely the charges in the particle beam in the present case, can be taken as providing a prescribed classical current, then the interaction Hamiltonian (\ref{hamiltonain1a}) can be written as
\begin{equation}\label{hamiltonian1b}
\mathscr{H}_{\stext{int}}(t) = - \sum\limits_{\ell}\int\limits_{0}^{\infty} \!d\omega \left[  J_{\ell}(\omega; t)\cc a_{\ell}(\omega) + J_{\ell}(\omega; t)a_{\ell}(\omega)\hc\right],
\end{equation}
where 
\begin{equation}\label{source1b}
J_{\ell}(\omega; t) \equiv \sqrt{\tfrac{2\pi\hbar}{\omega}}
\int\! d^{3}\bv{x}\,\, \bv{u}_{\ell}(\bv{x}; \omega)\cc \!\cdot\!  \bv{J}(\bv{x},t).
\ee
As long as the current density $\bv{J}(\bv{x},t)$ is considered a prescribed current density, rather than the result of self-consistent trajectories of charged sources moving in total fields, it need not be expressed explicitly in terms of canonical particle coordinates, and the above form for the interaction Hamiltonian thus holds even if the actual sources are arbitrarily relativistic.

We see that each mode in the medium still evolves independently if subjected to prescribed sources, so for simplicity we will drop all subscripts and indices cluttering the notation and look at a single mode, which evolves according to
\be\label{coherent_ham}
\mathscr{H}(t) = \hbar \omega\, a\hc a - \left[ \EuScript{J}(t)\cc a + \EuScript{J}(t)a\hc \right]
\ee
for a given $c$-number function $\EuScript{J}(t)$.  Note that this Hamiltonian looks exactly like the corresponding Hamiltonian in absence of the dielectric medium and with the same
prescribed currents, but here the definition and meaning of each dressed modal annihilation operator $a$ is generally somewhat different than in a vacuum; by design, the effects of the medium are incorporated (hidden) in the  spatial structure of the dressed modes, determining what kind of excitation (\ie, dressed photon) the operator $a$ annihilates.
If the initial state for this mode is described a Glauber coherent state up to some overall phase, \ie, if  $\ket{\psi(0)} =
e^{i\phi(0)}\ket{\alpha(0)}$ for some real number $\phi(0)$ and complex $\alpha(0)$, then we claim that in the Schrodinger picture, the state at later times is still given by some 
\begin{equation}\label{ket1}
\ket{\psi(t)} = e^{i\phi(t)}\ket{\alpha(t)},
\end{equation}
where $\ket{\alpha(t)}$ is a coherent state with standard phase.  Because the coherent states are over-complete, the state (\ref{ket1}) will be a solution to the time-dependent Schrodinger equation corresponding to the Hamiltonian (\ref{coherent_ham}), provided 
\begin{equation}\label{schrodinger1}
\bra{\beta } i\hbar\tfrac{\del}{\del t} \ket{\psi(t)} = \qamp{\beta}{\mathscr{H}(t)}{\psi(t)}
\end{equation}
holds for any multiple of any arbitrary time-independent Glauber state $\ket{\beta}$, \ie, for any state such that $\bra{\beta }a^{\dag} = \beta\cc \bra{\beta }$ and $\tfrac{d}{dt}\beta = 0.$  Using
the eigenvector conditions for the coherent states $\ket{\alpha(t)}$ and $\ket{\beta}$, the inner product in the form
\begin{equation}
\innerp{\beta}{\alpha} = e^{-\half \left(  \abs{\beta}^2 + {\alpha}^2 - 2\beta\cc \alpha \right)},
\end{equation}
and the over-completeness relation
\begin{equation}
\int \tfrac{d^2\beta}{\pi} \ket{\beta}\bra{\beta} = \mathcal{I},
\end{equation}
where $d^2\beta = d  \realpart(\beta) d\impart(\beta)$ and $\mathcal{I}$ is the identity operator on the relevant modal Fock space, we find that indeed $\ket{\psi(t)}$ is a solution for all subsequent time, provided only that $\alpha(t)$ satisfies
\begin{equation}\label{csea}
\tfrac{d}{d t} \alpha(t) = -i\omega \alpha(t) + \tfrac{i}{\hbar}j(t),
\end{equation}
and  $\phi(t)$ satisfies
\begin{equation}\label{cseb}
\tfrac{d}{d t} \phi(t) = + \smallhalf \tfrac{1}{\hbar} \left[ \EuScript{J}(t)\cc\alpha(t) - \EuScript{J}(t)\alpha(t)\cc \right]. 
\end{equation}
Note that the real quantity $\phi(t)$ is the overall phase of the quantum state vector for this mode, which does not appear in the corresponding density matrix, and should not be confused with the actual \textit{optical} phase of the mode itself, whose
expectation value may be taken as  $\arg[\alpha(t)]$, and is not directly related to $\phi(t)$.  Since the modal equation of motion (\ref{csea}) is identical to that associated with the classical evolution of $H(t)$ with the same sources, the expectation value
$\mean{a(t)} = \alpha(t)$ simply tracks the classical field dynamics driven by the given sources, starting from a classical field given by the initial expectation value $\mean{a(0)} = \alpha(0)$ of the quantum field $a$.  Since this is true for each of a complete set of radiation modes, the result is proved.

The essential idea is that the Heisenberg equations of motions for the electromagnetic field operators
are exactly analogous to the classical Maxwell equations, or equivalently to evolution equations for
harmonic oscillators (one for each EM mode), and so if the sources are 
the same prescribed $c$-number functions  in both cases, we expect the solutions to be as nearly identical as is allowed by quantum mechanics.  The average of the quantum mechanical solution coincides with the classical solution, while the uncertainties, quantified in terms of the variances and covariances of field quantities, remain just the minimum
allowable by the Heisenberg uncertainty principle (because the amplitude and phase of each mode are conjugate), and are conventionally ascribed to ``vacuum fluctuations.'' 

This is a crucial result to our main argument, and merits some elaboration.  Coherent states can be thought of as phase-space displacement of the vacuum, \ie, a translation of the vacuum fluctuations by some average amplitude
and phase angle, so that the resulting state retains the symmetric minimum-uncertainty or ``zero-point'' fluctuations, only centered about the new average amplitude and phase.  For classical, prescribed sources, an initial coherent state remains a
minimum uncertainty state, with the centroid just following the classical trajectory

This correspondence between the classical solution and the average of the quantum
mechanical state holds however dim the radiation; that is, regardless of how weak the sources are or
how small the resulting average photon density is.  The relative uncertainty increases, or in other words the SNR decreases, but the average behavior is still exactly classical.

Note that Glauber's Correspondence Principle holds for arbitrarily relativistic sources, and the ``prescribed'' classical particle trajectories can include the effects of charges moving in non electromagnetic classical force fields, external classical EM fields such as the wiggler magnetic field, mean space-charge fields, or even the instantaneous (non-radiative) Coulomb fields of other charges, as represented in the scalar potential in the generalized-transverse gauge, just no feed-back from the radiation field itself.  In general, any appreciable recoil during the radiation process, or subsequent scattering off of other particle's real (as opposed to virtual) radiation, can destroy Glauber coherence of the radiation field.  In particular, any gain in the pickup wiggler due to the FEL instability must be negligible, so that all radiation is dominated by spontaneous rather than stimulated wiggler emission.

\subsubsection{Optical Equivalence Theorem}

While the trajectories of the particles are describable by classical mechanics and assumed definite, they
will not be known exactly but only probabilistically.  If sources are classical but stochastic in this classical sense, \ie, definite in principle but unknown, then the radiation field will be in a statistical mixture of coherent states, each with positive statistical weight, \ie, a so-called proper mixture of coherent states,  with the weights interpretable as a standard probability distribution reflecting ignorance.

It turns out that \textit{any} density matrix can be written as a weighted sum of projections into coherent states\cite{glauber:1963c, glauber:1965, glauber:1966c, glauber:1969b, sudarshan:1963, sudarshan:1965, sudarshan:1965}, if we are sufficiently generous in allowing the weights to become negative or include singularities or generalized functions.  Specifically, for the case of discrete modes, we have the diagonal coherent-state representation 
\be
\rho_{\stext{rad}} = \!\!\int \!d^2 \alpha_{\ell_1}(\omega_1)d^2 \alpha_{\ell_2}(\omega_2)\dotsb P\bigl(\alpha_{\ell_1}(\omega_1), \alpha_{\ell_2}(\omega_2), \dotsc  \bigr)   \outerp{\alpha_{\ell_1}(\omega_1);  \dotsb }{  \alpha_{\ell_1}(\omega_1); \dotsb },
\ee
where $P\bigl(\alpha_{\ell_1}(\omega_1), \alpha_{\ell_2}(\omega_2), \dotsc  \bigr)$ is known as the Glauber-Sudarshan $P$-function, or normally-ordered quasi-distribution function.

Radiation from classical prescribed sources is described by states where the $P$ function is non-negative and normalizable, so may be interpreted as a classical probability measure over the corresponding coherent states.   In addition, the quantum measurement statistics of any state with such a nonnegative $P$ function can be reproduced in  a semi-classical model (\ie, quantum detectors, classical field) by using the analogous classical field.  Conversely, all known manifestations of truly quantum mechanical behavior in radiation, such as squeezing in number or quadrature, anti-bunching, violation of Bell's inequalities, etc., that cannot be reproduced in a classical or semi-classical model, involve states with negative or even infinitely negative $P$ functions, leading to Sudarshan's \textit{Equivalence Principle}\cite{sudarshan:1963, sudarshan:1965}, which asserts that non-negativity and normalizability of the $P$ function are jointly necessary and sufficient \footnote{Necessity is now widely recognized.  Sufficiency is sometimes still debated for very weak or occasionally other states.  But any discreteness or other quantum signatures in the measurements of such a dim radiation state can always be attributed to quantum mechanical features of the \textit{detectors} rather than the \textit{fields}, so we stand by sufficiency as well.} for classical behavior of the radiation field.  

The closest analogs to classical radiation  allowed by quantum mechanics correspond to coherent states or
their proper statistical mixtures, \ie, to positive, normalizable $P$ functions; and conversely, any state for which $P$ is
normalizable and everywhere positive, so that it can be interpreted as a classical probability distribution, is in its field statistics
and properties virtually indistinguishable from the corresponding stochastic classical field, regardless of how small the
average photon number.  Note that textbook requirement of large degeneracy (in the sense of large mean modal occupation, \ie, $\mean{\EuScript{N}_j(\omega_i)} \gg 1$ ) of the relevant participating modes is, strictly speaking, neither necessary nor sufficient for classical behavior of radiation.
However, in practice increasing the degeneracy does tend to make observation of remaining quantum signatures (\ie, statistics which could not be reproduced classically) more difficult.  Conversely, at low degeneracy, any state, including coherent states or thermal states, will be subject to large relative uncertainties in photo-counts, but this can be consistently attributed to the quantum nature of the atoms in the detector, because exactly the same statistics can be reproduced in a semi-classical theory.

In many respects coherent states can be thought of as equivalent to a (classical) superposition of a deterministic classical field (itself representable as a deterministic superposition of plane waves or other classical modes of the optical system in question) and an unavoidable amount of stochastic noise associated with quantum mechanical uncertainty.

Thus, we conclude that classical (but possibly stochastic) sources produce classical-looking (but possibly
stochastic) radiation, as might be expected after a bit of reflection. 

\subsubsection{Quasi-Distribution Functions and Quantum Characteristic Functions}

The Glauber-Sudarshan $P$ function is the most natural quasi-distribution function from which to assess the classical/quantum correspondence for radiation, but others are useful in different settings.   Consider the case of a single discrete mode for simplicity, where all subscripts will be suppressed to avoid notational clutter\footnote{The results here can be simply generalized in the obvious manner to multiple discrete modes, or even to continuum modes where the expressions will involve quasi-distribution functionals, but the notation involves functional derivatives and path integrals and gets rather cumbersome.}.  We can define the $s$-ordered quantum characteristic function\cite{glauber:1969,glauber:1969b} in terms of the von Neumann density matrix $\rho_{\stext{rad}}$ as
\be
\chi(\xi; s) = \trace\left[\rho_{\stext{rad}} D(\xi) \right]e^{s\smallhalf \abs{\xi}^2}
\ee
in principle for any $s \in \realsymbol,$ but in practice the most useful choices are $s = +1, 0, -1$, naturally corresponding, as we will see, to normal-ordering, symmetric-ordering, and anti-normal ordering of the mode operators.  As in classical statistics, the derivatives of the characteristic function are related to the appropriately ordered expectation values:
\be
\mean{\left[a^{\dag l} a^{m} \right]_s } = \left. \left(\tfrac{\del}{\del \xi}  \right)^{l}_{\xi\cc} \left(-\tfrac{\del}{\del \xi\cc}  \right)^{m}_{\xi} \chi(\xi ; s)\right\rvert_{\xi = 0},
\ee
where $\left[a^{\dag l} a^{m} \right]_s$ denotes the $s$-ordered product, so that, for example,  $\mean{  \left[a\hc a \right]} = \mean{a\hc a } + \smallhalf(1-s)$.

The corresponding quasi-distribution functions are obtained by Fourier transforms, as in the classical case:
\be
F(\alpha; s) = \tfrac{1}{\pi^2} \int \! d^2\xi \,\, \chi(\xi; s) e^{\alpha \xi\cc - \alpha\cc \xi},
\ee
so that $s$-ordered moments are calculated just like normal moments of ordinary random variables:
\be
\mean{\left[a^{\dag l} a^{m} \right]_s } = \int \!d^2\alpha\,\, \alpha^{\ast l} \alpha^{m} F(\alpha; s).
\ee
The quasi-distribution functions are always real, but not necessarily non-negative for $s > -1$, and may be highly singular. Each is is a smoothed version of the preceding via convolution with a Gaussian:
\be
F(\alpha, s-1) = \tfrac{2}{\pi} \!\int \! d^2\beta\,\, F(\beta; s) e^{-2\abs{\alpha - \beta}^2}.
\ee
The normally-ordered choice $s = 1$ corresponds to the Glauber-Sudarshan $P$-function introduced above; the symmetrically-ordered choice corresponds to the Wigner Function $F(\alpha; 0) = W(\alpha) = \tfrac{1}{\pi} \trace\left[D(\alpha)\hc \Pi D(\alpha) \rho_{\stext{rad}} \right]$, where $\Pi$ is a parity operator which flips the sign of the ``momentum-like'' quadrature component but leaves the ``position-like'' component unchanged; and the anti-normally ordered choice $s = -1$ corresponds to the nonnegative and nonsingular Husimi function, $F(\xi; -1) = Q(\alpha) =\tfrac{1}{\pi} \qamp{\alpha}{\rho_{\stext{rad}} }{\alpha} \ge 0$. 

From the $P$ function, we can directly determine whether the state can exhibit non-classical behavior.  The Husimi $Q$ function is useful to graphically represent the modal phase space ``occupied'' by the state, since it is always non-negative.  The Wigner function corresponds to symmetrically-ordered (or Weyl-ordered) observables which often are of theoretical or experimental concern, and is the distribution that people implicitly have in mind when they speak of the vacuum fluctuations or zero-point motion as being equivalent to  ``half a photon'' per mode.  To see this, note that the characteristic functions for a thermal, or blackbody state containing an average of  $\bar{n}$ photons are
\be
\chi_{\bar{n}}(\xi; s) = e^{-\smallhalf(2\bar{n} + 1 - s)\abs{\xi}^2},
\ee
so if $s = 0,$ we see that in a thermal state, the contribution from vacuum fluctuations (\ie, the part independent of $\bar{n}$) is exactly equivalent to the addition of half a photon.  While obviously the $P$-function of a coherent state corresponds to a Dirac delta function, the corresponding $W$ function for a coherent state is a Gaussian occupying about $\smallhalf \hbar$ of quadrature phase space.

\subsubsection{Summary of Field State Prior to Amplification}

In practice, rather than working directly with the resulting density matrix, it will be useful to defer the averaging over the classical uncertainty in the particle trajectories until the end of the calculation.  That is, because of the linearity of the quantum evolution and of any averaging over classical ignorance, we can propagate the \textit{conditional} state given the particle trajectories,  apply these fields to the particles in the kicker, and only then average over the initial particle distribution function, in order to determine the actual statistics for the beam cooling at each pass.

This conditional state of the radiation in the coherent wiggler mode, given the particle orbits, will be a pure state, in particular a multi-mode \footnote{Viz a viz the radiation of concern to us, each particle radiates into a single coherent wiggler mode consisting approximately of a diffraction-limited and Fourier-limited wave packet of length $N_u\lambda_u$, but the eigenmodes of the optical system were taken to be time-harmonic, so many of these modes are required to represent the coherent wiggler mode.  However, only part of the zero-point noise from each of these harmonic eigenmodes appears over the actual extent of the radiated wave-packet felt by any given electron, so we can still count roughly one mode in the  ``half photon per mode'' noise rule.} Glauber coherent state, such that before any amplification, its expectation value is
\be\label{premean}
\mean{\bv{E}(\bv{x},t)}_{pre} =  \sum\limits_n \mean{\bv{E}_n(\bv{x},t)}
\ee
is equal to the corresponding classical solution consisting of the sum of the contributions from all the particles, while its variance will scale like 
\be\label{prevar}
\mean{ \abs{\Delta \bv{E}(\bv{x},t)}^2 }_{\stext{pre}} \approx  \smallhalf \frac{\hbar \omega_0}{V_u},
\ee
where $V_u \sim \pi w_0^2 N_u \lambda_0 \approx \tfrac{N_u^2 \lambda_u \lambda_0^2}{8\pi}$ is the original coherent mode volume in real space (not phase space).

\subsection{Amplification of Pickup Wiggler Radiation}

The exact quantum mechanical evolution of a mixture over such multi-mode coherent states during the amplification process will of course depend on precise details of the laser amplifiers, but by considering idealized quantum mechanical amplifiers, the best-case statistical properties of the amplified radiation and the minimum amount of added amplifier noise can be derived using just a few very basic assumptions about the input-relations.  We will follow the very illuminating fully quantum treatment of Caves\cite{caves:1982}, who extended and clarified pioneering analyses (both semi-classical and quantum) in \cite{townes:1957,schawlow:1958, louisell:1961, haus:1962, gordon:1963, gordon:1963b,kogelnik:1964, lax:1966, glauber:1967, glauber:1967b} and elsewhere.  (See also \cite{yamamoto:1986, yuen:1996} and references therein for other insightful modern treatments of quantum amplifier noise).

First, it will be supposed that the amplifier output mode field operators depend only linearly on the input
mode operators, implying that saturation effects in the output are ignored, although it
need not be assumed that the internal dynamics within the gain medium or the interactions with the
pumping or any other supplementary field modes are actually fully linear. The amplifier is further assumed to be time-stationary, or frequency-preserving and phase-preserving; \ie, an input mode at a given frequency is mapped into an amplified output mode at the same frequency (typically just the same mode,
extrapolated forward along the paraxial optical path), and if the input is shifted in phase by some constant
offset, then the output will be shifted by the same amount, so the amplifier has no absolute internal
clock, nor any preferred phase bias.  For transit-time OSC, this time-stationarity is desirable, or even essential, because the information used by the cooling mechanism to reduce particle phase space is stored almost exclusively in the phases rather than amplitudes of the light, and efficient field/particle coupling in the kicker will require that the light frequency remains near the resonant frequency of the particles in the wiggler.

To develop the needed gain, the amplifier can be multi-stage (cascaded), but must be single-pass, \ie, non-regenerative, for two reasons.  Firstly, the amount of total particle time-of-flight delay that can be realistically introduced between pickup and kicker compared to the proper time between these points (\ie, the time it would take light to travel between them in vacuum) is not very great for heavier particles like muons traveling at near-luminal velocities, so there simply is not time to circulate the light before injecting it into the kicker coincidentally with the particles that radiated it.  Secondly, a regenerative amplifier operating at high pump power can easily be pushed over the oscillation threshold, where it converts from an amplifier preserving the phase and modulations of the input to a laser oscillating very cleanly but at essentially some random phase.

\subsubsection{Quantum Mechanics of Linear Optical Amplifiers}

In the Heisenberg picture, where time-evolution is ascribed to the operators, it is supposed that the amplifier output mode field operators depend only linearly on the input mode operators  The amplifier is further assumed to be time-stationary, that is, frequency and phase-preserving.  Consistent with the 1D dynamics adopted throughout the current analysis, it is also taken to preserve transverse coherence, in the sense of maintaining a one-to-one relationship between input and output transverse modes.  As demanded by quantum mechanics, the time-evolution of the field must be unitary, and the field observables must satisfy appropriate commutation relations and Heisenberg uncertainty relations at all times before, during, and after amplification.

It turns out that these basic assumptions about linearity and lack of phase bias, along with the axioms of quantum mechanics are sufficient to characterize the action of and best possibly noise figures for any such linear amplifier, without needing to know the specific details of the amplifier system, or explicitly following the evolution of all the DOFs in its gain medium.

In the Heisenberg picture, we take $a_{\ell}(\omega)$ to represent the continuum field annihilation operator for a mode \textit{before} amplification (\ie, the input mode operator) and $b_{\ell}(\omega)$ to represent the annihilation operator \textit{after} amplification (\ie, the output operator). The modes are specified by a continuum frequency $\omega$ as well as some discrete index $j$ meant to represent different transverse paraxial modes, or otherwise lift degeneracy.  The input modes satisfy the canonical bosonic commutation relations (\ref{ccomm2}) as must the output modes, \ie 
\bsub
\begin{align}
\comm{b_{\ell}(\omega)}{ b_{\ell}(\omega')\hc} &= \delta_{\ell\, \ell'}\delta(\omega - \omega')\\
\comm{b_{\ell}(\omega)}{ b_{\ell'}(\omega')^{\phantom{\dag}}} &= 0,
\end{align}
\esub
consistent with the quantum mechanical evolution remaining unitary.
By linearity, we mean that the input and output modes are related by affine transformation, so we are disregarding any sort of saturation effects in the gain medium.  The most general possible linear input-output relation is then
\be
b_{\ell}(\omega) = \sum\limits_{\ell'}\!\int \!d\omega' \left[ M_{\ell \ell'}(\omega, \omega') a_{\ell'}(\omega') + L_{\ell \ell'}(\omega, \omega')a_{\ell'}(\omega')\hc\right] + F_{\ell}(\omega)
\ee
Next we demand that the input-output relations are time-stationary, meaning both frequency and phase-preserving.
By frequency preserving we mean that input at frequency $\omega$ is mapped to output at frequency $\omega$, while by phase-preserving we mean that a phase rotation of the input leads to the same phase rotation of the output, and that if the input has time-stationary noise statistics, then so will the output.  These are desirable properties for any amplifier to be used for transit-time OSC, and together imply that
\bsub
\begin{align}
M_{\ell \ell'}(\omega, \omega') &= M_{\ell \ell'}(\omega)\delta(\omega-\omega')\\
L_{\ell \ell'}(\omega, \omega') &= 0.
\end{align}
\esub
Finally, we assume a one-to-one relationship between input and output transverse modes, so that the matrix $M_{\ell \ell'}(\omega)$ is unitarily diagonalizable at any frequency $\omega$.  Actually, without any loss of generality under our assumptions, and to avoid introducing new notation, we may suppose that the $a_{\ell}(\omega)$ and $b_{\ell}(\omega)$ have \textit{already} been chosen in the proper linear combination to correspond to the transverse eigenfunctions of the amplifier, so that $M_{\ell \ell'}(\omega)$ is already diagonal:
\be
M_{\ell \ell'}(\omega) = \delta_{\ell \ell'}M_{\ell}(\omega),
\ee
and the input-output relations for the amplifier become simply
\be
b_{\ell}(\omega) = M_{\ell}(\omega)a_{\ell}(\omega) +  F_{\ell}(\omega).
\ee
Obviously $M_{\ell}(\omega)$ is the amplifier transfer function for the $j$th mode of frequency $\omega$, such that
$M_{\ell}(\omega)\hc M_{\ell}(\omega) =  G_{\ell}(\omega)$ is the power gain for the mode.  In general, $M_{\ell}(\omega)$ can be an operator ($q$-number) commuting with all input mode operators:
\be
\comm{a_{\ell}(\omega) }{ M_{\ell'}(\omega') } = \comm{a_{\ell}(\omega)\hc }{ M_{\ell'}(\omega')^{\phantom{\dag}}} = 0.
\ee
The additive noise operators $F_{\ell}(\omega)$ must be present to ensure unitarity.  They represent independent degrees-of-freedom corresponding to some internal or auxiliary states of the amplifier, so can be taken commute with the input and output modes.  In fact, in order that the bosonic commutation relations are preserved, we can deduce the commutation relations
\bsub
\begin{align}
\comm{a_{\ell}(\omega) }{ F_{\ell'}(\omega')\hc} &=  \comm{a_{\ell}(\omega) }{ F_{\ell'}(\omega')^{\phantom{\dag}}} = 0,\\
\comm{F_{\ell}(\omega)}{ F_{\ell'}(\omega')^{\phantom{\dag}}} &= 0,\\
\comm{F_{\ell}(\omega)}{ F_{\ell'}(\omega')\hc} &= \delta_{\ell\, \ell'}\delta(\omega - \omega')\left[1 - M_{\ell}(\omega)M_{\ell}(\omega)\hc \right],
\end{align}
\esub
for the noise operators $F_{\ell}(\omega)$.

To complete the description of the amplifier, we must specify an initial state (density operator) for the scaling and noise operators.
In the Heisenberg picture, time evolution resides in the operators, so this operating state is fixed, and should be independent of any possible input signal.  That is, the density matrix of the total system (modes to be amplified plus all auxiliary modes internal DOFs of the amplifier) factorizes: $\rho_{0} = \rho_{a}\otimes \rho_{\stext{amp}}.$

No fundamental physical principle imposes any lower bound on the variances and covariances of the $M_{\ell}(\omega)$, so in the ideal case they can be taken to be $c$-numbers, implying no \textit{multiplicative} quantum noise in the output, that could arise from variations in the gain via fluctuations in the pump strength, for example.  Besides, such noise can easily be added back later, if needed.  With $G_{\ell}(\omega) = \abs{M_{\ell}(\omega)}^2 \ge 1$ and deterministic, note that we can write $F_{\ell}(\omega) = \sqrt{G_{\ell}(\omega) - 1} \, f_{\ell}(\omega)\hc$, where $f_{\ell}(\omega)$ is a bosonic annihilation operator for what may be interpreted as the effective internal amplifier mode that contributes the extra noise.  Nothing prevents an assumption excluding bias in the noise, such that  $\mean{F_{\ell}(\omega)} = 0$, and in fact our previous assumptions about phase preservation require it.

But the commutation relations together with a generalized Heisenberg Uncertainty Principle applicable to creation and annihilation operators constrain the second-order noise statistics.  For the symmetrized covariances, we find the inequality
\be
\smallhalf \mean{F_{\ell}(\omega)\hc F_{\ell'}(\omega') + F_{\ell'}(\omega') F_{\ell}(\omega)\hc} - \mean{F_{\ell}(\omega)\hc}\mean{F_{\ell'}(\omega')} \ge \tfrac{1}{2} \abs{\comm{ F_{\ell}(\omega)}{F_{\ell'}(\omega')\hc } }
\ee
In particular, for coherent-state input, the output noise statistics, in terms of symmetrized covariances become,
\be
\mean{ b_{\ell}(\omega)b_{\ell'}(\omega') } - \mean{ b_{\ell}(\omega) }\mean{b_{\ell'}(\omega')} = 0;
\ee
and
\be
\begin{split}
\tfrac{1}{2}  \mean{ b_{\ell}(\omega)\hc b_{\ell'}(\omega') +  b_{\ell'}(\omega') b_{\ell}(\omega)\hc }  &-  \mean{ b_{\ell}(\omega)\hc } \mean{b_{\ell'}( \omega') }  = \\
&G_{\ell}(\omega)\delta_{j \,j'}\delta(\omega-\omega')\left[  \tfrac{1}{2} + S_{\ell}(\omega) \right].
\end{split}
\ee
 The first term on the right-hand side represents the amplified ``vacuum fluctuations'' \footnote{We will follow standard convention and refer to the quantum mechanical uncertainties in the vacuum state or coherent states \textit{qua} displaced vacuum states as ``vacuum fluctuations'' or ``zero-point motion,'' even though the mathematical formalism clearly identifies them as predictive uncertainties, not physical fluctuations.} present in the input, equivalent in noise to ``half a photon per unit bandwidth'' before
amplification, while the second term represents the extra noise added by the amplifier (typically through amplified spontaneous
emission in the gain medium), in which the added-noise spectral density $S_{\ell}(\omega)$ must satisfy:
\be
S_{\ell}(\omega) =\tfrac{1}{2} \abs{ 1 - \tfrac{1}{G_{\ell}(\omega)} }  \sigma^{2}_{\ell}(\omega),
\ee
where $\sigma^{2}_{\ell}(\omega) \ge 1,$  with equality in the Heisenberg-limited case, so at large gain the added noise power is
the equivalent of at least one additional ``half photon'' per mode of input to the amplifier.

For a single mode, note that the fluctuations in photon number in the output are super-Poissonian.  If the input is coherent, the Fano factor starts off as unity: $\tfrac{ \mean{(\Delta \EuScript{N}_a)^2} }{ \mean{\EuScript{N}_a} } = 1$ but at output, becomes $\tfrac{ \mean{(\Delta \EuScript{N}_b)^2} }{ \mean{\EuScript{N}_b} } \approx G+1 \sim G \gg 1,$ where here $\EuScript{N}_a = a\hc a$ and $\EuScript{N}_b = b\hc b.$
The output state is chaotic, not coherent.

\subsection{Quasi-Distribution Functions}

For simplicity, we will temporarily drop all the indices and arguments and for the moment treat a single discrete input mode $a$, output mode $b$ and internal amplifier mode $f$, assuming a gain factor $G \gg 1$. The multi-mode results are exactly analogous, only the notation is quite cumbersome.  While we worked with symmetrically-ordered operators above, it is more convenient here to begin with normally-ordered characteristic functions and the corresponding Glauber-Sudarshan quasi-distribution function.

We assume the input mode $a$ is in a Glauber coherent state.  Actually, it should be described in some statistical mixture of Glauber states, weighted by the particle distribution function, but again, that averaging will be deferred until later -- in effect, we are here conditioning on supposed knowledge of the classical particle trajectories in the pickup wiggler.

Recalling that the combined density matrix for the input and amplifier factorize, and the input and internal operators all commute,
the normally-ordered characteristic function of the output mode is then
\be
\begin{split}
\chi_b(\xi, 1) &= \trace\left[ \rho_0 e^{\xi b\hc - \xi\cc b    }  \right]e^{\smallhalf \abs{\xi}^2} =
\trace\left[ \rho_a \otimes \rho_{f} e^{\xi (M\cc a\hc + F\hc) - \xi\cc (Ma+F) }  \right] e^{\smallhalf \abs{\xi}^2} \\
&= \trace\left[ \rho_a e^{\xi M\cc a\hc - \xi\cc Ma} \right]   \trace\left[ \rho_{f} e^{\xi \sqrt{G-1}\, f\hc -   \xi\cc \sqrt{G-1}\, f }  \right]  e^{\smallhalf \abs{\xi}^2}\\
&= \chi_a(M\cc \xi, 1) \, e^{-\abs{\xi}^2(G-1)} \, \chi_f(\sqrt{G-1}\, \xi , 1) 
\end{split}
\ee
Now, for a coherent state input, it is easy to see that  $\chi_a(M\cc \xi, 1) = \chi_{Ma}(\xi, 1)$, \ie, the first term is the characteristic function for a deterministically scaled coherent state with expectation $\mean{Ma} = M \mean{a}$.
The second term on the RHS is recognized as the normally-ordered characteristic function for a thermal (chaotic) state with an average of $\bar{\mathcal{N}}_n = G-1$ photons. The final term is the characteristic function for the internal mode operator $f$ after scaling by $\sqrt{G-1}$.  Considerations of the central limit theorem or of maximum entropy suggest that the added noise appearing through the initial state of the $f$ mode should, in the absence of further specific information, also be taken as thermal: 
$\chi_f(\xi, 1) = e^{-\bar{\mathcal{N}}_f \abs{\xi}^2}$, where $\mean{f} = \mean{f^2} = \dotsb = 0$, and we have defined $\bar{\mathcal{N}}_f \equiv \mean{f\hc f} \ge 0,$ so the characteristic function for the amplified field mode becomes
\be\label{fieldstate8}
\chi_b(\xi, 1)  = \chi_{Ma}(\xi, 1) \,e^{-(\bar{\mathcal{N}}_f +1)(G-1) \abs{\xi}^2}
\ee
According to the Fourier convolution theorem, we see that after amplification, the resulting Glauber-Sudarshan $P$-function will be a displaced chaotic state, or equivalently the sum of a multiple of the input coherent state and an independent additive chaotic noise term corresponding to about $(\bar{\mathcal{N}}_f +1) \left( 1 - \tfrac{1}{G}\right)$ thermal photons at input, where $\bar{\mathcal{N}}_f \to 0$ in the ideal limit.  It is easy to verify that the Fourier transform of (\ref{fieldstate8}), \ie, the $P$-function, is nonnegative and normalizable --- in fact it is just a Gaussian, centered on the appropriate multiple of the input state, and with a width characteristic of a thermal state.

\subsubsection{A Heuristic (Single-Mode) Picture}

Any one particle interacts in the kicker with radiation from itself particles within a coherence length, \ie, with radiation that is ``almost' confined to one mode.   Therefore, while keeping in mind that the quantum state of the radiation is not actually diagonal in any number state basis and that a more careful multi-mode analysis is strictly necessary because the spectral bandwidth is not vanishingly small and the effective number of longitudinal modes in the entire radiation field exceeds unity, we can nevertheless paint the following suggestive picture: before amplification, the radiation field from the pickup wiggler that would be seen by any one particle ``consists'' of, on average, the equivalent of:  $O(\alpha)$ self-radiated photons constituting the cooling signal; $\smallhalf$ photon of ``vacuum fluctuations,'' and $O(N_s \alpha) \gtrsim 1$ photons radiated from other particles in the sample.  The relevant SNR is approximately $\tfrac{\alpha }{ \alpha N_s + 1/2}.$

After amplification in an ideal, quantum mechanical phase-preserving linear amplifier, with overall gain factor $G \gg 1$, the field seen by the same particle is essentially classical but stochastic, containing the equivalent of $O(G\alpha)$ photons with proper phasing on average for cooling, $\smallhalf G$ randomly-phased photons from the amplified vacuum fluctuations, at least another $\smallhalf G$ photons from the noise added by the internal amplifier mode or modes, and $O(GN_s\alpha)$ from amplification of the other particle's radiation, with approximately $O(G)$ photons at each of about $O(N_s \alpha)$ different random phases.  The SNR becomes approximately $\tfrac{\alpha}{\alpha N_s + 1}$, comparable to, but slightly worse than, the pre-amplified SNR.

Quantum amplifiers do not improve the SNR, but ideally amplify the input without serious degradation of the SNR, up to some sufficiently high level, where the field can be be measured,  manipulated, or employed  by course-grained classical means without any significant further loss of SNR by quantum back-action.   But the essential point is that neither the pre-amplified nor post-amplified SNR  is ``catastrophically'' affected by quantum mechanical fluctuations or uncertainties inherent in emission, or amplification,  through multiplicative noise,  but rather the effects are just roughly equivalent to having an extra $O(\alpha\inv)$ particles in the sample.  Of course, this equivalence to extra sample particles is not precisely true, because the noise photons added by the amplifier will not have exactly the same spectrum or quasi-distribution function as the photons actually radiated in the wiggler, but if $N_s \gtrsim O(10)$, the Central Limit Theorem suggests that the net field from all the other particles in the sample will also be close to a Gaussian chaotic state, as will the sum of the sample-particle and noise contributions.

\subsubsection{The Bottom Line on Quantum Amplification and the Amplified Field}

Conventional linear phase-preserving amplifiers do not first measure then multiply photon number states, but instead linearly amplify the entire input field, while unavoidably adding additional noise.

In each mode, the additive noise is equivalent at input to the original vacuum fluctuations, themselves equivalent to about ``half a photon,'' plus another independent set of vacuum fluctuations, another ``half a photon.''

As Caves\cite{caves:1982} aptly put it, ``quantum mechanics extracts its due twice.''  After amplification in a high-gain amplifier, the two non-commuting quadrature components of a field mode may be measured classically with little further degradation in the SNR.  But simultaneous measurement of non-commuting variables must involve auxiliary variables that do commute, and whose uncertainties add to the uncertainty already present \footnote{As a corollary, we deduce that if somehow only one quadrature component is amplified, then no \textit{additional} half photon of noise at input need be added, beyond that represented in the original ``vacuum fluctuations.''}.

Once linearity is demanded of the amplifier, it must amplify input vacuum fluctuations along with both quadrature components of any signal, and indeed cannot distinguish between what we, the cooling system, or the beam particles might consider signal versus noise.  There are of course states of the combined system consisting  of the input fields modes and the amplifier for which the additional noise in the form of a second set of vacuum fluctuations is not present in the output, but they will necessarily  involve correlations between the input and internal modes, but because any internal amplifier modes must be prepared independently of the input, their zero-point motion will be amplified and added to the output (physically, ultimately through the mechanism of spontaneous emission.)

But apart from this doubled additive noise, quantum mechanics does not require anything like the speculated multiplicative Poissonian noise due to ``quantum jumps'' that we found so deleterious in our naive model.

Despite naive fears to the contrary, the signal \textit{does} contain the needed phase information, albeit
corrupted by a large amount of noise which, however, is random in phase (unbiased), independent of the
input state, and purely additive.

The noise added by the amplifier can be characterized by an effective temperature assuming it roughly thermal in distribution, or as an equivalent number of photons at input, or perhaps most approximately but most usefully, as an equivalent number of extra particles in the sample.  Because the cooling signal in the coherent
portion of the wiggler radiation spectrum was approximately $O(\alpha^{-1})$ photons on average before amplification, while the total noise power at input is about $O(1)$ photons, the amplifier noise is expected to be the equivalent of approximately $N_{n} \sim O(\alpha^{-1})$ extra particles in the sample.  While not nearly as detrimental to cooling as the noise predicted from the quantum jump model, this will eventually saturate improvements  expected from further beam stretching, and limit the ultimate rates of cooling and the equilibrium emittances and energy spreads achievable.

When the gain is large, the radiation density matrix for the amplified field, conditioned on a given set of particle trajectories, is described as a biased chaotic state, essentially indistinguishable from a classical but stochastic light field consisting of the superposition of the deterministically-scaled classical pickup radiation, together with some admixture of chaotic light characterized by a certain noise temperature proportional to the overall gain.
Because the Glauber-Sudarshan $P$ function of such a displaced chaotic state is everywhere nonnegative and well-behaved, the field can be interpreted classically.

In fact, the post-amplified field values can be taken as Gaussian $c$-number random variables with means 
\be\label{postmean}
\mean{\bv{E}(\bv{x},t)}_{\stext{post}} \approx \sqrt{G} \mean{\bv{E}(\bv{x},t)}_{\stext{pre}}
\ee
and variances
\be\label{postvar}
\mean{ \abs{\Delta \bv{E}(\bv{x},t)}^2 }_{\stext{post}} \approx G \mean{ \abs{\Delta \bv{E}(\bv{x},t)}^2 }_{\stext{pre}}
\ee
so the SNR is approximately preserved.
Calculation of the cooling dynamics can hereafter proceed classically, only with the added stochastic contributions from the quantum vacuum fluctuations and the amplifier noise included.

\section{Cooling Dynamics for a Simplified Model}

After using the hemi-classical radiation model and this very general formalism for linear amplifiers,
determination of the interaction of the particles and amplified optical field within the kicker and calculation of the resulting statistics for cooling kicks on each pass can proceed essentially classically.  
Because we are here primarily interested in questions of principle, and endeavor to demonstrate that quantum effects properly considered will not catastrophically affect cooling rates, the cooling statistics can be worked out  explicitly, albeit for a highly idealized, one-dimensional model, wherein we make a number of simplifying assumptions in order to achieve a greater measure of tractability.

The average beam energy $\gamma_{0} \gg 1$ is assumed highly relativistic, and the energy deviations 
$\delta\gamma_{j} = \gamma_{j} - \gamma_{0}$ are assumed sufficiently small, so that $\abs{\delta \gamma_{j}} \ll \gamma_{0},$ and $\abs{\gamma_{j} -\gamma_{k}} \ll \gamma_{0},$  and also $\abs{v_{z _j}(t) - v_{0}} \ll c,$ where $v_{0} = c\beta_{0} \approx c\sqrt{1 - \tfrac{1}{\gamma_{0}^2}}$ is the mean longitudinal velocity outside the wigglers.  All particles then approximately have the same longitudinal velocity and the same magnitude of transverse quiver momentum inside the wigglers, and their trajectories can be be determined from conservation of energy and of transverse canonical momentum.  Only longitudinal cooling is considered here; the possibility of cooling of transverse DOFs, by making part of the time-of-flight delay proportional to betatron errors, and then exploiting dispersion in the lattice, is here omitted for simplicity; the calculation is similar if somewhat more involved.  Because the beam is highly relativistic, the change in longitudinal momentum is related to the energy kick by 
\be
\tfrac{\Delta p_{z_j}}{p_{z_{j}}}  \approx  \tfrac{\Delta p_{z_j}}{p_{0}} \approx  \tfrac{\Delta
\gamma_{j}}{\gamma_{0}} \approx \tfrac{\Delta \gamma_{j}}{\gamma_{j}}
\ee
where $p_{0} \approx m c \gamma_{0} \beta_{0}$ is the average
longitudinal momenta of the beam particles.

In the kicker, the wiggler fields are much stronger than the amplified  radiation fields from the pickup and vastly stronger than any newly-emitted spontaneous radiation.   So just classical wiggler fields together with the particle initial conditions at the wiggler entrance are assumed to determine particle spatial trajectories, with negligible perturbations from the radiation field or space-charge fields.
Both wiggler fields are assumed to be planar and perfectly transverse, with no transverse variation over the extent of the beam, and exactly sinusoidal in their longitudinal variation, requiring $\lambda_u \gg \sigma_{\perp}$ and $\tfrac{a_u}{\gamma_0} \ll 1$, both satisfied with orders-of-magnitude to spare.
That is, the wiggler magnetic fields in both kicker and pickup are assumed to be of the form $\bv{B}_{u}(\bv{x}) \approx \unitvec{y} \sin \bigl(k_{u}(z-z_{u})\bigr)$ over the extent $z_{u} \le z \le z + N_{u}\lambda_{u}$ of each wiggler, with fringe fields neglected.

Off-axis effects and all transverse motion of particles are ignored accept the idealized $1$D quiver motion in the wiggler field, as determined by canonical momentum conservation.  Higher harmonics of particle motion the occur in planar wiggler are ignored, and the time-averaged longitudinal components of the unperturbed orbits are deduced from the transverse quiver and conservation of kinetic energy in a magnetic field.  The effects of space-charge forces and any plasma waves or other collective effects on the beam distribution are expected to be negligible and are ignored.  Any external focusing is ignored as well, although in practice these forces may not be small.  

Specifically, with the effects of energy variation and possible off-axis injection ignored,  the transverse
motion for each particle inside the wiggler is determined within our approximations by the conservation of transverse canonical momentum conservation in the wiggler field only:
\begin{equation}\label{undulation}
\beta_{\perp_j}(z) \approx \frac{1}{\gamma_{0}}\tfrac{q}{\abs{q}} a_u \sin\bigl(k_{u} (z-z_u)\bigr) \hphantom{yyy} \text{ for }\hphantom{yy}  z_u \le z \le z_u + N_u \lambda_u,
\end{equation}
where we can use $\gamma_0$ rather than $\gamma_j$ because the errors will be of higher order than terms retained in our approximate calculation.

Because variations in longitudinal velocity between different particles are neglected, and the fast
oscillating component of this velocity due to motion in a planar wiggler is ignored, for the  longitudinal motion one finds
\begin{equation}\label{longvel}
\beta_{z_j}(t) \approx \beta_0 \tfrac{1}{1 +
\lambda_{0}/\lambda_{u}},\end{equation}
so that $\tfrac{1-\beta_{z_j}}{\beta_0}  \approx \tfrac{\lambda_{0}}{\lambda_{u}}  \ll 1,$ and the time spent in the kicker wiggler is then $T_{j} \approx \tfrac{N_u\lambda_{u}}{c \beta_0}$ for any particle.

Spectral broadening or other effects on the spontaneous emission spectra from the pickup, due to off-axis injection or the angular deviations associated with the transverse quiver motion are ignored, and the expected (\ie, classical) emission from each particle is assumed to consist of exactly $N_{u}$ plane-polarized, sinusoidal oscillations at central frequency $\omega_{0}$, emitted on-axis, with amplitude $E_{0}$, with expected phase determined by the random longitudinal position of the particle in the beam, and extending transversely over some spot size $w_0$ larger than the transverse beam size and the extent of a single particle's quiver orbit, \ie, for which  $w_0 \ge \sigma_{b \perp}$ and $w_0 \ge \tfrac{a_u}{\gamma_0} \tfrac{\lambda_u}{2\pi}$.  At best, the latter conditions are probably only marginally satisfied with the present parameters,  but for simplicity we do neglect for now any transverse structure or imperfect transverse coherence in the radiation fields.

A single plane-polarized fundamental transverse (say $\ell = 1$) mode of the pickup radiation is assumed to have been amplified by an ideal laser with a flat gain $G >1$ profile and negligible linear or nonlinear dispersion over the bandwidth $\Delta \omega_{L} \gtrsim \Delta \omega$.

Outside the amplifier, all radiation is assumed to propagate purely longitudinally according to the vacuum dispersion relation.  After amplification, the field is stochastic due both to the original beam shot noise and the added amplifier noise,  but is essentially classical in its statistics and dynamic behavior, consistent with hemi-classical quantum optics detailed above.

The bypass lattice is assumed to introduce a time-of-flight variation linearly proportional to energy deviation $\delta\gamma_{j} = \gamma_{j} - \gamma_{0}.$  The energy change in the kicker is calculated to leading order in $\frac{1}{\gamma}$ using the \textit{unperturbed} quiver orbit of the particle in the kicker field.  Any particular cooling kick on a single pass for a single particle will also be relatively small compared to the particle's total kinetic energy, \ie, $\abs{\Delta \gamma_{j}} \ll \gamma_{0} - 1,$ but \textit{not} necessarily very small compared to the energy deviation $\delta \gamma_{j}$. 

The bunch as a whole is assumed to contain many particles, $N_{b} \gg 1,$ and after stretching 
the beam density $\rho_b$ is assumed to be uniform over the entire stretched length, which is presumed very long compared to the sample length, \ie, $L_{b} \gg N_{u} \lambda_0,$ so that the relative size of any sample is small: $N_{s} \ll N_{b}$.   (A Gaussian bunch with $\sigma_{b\parallel} \sim L_{b}$ is also analytically tractable).  Each sample will still contain a moderate number of particles, typically $N_s \sim O(10^2)$.

Nearby particles will experience similar kicks and will become partially correlated, but if the sample sizes are small in relation to the beam as a whole, this correlation effect is negligible when averaged over all particles in the beam.  So if good mixing between cooling kicks is assumed, so the noise effects in a given particle's kicks can be taken to be statistically independent from kick to kick, and the full kicks (including coherent and incoherent parts) can be taken to be effectively uncorrelated for different particles, then only the single-particle, reduced distribution function at one moment of time need be tracked.  In fact,  the energy distribution will be assumed to start and remain Gaussian, so that only first-order and second-order moments need by retained.

Many of these assumptions are fully justifiable in our parameter regime, while others are rather drastic 
and will eventually need to be relaxed in a more realistic calculation, but should be adequate for our present goal, a proof-of-principle that quantum noise is no deal-breaker for the scheme.

\subsection{Bunch Stretching and Compression}

We model the initial bunch stretching and final bunch re-compression very simply, as symplectic maps which ultimately leave the transverse phase space distribution unchanged, while preserving the action of the longitudinal phase space. 

Recall that the stretching is to be accomplished by drift in a ring with a high-momentum-compaction-factor, in order to disperse the beam based on energy spread, followed by use of a LINAC to remove the resulting head-to-tail correlation between longitudinal position and energy.  After cooling, the process is reversed to re-compress the beam, by first applying an appropriate energy chirp with the LINAC then allowing the beam to drift in the compaction ring, where now particles in the tail will tend to catch up to those in the head of the beam.  Obviously, the transverse phase space cannot remain unaffected or even uncorrelated with the longitudinal phase space at intermediate times, \textit{during} the actual stretching or compression, because particles with greater longitudinal momenta are being forced onto longer orbits by highly-dispersive beam optics, but it can hold to a reasonably good approximation \textit{after} the expansion or contraction is completed.  We assume that the stretching/compression ring is sufficiently long to avoid ``wrap-around'' effects, which will lead to striations in longitudinal phase space.  The dispersive drift shears the occupied phase space horizontally, while the ramped kick from the LINAC can shear it more or less vertically.  Assuming all the head-to-tail energy variation is removed by the LINAC, and correlations between transverse and longitudinal DOFs that build up at intermediate times are undone by proper lattice design, then from Liouville's theorem, supposing $t_b$ and $t_a$ are some times before and after any stretching, then
\be
L_b(t_a)  \delta\gamma(t_a) \approx L_b(t_b) \delta \gamma(t_b).
\ee
If the bunch length satisfies $L_b{t_a} \gg L_b(t_b)$, then the energy spread satisfies $\delta \gamma(t_a) \ll \delta\gamma(t_b)$.
Not only does stretching lower the beam density $n_b$ and the resulting sample size $N_s,$ but it also (reversibly) reduces the range of energy variation $\delta \gamma$, which relaxes design constraints on the bypass lattice.  If subsequent OSC then (irreversibly) reduces the energy spread by some factor $\mu$ by time $t_c$ without affecting the overall bunch length, \ie, $\delta\gamma(t_c) \approx \mu \delta\gamma(t_a)$ while $L_b(t_c) \approx L_b(t_a)$,  and then the bunch is re-compressed  to its original length by the final time $t_f$ by reversing the stretching procedure, then 
\be
\delta \gamma(t_f) \approx  \delta\gamma(t_c) \tfrac{L_b(t_c)}{L_b(t_f)} = \mu \,\delta\gamma(t_b),
\ee
so the final energy spread after cooling and re-expansion remains reduced by the same \textit{factor} $\mu$ compared to the initial energy spread of the incident (uncompressed) beam.

The stretching/compression phase does raise some non-trivial practical issues that appear to have been overlooked.  If $N_b \sim O(10^9)$ and $L_b(t_a)  \sim O(10^{-1}\; \text{m})$, then to get down to sample sizes of $N_s \sim O(10^2)$ for $N_u \sim O(10)$ and $\lambda_0 \sim O(10^{-6}\; \text{m})$ requires a stretching factor of $L_b(t_b) / L_b(t_a) \sim O(10^3)$.  But the time required for a relativistic beam to traverse a compaction ring of circumference $C_0 \gtrsim L_b(t_b) \sim O(10^2\; \text{m})$ is already about $t \sim O(10^{6} \;\text{s})$.  The problem is that if we try to damp for several $e$-foldings such that $\mu = O(10^{-2})$, then the when it comes time to re-compress the bunch after cooling, we naively expect the compression to take $O(\mu\inv) \sim O(10^2)$ longer in the same lattice (with the same dispersion function),  because the energy deviations are proportionally smaller.  So the stretching and compression is not entirely trivial, and, unfortunately, perhaps different rings must be used for the compression and re-stretching, but we leave lingering questions for another day. 

\subsection{Energy Kick Statistics for Individual Particles}

Within this simple model, the first-order and second-order statistics of the cooling kicks for individual particles can be worked out analytically by lengthy but straightforward calculations, at least to leading order in the various small quantities, so we will summarize the results without presenting the unilluminating mathematical details and intermediate steps.

\subsubsection{First-Order Statistics}

Within the kicker, recall that the effects of both beam self-fields and any newly-emitted spontaneous radiation can be assumed negligible compared to those of the wiggler fields and amplified radiation from the pickup wiggler.  The latter fields, while highly amplified, still have a small effect on the actual \textit{spatial} trajectories followed by the particles, as compared to the effect from the static wiggler field, so each single-particle orbit can be assumed to be determined by initial conditions at the entrance to the kicker and by the Lorentz forces produced by the wiggler field only.  Of course, the whole point of the kicker interaction is to change the \textit{energy} of the particle, and any such energy kick must arise from work done by a component of the optical electric field $\bv{E}(\bv{x},t)$ in the kicker parallel to the particle velocity.

The total energy change inside the kicker for the $j$th particle during a given pass
through the cooling section can then be estimated as:
\begin{equation}\label{deltagamma1}
m c^2 \Delta \gamma_{j} \approx q\!\!\!\!\!\!\!\!\!\int\limits_{t_{j}+\Delta t_{j}}^{t_{j} +
\Delta t_{j}  + T_{j}}\!\!\!\!\!\!\!\!\!\! dt\;  \bv{v}_{\perp j}(t) \cdot \bv{E}\bigl(\bv{x}_{j}(t), t \bigr),
\end{equation}
where $q$ is the (signed) charge of the particle; $\bv{x}_j(t)$ is the
corresponding particle trajectory inside the wiggler, and $\bv{v}_{j}(t) = \tfrac{d}{dt}\bv{x}_{j}(t)  = v_{z_j}(t)
\hat{z} + \bv{v}_{\perp_j}(t)$ is the particle velocity, both of which can be approximated assuming the dynamics are
governed only by the wiggler field; $\bv{E}(\bv{x}, t)$ is the electric field of the amplified pick-up radiation as it
propagates through the second wiggler; $t_{j}$ was the arrival time of the particle at the entrance of the pickup wiggler;
$\Delta t_{j}$ is the total time delay between arrival at the first wiggler and arrival at the second wiggler, depending on the
time-of-flight delay introduced in the bypass lattice; and $T_{j}$ is the duration of time spent in the second wiggler.  That
is, the energy kick due to the interaction of the optical electric field with the transversely quivering particle can be calculated in a perturbative fashion, in which to leading-order the effects of the changing energy on the particle orbit are neglected.

Neglecting various small terms, the conditional average energy kick for the $j$th particle, given values for the reference energy and of its own pre-kick energy deviation, but averaged over all intrinsic quantum mechanical uncertainty in the radiation as well as over the classical shot noise, \ie, random arrival times of other particles in the its sample, is given by
\be\label{avgkick}
\innerp{ \Delta\gamma_{j} }{ \gamma_{j}, \gamma_{0} } 
 \approx \tfrac{q a_{u} N_{u} \lambda_{u} \sqrt{G}E_{0}}{2 m c^{2}\beta_{0}\gamma_{0}}\Theta\left(1 -
\tfrac{\abs{\phi_{j j}}}{2\pi N_{u}}\right)\left[1 - \tfrac{|\phi_{jj}|}{2\pi N_{u}}\right]\sin(\phi_{j j}),
\ee
where $\Theta(s)$ is the Heaviside step function, $\phi_{j j} = \phi_{j j}(t_{j},\gamma_{j},\gamma_{0})$ is the phase of the $j$th particle's amplified self-field
as seen by that same particle at the entrance to the kicker, while $t_{j}$ is the time of arrival of particle $j$ at the pickup.

Note that $\phi_{j j}$ is determined by the particle's time-of-flight
through the pickup wiggler and bypass lattice compared to the delay of the pickup radiation during propagation and amplification as well as any further phase-shift introduced by the laser amplifier system.
If the particle delay in the bypass lattice is arranged such that for sufficiently small energy deviations $\delta\gamma_j$, the relative delay leads to $q\sin(\phi_{j}) \propto
-\delta\gamma_{j} + O(\delta\gamma_{j}^{3})$, then the particle will experience a \textit{restoring force} on
average, tending to push the energy back toward that of the reference orbit.
Ideally, the probable range of variation in $\delta\gamma_{j}$ before the cooling kick
is mapped into just $\pm \tfrac{\pi}{2}$ of phase delay, so that the magnitude of the damping force grows monotonically with the
field gain $\sqrt{G}$ and the magnitude $\abs{ \delta\gamma_{j} }$ of the energy deviation.

\subsubsection{Second-Order Statistics}

After some more algebra, the corresponding conditional expectation of $\Delta\gamma_j^{2}$ becomes approximately:
\be\label{secondmoment}
\begin{split}
\innerp{ (\Delta\gamma_{j} )^{2} }{ \gamma_{j}, \gamma_{0}}  &\approx
\left(\tfrac{q a_{u} N_{u} \lambda_{u} \sqrt{G}E_{0}}{2 m c^{2}\beta_{0}\gamma_{0}}\right)^{2}
\Biggl[ \Theta\!\left(1 - \tfrac{\abs{\phi_{j j}}}{2\pi N_{u}} \right)\left(1 -
\tfrac{\abs{\phi_{j j}}}{2\pi N_{u}}\right)^{2} \sin^{2}(\phi_{j j})  \Biggr.\\
&+ \tfrac{N_b}{L_b} \tfrac{N_u\lambda_0}{3} \left(1-\tfrac{3}{8\pi^{2}
N_u^2}\right) \Theta\!\left(1 - \tfrac{\abs{\phi_{jj}}}{k_{0}L_{b}}\right)\left(1 -
\tfrac{\abs{\phi_{j j}}}{k_{0}L_{b}}\right)\\
&+
\left\{ \tfrac{1}{N_{\stext{ph}}} \tfrac{1}{\pi^2} \tfrac{N_u^2}{b_L(b_L^2 - 4 N_u^2)} \sin^{2}(\pi b_L)
 +\right.\\
 &\Biggl. \left. +\tfrac{1}{N_{\stext{ph}}} \tfrac{1}{8\pi} \left\{ 2 \Si\left(2\pi b_L\right) -
\Si \left(2\pi[b_L - 2 N_u]\right) -
\Si \left( 2 \pi [b_L + 2 N_u] \right)\right\} \right\} \Biggr],
\end{split}
\ee
where $b_{L} = \tfrac{1}{2} N_{u} \tfrac{\Delta \omega_L}{\omega_0}$ is a measure of the amplifier
bandwidth (FWHM) relative to the coherent bandwidth of wiggler radiation, and is $O(1)$, while $\Si(x) =
\int\limits_{0}^{x}\! du\,\frac{\sin(u)}{u}$ is the sine integral. Note that $Si(-x) = -\Si(x),$ and for $\abs{x} \lesssim \tfrac{\pi}{2}$, $\Si(x) \approx x - \tfrac{1}{18}x^3$, while for $x \gtrsim \tfrac{\pi}{2},$ $\Si(x)$ undergoes decaying oscillations, peaking at $\Si(\pi) \approx 1.852$ and asymptotically approaching $\Si(x) \to \tfrac{\pi}{2}$ as $x \to \infty.$
The first term on the RHS of (\ref{secondmoment}) is due entirely to self-fields, and is just the square of the first-order conditional average; the second term is the contribution from shot noise, reflecting heating from other particles within the sample, while the final terms reflect the minimal additive amplifier noise consistent with quantum mechanics.

In principle, the conditional two-particle correlations $\innerp{ \Delta\gamma_{j} \Delta\gamma_{k} }{
\gamma_{j}, \gamma_{0} }$ are also needed to determine the evolution of the beam energy spread, but with good mixing, the effect of these cross-terms will be negligible when averaged over \textit{all} the particles in the beam, so we need not actually calculate them explicitly.

\subsection{From Single Particle Statistics to Beam Properties:\\ Evolution of the RMS Beam Energy and Energy Spread}

Because the SNR is so low, the magnitude and even direction of any given particle's energy kick on any one pass are subject to large uncertainty, and cannot be estimated reliably.  In conventional RF stochastic cooling, relatively small kicks delivered over many passes lead via a time-averaging or smoothing procedure to a continuous-time, Fokker-Planck equation describing the evolution of the single particle distribution function.  Any one kick for any one particle cannot be predicted with much precision, but the cumulative effect of many kicks enjoys the usual advantages of the law of large numbers, and can be reliably determined.

In ultra-fast OSC, the beam must be cooled with a relatively small number of relatively large kicks, so
time-averaging is less justified, and less effective at improving predictability.  However, particles phase space deviations remain largely uncorrelated during the cooling, so while prediction of \textit{individual} particle behavior is unreliable, averaging over \textit{all} particles in the beam is expected to give $O\Bigl( N_b^{-1/2} \Bigr)$ improvement in predictive accuracy. 

In analyzing the stochastic evolution of beam properties, we should be careful to distinguish between \textit{arithmetic} means over all beam particles, so as to define \textit{intensive} beam properties such as the average density, velocity, energy, or energy spread, or (classical and/or quantum mechanical) \textit{statistical} or \textit{ensemble} averages, or expectation values, over our uncertainty in either individual particle or collective observables.  

If $t_{-}$ is some time just before the beam enters the cooling section, at which point the average perp-particle energy is $\gamma_{0},$ and $t_{+}$ is a time just after it has passed through the cooler but before it reaches the next cooling section, then the net change in the mean particle energy can be predicted by further averaging the conditional averages
(\ref{avgkick}) over the energy distribution for individual particles and over all particles in the bunch.  The mean energy per particle of all beam particles beam at any time $t$ is 
\be
\bar{\gamma}(t) = \frac{1}{N_{b}}\!\sum\limits_{j = 1}^{N_b}\gamma_{j}(t),
\ee
which remains a ``random variable.''  Since all the beam particles are assumed identical, The statistical expectation value of this mean per-particle energy after a cooling pass is
\begin{equation}\label{expenergy}
\mean{ \bar{\gamma}(t_+) } = \mean{ \gamma_{j}(t_+) }  = \mean{ \gamma_j(t_{-}) }  + \mean{ \Delta\gamma_j },
\end{equation}
where
\begin{equation}\label{expenergy2}
\mean{ \Delta\gamma_{j} } = \mean{ \innerp{ \Delta\gamma_{j} }{ \gamma_{j}, \gamma_{0} } }
\end{equation}
is the conditional average (\ref{avgkick}) averaged further with respect to
the remaining random variables, \ie, with respect to the probability distribution for pre-kick particle energies.  Ideally, the cooling will not change the mean beam energy appreciably but will only redistribute energy from more energetic to less energetic particles: 
\begin{equation}\label{expenergy3}
\mean{ \bar{\gamma}(t_+) } = \mean{ \gamma_{j}(t_+) } \approx  \mean { \gamma_{j}(t_-)}  = \gamma_{0}.
\end{equation}
In order to determine either the expected decrease in energy spread due to cooling, or the uncertainty in our prediction (standard error) for the per-particle beam energy, second-order statistics for $(\Delta\gamma_{j})^2$ are also needed, but these two quantities should not be conflated: our uncertainties and intra-beam fluctuations are not the same thing.  Since statistics for the different beam particles are approximately independent, we expect that the standard error of prediction for $\bar{\gamma}(t)$ will be $O(\tfrac{1}{\sqrt{N_b}})$ better than that for any one particle.

Evolution of the mean per-particle energy has nothing to do with actual cooling \textit{per se}: beam slowing is not the same as beam cooling.  To assess the latter, we need to consider second moments.

First consider the unconditional variance of the $j$th particle's energy, defined by:
\begin{equation}
\label{var1} \sigma_j^{2}(t) =
\mbox{Var}[\gamma_j(t)] = \mean{ \left[ \gamma_{j}(t)
-\mean{\gamma_j(t)} \right]^{2} },
\end{equation}
so that
\be\label{var2}
\begin{split}
\sigma_j^2(t_{+}) &=  \sigma_j^2(t_{-}) + \left[ \mean{ \innerp{ (\Delta\gamma_j)^{2} }{ \gamma_j, \gamma_{0}} } -  \mean{ \Delta\gamma_j }^{2} \right] \\
 & + 2\left[     \mean{ \innerp{ \Delta\gamma_j }{ \gamma_j, \gamma_{0} } \delta\gamma_j(t_{-}) }  -
\mean{\Delta\gamma_j} \mean{\delta\gamma_{j}(t_{-}) } \right]
\end{split}
\ee
If  $\mean{ \Delta\gamma_j } \approx 0$, then this can be rearranged and simplified to yield:
\begin{equation}\label{var3}
\sigma_j^2(t_{+}) -  \sigma_j^{2}(t_{-}) =  2\mean{ \Delta\gamma_{j}\delta\gamma_{j}(t_{-}) } + \mean{ (\Delta\gamma_{j})^{2} },
\end{equation}
where $\mean{ (\Delta\gamma_{j})^{2} } = \mean{ \innerp{ (\Delta\gamma_{j})^{2} }{ \gamma_{j},
\gamma_{0}} }$, and $\mean{ \Delta\gamma_{j} \delta\gamma_{j}(t_{-}) } = \mean{ \innerp{ \Delta\gamma_{j} }{\gamma_{j}, \gamma_{0} } \delta\gamma_{j}(t_{-}) }$.
Since for small energy deviations, the kick will satisfy $\Delta\gamma_{j} \sim -\delta\gamma_{j}$ by design, the first term on the right-hand side of (\ref{var3}) is negative and reflects the impact (on our uncertainty, not necessarily physically, on the particles) of cooling due to the coherent self-interaction, while the second term is always positive and represents the effects of heating.  However, it is important to note that  $\sigma_j(t_{+})$ actually quantifies changes in uncertainty in the $j$th particle's energy after being subjected to the stochastic kick, and is not equal to the particle's actual but unknown energy deviation $\delta\gamma_{j}(t_{f})$ or to the unknown change in this energy via the kick, although it is numerically equal to the expectation value of the squared physical energy deviation.

But in the present context, cooling means reducing the actual extent of energy deviations amongst the beam particles, not reducing our uncertainty in what the particle energies might be, so (\ref{var3}) is not necessarily equal to the real quantity of interest for cooling.  

Longitudinal cooling can be assessed by changes in some measure of the actual fluctuations about the mean energy, such as the RMS energy spread $\delta\gamma(t)$ of particles in the bunch at time $t$, defined by:
\be\label{energyspread}
\delta\gamma(t) = \left[\tfrac{1}{N_{b}}\sum\limits_{j =1}^{N_{b}}\left(\gamma_{j}(t) -
\bar{\gamma}(t)\right)^2\right]^{1/2}.
\ee
Because the kicks are stochastic, the $\delta\gamma_{j}(t_+)^2$, and hence $\delta\gamma(t)$ cannot be known exactly, but the RMS expectation value of the latter can, and is approximately given by:
\begin{equation}\label{energyspread2}
\mean{ \left[\delta\gamma(t)\right]^2 } \approx  \sigma^2_j(t).
\end{equation}
That is, on average, the RMS energy spread of the bunch is approximately equal in value to the single-particle energy variance, as might be expected from a bunch comprised of particles which are assumed statistically independent.  But while the magnitude and direction of any one particle's kick are subject to large uncertainty and cannot be estimated reliably, averaging over all the beam particles ensures that (\ref{energyspread2}) will be quite accurate as a point estimate for the overall RMS energy spread within the beam.

The exact calculation of this standard error would require knowledge of still higher-order moments, but assuming a Gaussian moment closure (justified by the Central Limit Theorem under our present assumptions), we can obtain the estimate $\var{ \delta\gamma(t_+)^2 }  \sim O\left(\frac{2}{N_{b}}\right)  \sigma_j^2(t_+),$ so the \textit{relative} uncertainty of our estimate of the RMS energy fluctuations in the beam is expected to be $O\left(\tfrac{1}{\sqrt{N_b}}\right) \ll 1$.

Defining the instantaneous cooling rate $\tau_{c}^{-1}$ for energy spread as:
 \begin{equation}\label{coolingrate1}
\tau_{c}^{-1}(t) = -\tfrac{1}{2}\frac{d}{dt}\ln\left[ \mean{ \left(\delta\gamma(t)\right)^2 } \right],
\end{equation}  
a formal expression for the approximate energy cooling rate averaged over the current cooling pass can be found:
\be\label{coolingrate20}
\begin{split}
\tau_{c}^{-1} &\approx -f_{0}\frac{   \mean{ \delta\gamma(t_{+})^2  }  -
\mean{  \delta\gamma(t_{-})^2} }{   \mean{ \delta\gamma(t_{-})^2 } }\\
&= f_{0}\left[  \frac{\mean{ \innerp{ \Delta\gamma_{j} }{
\gamma_{j}(t_{-}) , \gamma_{0} } \left[-\delta\gamma_{j}(t_{-})\right] } }{ \mean{ \delta\gamma(t_{-})^2 } } -
\frac{1}{2}\frac{\mean{ (\Delta\gamma_{j})^2 } }{ \mean{ \delta\gamma(t_{-})^2} } \right],
\end{split}
\ee
where again $f_{0}$ is the frequency of cooling kicks experienced by the bunch.  Since $\Delta \gamma_j \sim \sqrt{G},$ Note that this is exactly of the form of equation (\ref{cooltime}).

If the particle energy distribution is assumed to start and remain Gaussian, and the time-of-flight delays are assumed to be linear functions of the energy deviations, then $(\ref{coolingrate20})$ can be calculated analytically.  Assuming optimal choices for gain and time-of-flight delays, the expressions become essentially identical in form to (\ref{cooling_rate1}), with $N_{s} \approx \tfrac{1}{3}\frac{N_{b}}{L_{b}}N_{u}\lambda_{0}$, and with $N_{n}$ assuming a quite complicated explicit form, but for which typically $N_{n} \sim O(\alpha\inv)$. in practice.  If the Gaussian approximation is not valid, then the distribution functions can be evolved numerically, but in this case higher-order moments may be needed, which are increasingly cumbersome to calculate.

The cooling rate is essentially that predicted classically, except for some unavoidable noise from the quantum amplifier which appears additively, roughly as the equivalent of an extra $O(\alpha\inv)$ or so particles in the sample, not as a multiplicative degradation as in the rate (\ref{cooling_rate2}) in the Poissonian emission model, or worse, as in (\ref{cooling_rate3}) with both quantum jumps and Heisenberg phase noise.

To estimate the uncertainty in this predicted cooling rate, we would need to know fourth-order moments of the energy kicks and energy deviations, which can be laboriously worked out if needed, but again assuming Gaussian distributions, the estimated rate can be shown to be statistically reliable as a characteristic of the behavior of the beam as a whole while undergoing cooling, because of the now familiar $O\left(\tfrac{1}{\sqrt{N_b}}\right)$ scaling.

However, note that (\ref{coolingrate20}) should not necessarily be interpreted as the reciprocal of a true exponential decay time, because the continuum approximation needed is not fully justified in the ultra-fast OSC regime.  With traditional, slower stochastic cooling methods, typically the effects of many ($O(10^{3})$) kicks must be accumulated before any appreciable reduction in phase space occurs whatsoever,  so over such long times scales (compared to $f_{0}^{-1}$), cooling kicks can be assumed to be effectively smeared out more or less continuously in time, and the dynamics can be approximated by a Focker-Planck type differential equation.  Here, each individual kick is much bigger, but fewer are imposed, and accurate cooling estimates really require evaluation of the actual discrete-time difference equations.

\subsection{Final Cooling in Muon Accelerator}

With the classical model for OSC essentially confirmed, apart from some lower bound established on the additional additive amplifier noise that has its origins in quantum mechanics, rough estimates for cooling times achievable with given power, or for the power required to achieve given cooling rates, can be determined.  The demands of microsecond-scale cooling exceed present technology, but slower cooling would require more modest systems.  Because our rough estimates agree with those presented in \cite{zolotorev:1994, zholents:2001}, we just quote the results.

Cooling a $\gamma \sim O(10^{3})$ beam with $N_{b} \sim O(10^{9})$ particles per bunch, on a time-scale $\tau_{c} \sim O(10^{-7}\; \mbox{s})$ in order to achieve an $O(10^{-3})$ relative decrease in longitudinal emittance would require stretching to a size where $N_{s} \sim O(10^{2})$, and would require about $O(10)$ lasers amplifiers, each producing $O(10^{2}\mbox{ W})$ of average power at $O(10^{2} \mbox{ Hz})$ repetition rate, which is beyond the current capabilities of existing amplifiers, but perhaps achievable in the next few decades.

\subsection{Some Obvious Limitations and Extensions}
 
\subsubsection{Transverse Cooling}
 
 Although for simplicity we do not treat transverse cooling here, we should point out that it is possible, and in fact can be performed simultaneously with longitudinal cooling, at the expense of a certain slow-down in the cooling rates if the same cooling sections are used for both longitudinal and transverse DOFs.
 
 As mentioned above, the trick is to exploit dispersion.  For sufficiently small deviations from the desired reference orbit, the full trajectory of a transverse coordinate (say the $x$ coordinate) for a particular particle in the kicker is given by:
 \bsub
 \begin{align}
 x_j(z) = \bar{x}_0(z) + x_{\beta_j}(z) + \tfrac{\delta\gamma_j}{\gamma_0} D_x(z),\\
 x'_j(z) = \bar{x}'_0(z) + x'_{\beta_j}(z) + \tfrac{\delta\gamma_j}{\gamma_0} D_x'(z),
 \end{align}
 \esub 
 where: $\bar{x}_0(z)$ is the reference orbit in the cooling section (including the expected quiver motion); $x_{\beta_j}(z)$ represents the typically slowly-oscillating (compared to the quiver) betatron orbit (transverse deviations from the reference orbit in the absence of dispersion), determined by the beam optics of the external transport lattice and any focusing fields associated with the wiggler itself, together with the particle's initial conditions; $\tfrac{\delta \gamma_j}{\gamma_0}$ is the relative energy deviation from the reference orbit; and $D_x(z)$ is a dispersion function depending on the lattice and the insertion device (\ie, kicker wiggler), reflecting the fact that particles with different energies are bent differently in magnetic fields, while here primes indicate derivatives with respect to the longitudinal position, which is just taken as $z$ in the straight kicker section.

If the delay in the bypass between pickup and kicker is made proportional (in part) to the betatron amplitude, \ie, with a component proportional to either $x_{j\, \beta}$ or $x'_{j\, \beta}$,  then the resulting energy kick $\Delta \gamma_j$ will also depend on these errors in the transverse orbit.  If the kicker is located in a region of appreciable dispersion, then as $\delta \gamma_j$ changes due to work done by the transverse fields of the amplified radiation on the particle, as it quivers in the kicker, $x_{j\, \beta}(z)$ and $x'_{j\, \beta}(z)$ must also change in compensation, since $x_0(z)$ and $x_0'(z)$ are fixed, and the overall transverse position $x_j(z)$ and angle $x_j'(z)$ must remain continuous.  If the phasing is arranged properly, in this way the particle can be nudged onto a smaller-amplitude betatron orbit, which  will be apparent when it emerges into a region of smaller dispersion.

\subsubsection{Neglected Classical Optical Effects}

For the proof-of-principle argument here, we have assumed flat amplitude gain, and $1$D radiation fields propagating in vacuum without diffraction or dispersion.  Since the structure and phase of these fields is essential to the success of the transit-time OSC scheme, more careful calculations will eventually be needed to make tighter quantitative predictions.

\paragraph{Dispersion}\pskip
Since any cooling depends crucially on careful phasing, any dispersion in the amplifier or the optical transport system will diminish the effectiveness of the OSC.  The times-of-flight can of course be adjusted to compensate for any overall average delay for the frequency band, linear group-velocity dispersion and any nonlinear phase-modulation within the amplifier are expected to be deleterious to some degree.  Such dispersion can come from the usual linear index-of-refraction of the medium or the optics, or from the phase shifts that are proportional to the gain (amount of stimulated emission in the medium), which in practice is frequency-dependent.

\paragraph{Diffraction/Transverse Effects}\pskip
We have here effectively assumed that the amplified wiggler radiation consists of transverse plane waves, but even for well-collimated particle beams where $\sigma_{\perp} \ll w_0$, certain paraxial effects may become quite important.

Recall that the transverse polarization component of a fundamental Gaussian vacuum paraxial mode, assumed to be plane polarized and propagating along the $z$ direction with focus located at $z = 0,$ can be written as:
\be
E_x(x,y,z;t) = E_0\, \tfrac{w_0}{w(z)} \,e^{-i\left[k_0z - \omega_0 t - \eta(z)\right] - r^2\left[ \tfrac{1}{w(z)^2} + \tfrac{ik_0}{2R(z)} \right] }, 
\ee
where $E_0$ is the overall complex amplitude, $k_0$ is the carrier wavenumber, $\omega_0 = c k_0$ is the carrier frequency, $r = \sqrt{x^2 + y^2}$ is the transverse distance off-axis, $w(z) = w_0 \left[1 + \left(\tfrac{z}{Z_{\stext{R}}}\right)^2 \right]^{1/2}$ is the local waist, $R(z) = z\left[ 1 +  \left(\tfrac{Z_{\stext{R}}}{z}\right)^2 \right]$ is the radius of wavefront curvature, $\eta(z) = \arctan\,\left( \tfrac{z}{Z_{\stext{R}}}\right)$ is the Guoy phase, and $Z_{\stext{R}} = \smallhalf k_0 w_0^2$ is the Raleigh range, where $w_0$ is the waist size at focus.

Issues of coupling efficiency suggest that both kicker and pickup wigglers should have identical values for $\lambda_u$ and (nearly) identical values for $N_u$, while the Raleigh range of the amplified radiation in the kicker wiggler should be comparable to that of the radiation emitted in the pickup, by proper choice of collection and transport optics.  As the latter is approximately $Z_{\stext{R}} \sim \tfrac{1}{8\pi} N_u \lambda_u$,  diffraction will likely just become appreciable over the interaction distance $N_u\lambda_u$ in the kicker. 

The direct effects of the wavefront curvature are probably safely ignored, and the intensity fall-off due to the expansion of the transverse waist can be accommodated, although somewhat higher amplifier powers than estimated in the $1$D theory might be needed.  Of most concern is  the Guoy phase shift.  If the laser focus is placed near the center of the wiggler to help minimize amplifier power requirements, the Guoy phase will vary rapidly from close to $-\tfrac{\pi}{2}$ to $+\tfrac{\pi}{2}$ inside the kicker, and despite best efforts to ensure careful relative timing, the energy kick may tend to average away.  This needs to be studied carefully.  If the focus must be moved off-center or, worse, perhaps placed outside the kicker, by a distance comparable to the Raleigh range, in order to avoid a the region of rapid phase advance near the focus, then power requirements for OSC will become even more severe.

\paragraph{Near-Field Effects}\pskip
With quantum-mechanical fears allayed, an accurate quantitative calculation will require accurate knowledge of the classical field.  The paraxial Gaussian form may not be sufficiently accurate.   For example, the solenoidal (\ie, functionally-transverse, or divergence-free) part of the classical electric field (and hence the expectation value of the hemi-classical ``photon'' field) radiated by a single beam particle in the pick-up can be expressed in the well-known Li{\'e}nard-Wiechert form:
\be
\begin{split}
\bv{E}_j(\bv{x},t) &= q \left[ \tfrac{\unitvec{n}_j - \bv{\beta}_j}{\gamma_j^2\left( 1- \bv{\beta}_j\!\cdot\!\unitvec{n}_j \right)R_j^2}    \right]_{t'_j}  - q \left( \tfrac{\unitvec{n}_j}{R_j^2} \right)_t\\
&+ \tfrac{q}{c} \left[ \tfrac{ \unitvec{n}_j \times \left\{ (\unitvec{n}_j - \bv{\beta}_j) \times \dot{\bv{\beta}}_j   \right\}  }{\left( 1- \bv{\beta}_j\!\cdot\!\unitvec{n}_j \right)^3 R_j}  \right]_{t'_j}
\end{split},
\ee
where $\bv{R}_j = \bv{x} - \bv{r}_j(t)$, $R_j = \abs{\bv{R}_j}$, $\unitvec{n}_j = \tfrac{\bv{R}_j}{R}$,  $\bv{r}_j(t)$ is the particle trajectory, $\bv{\beta}_j(t) = \tfrac{1}{c}\dot{\bv{r}}_j(t) = \tfrac{d}{dt}\bv{r}_j(t)$ is the normalized particle velocity, $\dot{\bv{\beta}}_j(t)= \tfrac{d}{d t}\bv{\beta}_j(t)$ is the normalized acceleration,  and $t'_j$, defined implicitly by $t =t'_j+ \tfrac{R_j}{c}$, is the retarded time.

The far-field radiation pattern is determined by the asymptotic form of the final term only, which is proportional to the particle acceleration, and falls of like $O\left(\tfrac{1}{R}\right)$ in distance from the source.  The form of the far-field brightness function (\ie, optical Wigner function) of synchrotron radiation from relativistic particles moving in bending magnets, for which $\tfrac{a}{\gamma_0} \ll 1$,  has been worked out\cite{kim:1989} in the paraxial case for regimes in which $a_u \ll 1 ll \gamma_0$ or $1 \ll a_u \ll \gamma_0$ for essentially any $N_u > 1$, or for essentially any $a_u \ll \gamma_0$ if $N_u \sim 1$ or $N_u \gg 1$.

In the presumed OSC regime, where $N_u \sim O(10)$, while $a_u \gtrsim 1$, an exact analytic form is lacking, and the velocity fields may not be negligible.  Since the radiation is emitted over the length $N_u \lambda_u,$ we would naively expect that a Fraunhofer-type approximation to the angular spectrum can be deduced from the acceleration fields and will be valid if the radiation is observed at some suitably large distance $R$ from the wiggler, such that $R \gg \lambda_0,$ $R \gg N_u\lambda_u$, and $R \gg \tfrac{(N_u \lambda_u)^2}{\lambda_0}$.  While the first condition is entirely trivial, the second is unlikely to hold, and the third will certainly not hold.  In order to have time and room to amplify the radiation and feed it back onto the particle beam, the wiggler radiation emitted over $O(5\; \mbox{m})$ will be collected, collimated, and imaged into the amplifier at a distance only about $O(1 \mbox{m})$ from the end of the pickup, only a few or maybe ten times $\lambda_u$ downstream, where near-fields will still be present, and their effects are not entirely clear.  Such considerations will be explored in subsequent research.

\section{Discussion}

Despite fears that quantum uncertainties would prove devastating for optical stochastic cooling based on the amplification of very weak pickup signals, the technique of OSC can in principle work despite these quantum uncertainties.  What, then,  precisely, went wrong with the naive reasoning?

\subsection{Why ``Naive'' Quantum Treatments Fail}

Particles do \textit{not} emit wiggler radiation described by photon number states
or any statistical mixtures of of such states.  They do not radiate into any states resembling a number state, but rather a Glauber coherent state, or a classical statistical mixture of such states, in which photon number is not sharply defined, but phase information is partially available and joint quantum mechanical uncertainties in number/phase or in the quadrature components are jointly minimal.
These states can have arbitrarily low expectation values for photon number or EM energy, but still be every bit as ``real' and ``present'' on each pass as any number state.

Particles do not radiate in a series of stochastic quantum jumps, only once every $O(\alpha\inv)$ passes through the pickup wiggler.  The radiation from each particle is present on every pass, even though its average energy may correspond to much less than one photon, and is available to be amplified by a quantum mechanical interaction, provided no intervening measurement first projects the state by actually trying to count  photons.  Moreover the wiggler  radiation possesses, on average, exactly the amplitude and phase it would possess classically, so the ``coherent signal'' information necessary for transit-time cooling is effectively always present, only corrupted by the equivalent of some additional additive (not multiplicative) noise needed to satisfy the uncertainty principle, noise that physically can be traced back to ``vacuum fluctuations'' and amplified spontaneous emission within the gain medium.

A linear amplifier does \textit{not} function by multiplying photon number, either coherently or by projective measurement, but rather acts unitarily on the full quantum mechanical state of the EM field, scaling the field amplitude and adding some extra noise.

In the pickup, the particles do not radiate whole numbers of photons, and in the kicker the particles do not measure, absorb, or otherwise sense photon number, but rather respond directly to the EM fields.

Nothing actually prepares, emits, multiplies, evolves towards, measures, or counts photon number states, so the ``discrete'' nature of photons is not terribly important.

A careful analysis reveals that the naive quantum mechanical fears were misguided: the 
cooling rate which accounts for quantum mechanics essentially agrees with the classical rate calculation, except that the amplifier noise, which can in principle be made arbitrarily small classically, is constrained by the quantum mechanics to some non-zero lower bound.

This lower bound reflects the fact that the total fluctuations of the amplified field include the amplified input noise, which just consists of minimum-uncertainty vacuum fluctuations for coherent states, as well as added amplifier noise consisting of an independent set of amplified vacuum fluctuations.
Input noise is amplified along with the signal because of linearity: a linear amplifier cannot know what part of the input state will later be regarded as noise or as signal. The additional noise is added as a consequence of the uncertainty principle: a phase-insensitive linear amplifier amplifies both quadrature components of a field mode, which do not commute, so at least one extra mode must be involved.

Therefore quantum mechanical noise does not prevent fast optical cooling, but it does limit how small the incoherent heating terms can be made. The resulting expressions and estimates for the cooling times agree with classical calculations, provided the amplifier noise is accounted for through an effective value for $N_{n}.$

\subsection{Synchrotron Radiation Damping}

If the naive ``quantum jump'' model of discrete Poissonian photon emission failed so spectacularly for the case of OSC, why does it work so well for synchrotron radiation damping (SRD) of electrons in storage rings?  In such a damping scheme, relativistic electrons of normalized energy $\gamma$ orbiting in a ring of radius $r_0$ spontaneously emit radiation in a continuum of frequencies up to about $\omega_c \approx \tfrac{3}{2}\gamma^3 \tfrac{c}{r_0}$, and confined to an angular cone of about $O(\tfrac{1}{\gamma})$ with respect to the electron orbit.  Each electron emits the equivalent of about  $N_{\stext{ph}} \sim \tfrac{5 \pi}{\sqrt{3}} \gamma \alpha$ photons per revolution, at an average energy $\mean{ \hbar \omega } \sim \tfrac{8}{15\sqrt{3}} \hbar \omega_c$.
For example, assuming an $O(\mbox{GeV})$ electron beam, corresponding to electrons with $\gamma \sim O(10^4)$ circulating in rings with $r_0 \sim O(10^2\mbox{ m})$, each electron radiates the equivalent of about $N_{\stext{ph}} \sim O(6 \cdot10^2)$ photons per revolution, at wavelengths down to about $\lambda_c = \tfrac{2\pi c}{\omega_c} \sim O(10^{-10}\; \mbox{m})$, corresponding to $O(\mbox{keV})$ X-rays.

On average, each electron loses momentum along its direction of motion (parallel to its betatron orbit) due to the recoil from this synchrotron emission.  If they are then re-accelerated longitudinally, along the reference beam-line, transverse emittance is this reduced.

Ultimately cooling is limited by quantum mechanical ``randomness'' in the emission, which, consistent with fluctuation-dissipation relations, leads to an incoherent heating term in addition to the damping term, which limits the cooling rate as well as the ultimate emittance that can be achieved in equilibrium.  In a pioneering and influential study, M. Sands \cite{sands:1970} successfully modeled these effects by using what we have called the ``quantum jump'' model, assuming that electrons emit radiation in the form of a whole number of quanta in  random Poissonian fashion.

In hindsight, it is clear that the OSC and SRD occur in completely different regimes and rely on rather different physics.  In the OSC, each particle emits on any pass the equivalent of very few photons of small momentum, so direct radiation reaction effects are completely negligible.  In SDR, each electron emits the equivalent of many photons of high momentum, and recoil effects in the virtual Compton scattering processes are significant -- in fact, the large cumulative effects of recoil over several turns are precisely what is of principle concern.

In OSC, the radiated field is collected, coherently amplified, and fed back onto the particles that emitted it, to induce an energy kick far larger than any direct losses due to emission, so the quantum state of the radiation matters greatly, while in SDR there is no feedback other than the individual electron recoils.
Once any emitted radiation propagates away, it becomes irrelevant as far as the individual particles are concerned, and in fact one could in principle ``trace out'' the field degrees-of-freedom to obtain a non-Hamiltonian, diffusive evolution equation for the particles only.

In neither case do we actually expect the particles to radiate a whole number of quanta in discrete emission events, but in the case of SDR, whether the radiation is described by a number state or coherent state or some other state is largely irrelevant, as long as we get the low-order moments right.  Many different microscopic stochastic models can lead to the same Fokker-Planck equation.  In effect, the Sands technique gets away with modeling a diffusion equation by one particular choice of a random walk model for emission which does not faithfully describe the actual emission process or the resulting radiation phase space, but mimics the proper statistics for the electron phase space, which is of primary interest.

If simple scaling arguments or back-of-the-envelope calculations predict that the discrete nature of the \textit{energy} eigenstates of the EM fields might be important, then rather than justifying the quantum jump model, this only gives more reason to adopt a fully quantum treatment of the radiation, at least if radiation properties or statistics are themselves needed, because quantum mechanics evolves the radiation modes or any system unitarily, not by a series of quantum jumps.  If the radiation itself is never examined, but only the accumulated effects of its back-action on the particles during emission or absorption, then sometimes the Poissonian model will be useful, as with the Sands treatment of synchrotron radiation damping, or with the more sophisticated Poissonian emission model used by Friedland \cite{friedland:1984} to tease out the small-signal gain in FELs.

\subsection{Is the Field Evolution Really Unitary Throughout?}

The metaphor of corpuscular photons emitted in quantum jumps may be alluring, but can be deeply misleading in the analysis of stochastic cooling as well as certain other applications of wiggler radiation.  Benson and Madey\cite{benson:1985} assert:
\begin{quote}
\ldots the wiggler is capable of reducing the wavefunction [of a charged particle] by causing it to emit a photon.  The electronÕs position can therefore be deduced from subsequent measurements.
\end{quote}
Reluctant as we are to disagree with pioneers of FEL physics, we are of the opinion that, as a reasonably accurate description of our original, naive opinions concerning quantum mechanical effects of the wiggler radiation, this statement is completely incorrect in the current context.  We contend that the wiggler magnet itself does not in any way collapse or reduce any particleÕs wavefunction --- only a subsequent photon-counting or similar optical measurement will do that, which never happens in the context of OSC.  Until a reversible macroscopic measurement of particles or radiation is made, the EM fields and beam particles evolve unitarily.  Besides, with many electrons in a sample length, observation of a photon can tell us very little about the position or momentum of any one electron, so ``subsequent measurement'' only partially project a given particle's wavefunction.  Nor does the wiggler cause the particle to emit a photon, if by photon we mean either a number state of the electromagnetic field, or even just a state of definite energy and/or momentum.  In most practical regimes of interest, we have seen that the particles in a wiggler will generally emit radiation described by statistical mixtures of Glauber coherent states, which behave far differently (more classically) than number states.  If recoil or other reaction effects
become important, than the full quantum state of the radiation and particles may become highly entangled, or may exhibit other features associated with non-positive Glauber-Sudarshan quasi-distributions that cannot be reproduced classically, but there is no reason to think the states of the field will be close to eigenstates of either the number or energy operators or would be an even approximately diagonal mixtures of these states.

If photon-counting experiments are chosen to be performed, then whole numbers of photons will of course be measured, but this has little to do with what state describing the radiation \textit{before} measurement.  Of course the calculations of the field dynamics and field statistics can be performed in number state basis, or any other basis we choose, but the density matrices will not generally be sparse or nearly-diagonal in any sense, so the $P$-function representation is particularly natural and convenient.

Other fields learned these lessons earlier.  In \cite{gordon:1963}, Gordon, Louisell, and Walker conclude: 
\begin{quote}
We find that a classical description of the input fields and of the amplification process is completely valid provided we take correctly into account the response of the amplifier to the input zero-point fields.  This result is valid for inputs of arbitrarily small power.
\end{quote}
This is basically correct, apart perhaps from a possible factor of two clarified by Caves \cite{caves:1982}, arising, as we have seen, from the extra set of amplified vacuum fluctuations which must necessarily be added to the field for amplifiers with gain on non-commuting field observables, in addition to the original amplified vacuum fluctuations, which must be amplified along with the signal, because the amplifier cannot know what part of the input field is to be regarded as signal and what part is considered noise.

If the wiggler magnet is not collapsing the wave function of an electron and causing it to emit a photon, perhaps the amplifier is?  For large gain, whether the action of the amplifier is construed to be a measurement or not depends on your favorite interpretation of the quantum mechanics.  Some might argue that it can be regarded as an effective  measurement because it amplifies the field to levels that  ``we can lay our classical hands on,'' \ie, that can be classically observed, measured, and manipulated without significant further disturbance, measurement back-action or degradation of the SNR.  Others might object, pointing out that no agent makes an irreversible, macroscopic record of any information in the radiation fields, at least not until after they are re-applied to the particles.  Either way, it makes little difference to our argument.  Even if you construe the amplifier to be performing a measurement of the radiation, it is doing so in a coherent state basis, \textit{not} a Fock state basis.

What about the environment at large?  Even in a state-of-the art beam-line, with high vacuum, vibration control, etc., uncontrollable interactions with the numerous degrees-of-freedom in the environment will inevitably occur, leaving an imprint of information in various quantum entanglements.  Without observational access to all of these environmental DOFs, the reduced density matrix for the system alone (particles and fields) will appear impure, and subsequently the system can behave as if the environment has effectively performed a measurement but just not told us the answer.

This is the basic argument behind environmentally-induced decoherence\cite{zurek:1981, zurek:1982, zurek:1993b, paz:1993}, which we have already invoked in arguing for the classical behavior in the electron beam itself.  While quantum-mechanical decoherence in the radiation fields will undoubtedly proceed to some degree, it actually only helps our argument.  For simplified but sensible models of ``the environment'',  it is again the coherent states, not the Fock states, that survive the ``predictability sieve''  and emerge as the preferred ``pointer-basis'' into which the decohered radiation states will appear to be projected\cite{kubler:1973, zurek:1993b, zurek:1993c, gavish:2004}.  This should not be too surprising.  Decoherence seems to favor classical-looking states, and has in fact been suggested as a reason for emergence of a classical world.  Conversely, a certain robustness or stability in the face of inevitable sources of decoherence is a key feature of classical-like states.  Coherent states, or ignorance mixtures thereof, are according to the pioneering work of Glauber and Sudarshan and countless investigators since then, the closest thing to classical radiation fields allowed by quantum mechanics.

Conversely,  If we want to see discrete photons, we have to perform photon-counting experiments, which are usually (but not always) destructive, and usually do not leave the remaining field in a number state.  If for some reason we really want the ``amplifier'' to multiply photon number states, rather than scale coherent states, then we must use photon-number-amplifiers (PNA)\cite{yuen:1996} rather than linear amplifiers, devices which have been proposed but which would be difficult in practice to operate with high fidelity.  (Of course, using either unitary PNA or non-unitary photo-detection would be disastrous for OSC, as our naive model predicted).  Fock states or other non-classical states with negative Glauber-Sudarshan quasi-distribution functions are just rather delicate and temperamental creatures, difficult to create and maintain, and fragile to the touch of the environment.  The bane of many investigations in atomic physics or quantum optics, this fragility of quantum coherence only works to our advantage here.

 \subsection{Can the Quantum Mechanical Noise Limits be Achieved?}

We have found that quantum mechanics imposes a firm lower bound on the amount of noise that must be added by any linear amplifier in order to maintain consistency with the uncertainty principle and with unitarity, but have not addressed how difficult it may be to approach this ideal noise floor.

In practice, there are also equilibrium and non-equilibrium thermodynamic limitations.  The gain medium, as well as the various other optical elements are coupled to thermal baths, so the ``initial'' state of the radiation field prior to the arrival of the pickup signal is typically not a vacuum, but a thermal blackbody distribution, with an average of  
\be
\mean{\EuScript{N}_{\ell}} = \frac{1}{e^{\tfrac{\hbar \omega_{\ell}}{k_{\stext{B}}T}} - 1}
\ee
photons in each mode $\ell$ of frequency $\omega_{\ell}.$
At high temperatures, \ie, $k_{\stext{B}}T \gg \hbar\omega_k$, the average occupation $\mean{\EuScript{N}_{\ell}} \to \tfrac{k_{\stext{B}}T}{\hbar \omega_{\ell}} \gg 1$, and the amplifier will have an effective noise at input far higher than the zero-point vacuum energy.  Conversely, for low temperatures, where $k_{\stext{B}}T \ll \hbar\omega_{\ell}$, the average photon number becomes $\mean{\EuScript{N}_{\ell}} \to e^{-\tfrac{\hbar \omega_{\ell}}{k_{\stext{B}}T} } \ll \smallhalf$.  For $\lambda \sim O(1\; \mu\text{m})$, or equivalently $\hbar \omega_{\ell} \sim O(1\;\mbox{eV})$, and any temperature $T \lesssim O(10^4\mbox{K})$, the thermal occupation is negligible, and the blackbody state will be virtually indistinguishable from vacuum as input. 

Along with large gain bandwidths, this is another tremendous advantage optical-frequency stochastic cooling might enjoy over traditional schemes.  Merely from purely thermal considerations, it would be extremely difficult to construct and operate amplifiers anywhere close to  the quantum limit at radio or microwave frequencies, but this is possible at optical frequencies, thanks to the exponential dependence of the Boltzmann factor.  However, with very strong pumping, some active or passive cooling may be needed to ensure quiet and efficient operation.

However, while the surrounding radiation field may start out in thermodynamic equilibrium, the amplifier itself is not, but instead subject to various energetic flows from the pump as well as the input and output signals.  In fact a non-equilibrium state is of course necessary in order to achieve gain.  A finite pump strength leads to a finite population inversion which in turn leads not only to smaller gain, but to ``excess'' spontaneous emission beyond that needed to satisfy the uncertainty principle.

Using semi-classical arguments, it can be shown that the gain for narrow-band propagation along $z$ in a typical inverted medium can be written
\be
G =e^{\Gamma z} \gg 1,
\ee
where
\be
\Gamma \approx \tfrac{n_2 \eta_{23} \lambda^2}{ 4 P_{21} n_{\stext{ir}}^2 \tau_{0} \Delta\omega},
\ee
is the gain profile of the medium, and where $\Delta \omega$ is the bandwidth, $n_{\stext{ir}} \ge 1$ is the index of refraction of the gain medium, $\lambda$ is the wavelength for the resonant lasing transition between levels $2$ and $1$, $\eta_{23} \le 1$ is the quantum efficiency for pumping, $\tau_{0}$ is the characteristic fluorescence lifetime, or time-scale for spontaneous emission, $n_2$ is the average occupation density of the upper level $2$, and 
\be
P_{21} = \frac{n_2}{ n_2 - \tfrac{g_1}{g_2} n_1} \ge 1
\ee
is the population inversion parameter, in which $n_1$ is the average occupation density of the lower level $1$ and $g_1$ and $g_2$ are degeneracy factors for these levels.  

The effective number of noise photons then becomes
\be
\bar{\EuScript{N}}_{\stext{noise}} = \smallhalf + (P_{21} - \smallhalf)\abs{1 - \tfrac{1}{G} },
\ee
which approaches the ideal lower bound as $G \to \infty$ and $P_{21}  \to 1^+$, the latter implying total population inversion.

A back-of-the-envelope calculation indicates that required pump power in a multi-stage Ti:sapphire solid state laser amplifier suitable for OSC might correspond to $G \sim O\left(10^4\right)\gg 1$ and  $P_{21} \sim O\left(\tfrac{7}{4}\right)$, which corresponds to a noise power only $O(25\%)$ higher than the quantum limit.  In order to incorporate higher-than-ideal noise levels, we can just boost the effective noise temperature in the displaced thermal state describing the wiggler radiation after amplification.  Again, the effects will be similar to those obtained by adding more particles to the sample.

Quantum mechanics neither requires nor prevents an actual amplifier from exhibiting some multiplicative noise in addition to the required additive noise, for example due to pump fluctuations, but it will not be of the drastic all-or-nothing character that proved so deleterious to cooling in the Poissonian emission model.  Since the thermal or chaotic states are already the sate of maximum entropy for given moments, we expect that they can still be used in the case of multiplicative noise in the gain, just with the second-order moments suitably increased to account for the extra noise.

\subsection{Comparison to the Heifets-Zolotorev Model}
 
Independently, Heifets and Zolotorev\cite{heifets:2001} developed  a fully quantum $1$D model of transit-time optical stochastic cooling\footnote{M. Zolotorev should be credited with instigating both approaches to the problem.  After he suggested this investigation to us, and our answers were not forthcoming with sufficient alacrity, he went in search of a theoretical mercenary with bigger mathematical weapons at his disposal.  Our simpler treatment based on coherent states did uncover a mistake in preliminary calculations by Heifets and Zolotorev.}.  They follow the combined von Neumann density matrix for the particle bunch and the radiation through the various components of the cooling system, including the pickup wiggler (still with classical fields), the bypass for the particles and the amplifier for the radiation, and then the classical fields of the kicker wiggler.  While both particles and fields are treated quantum mechanically, a simplified one-dimensional, single-mode (with fixed wavenumber $k_z$) model of the interaction inside the wiggler is used to calculate the initial emission by the particles and the final energy kick, and a simple model of the optical amplifier is employed, based on a completely-inverted population of two-level atoms.

To solve analytically for the evolution, a number of expansions are made in various small parameters, particularly $\tfrac{N_u}{\gamma_0} \tfrac{\lambda_c}{\lambda_0} \ll 1$ , $\EuScript{N}_{\stext{ph}} \ll 1$ (the mean number of coherent photons emitted per particle in the pickup), $\tfrac{N_s}{N_B}$, and $\tfrac{\hbar \omega_0}{mc^2\delta \gamma}.$

After a minor \textit{tour de force} in the manipulation of special functions, they obtain results which appear to be very similar to those of our hemi-classical theory: namely, assuming the coherent field mode is initially in vacuum, the pickup radiation from a single particle will be described by a Glauber coherent state, with average photon number $\EuScript{N}_{\stext{ph}} \sim \alpha.$  The reduced density matrix of the amplified radiation is that of a chaotic, or thermal-like state, with the equivalent of one photon of amplifier noise at input (half for the original  ``zero-point motion,'' and half again from an independent set of atomic fluctuations,  because the amplifier provides gain for both non-commuting quadrature components), although their combined particle-field density matrix contains the small but vital correlations (what classically we called the coherent component of the signal) needed to cool the particles.  The cooling rates are close to the classical results except with the equivalent of $O(\alpha\inv)$ extra particles in the sample.

Even for a simplified dynamical model, the fully quantum results are impressive, but only seem to confirm what we have established, namely that a hemi-classical formalism (which in the end really just amounts to a classical but stochastic model) is sufficient, and in principle will be more easily generalized to account for finite bandwidth, off-axis, or other neglected effects (which, however, are beyond the scope of our current investigations).
 
\section{Conclusions: Summary and Future Directions}

Assuming optimal gain, the rate of any stochastic cooling scheme can be improved by increasing the bandwidth of the pickup-amplifier-kicker system or by decreasing the number of neighboring particles (which actually cause heating) in a sample length, which varies inversely with the bandwidth.  Large bandwidths available in solid-state or certain parametric amplifiers at optical wavelengths are very attractive for stochastic cooling, if various technological challenges can be met.  In the transit-time OSC scheme, spontaneous undulator radiation from the pickup wiggler is amplified while particles are given appropriate phase delays in bypass, then amplified radiation acts back on particles within the kicker wiggler to reduce momentum spread and/or betatron amplitude.

The pickup self-radiation of each particle (responsible for cooling) is very small --- on average, only $O(\alpha)$ photons per particle in coherent mode, suggesting quantum effects may be important  In a naive picture, akin to SandÕs treatment of synchrotron radiation damping, where particles emit whole photons in discrete Poissonian quantum jumps,  it would seem particles rarely experience cooling self-radiation but almost always feel large heating cross-radiation, leading to drastically slower cooling or over-correction
when the cooling photon finally appears.

A more careful analysis has indicates the feared quantum catastrophe is a red herring:
particles radiate into Glauber coherent states, or statistical mixtures thereof, and amplifiers amplify coherent states, not Fock (photon number) states.  The cooling signal is really present on every pass.
Particles emit radiation in a quantum mechanical state in which photon number and field energy are not sharply defined, but in which the expectation values of amplitude and phase are given by the classical values, rather than in some sort of quasi-classical Poisson process where whole photons are emitted in a series of discrete ``quantum jumps.''  A low-noise amplifier will operate quantum mechanically, transforming  the quantum state of the fields according to
Hamiltonian dynamics, and does not act to first projectively measure then multiply photon number.  After amplification, the amplified radiation in the kicker is essentially classical, and particles respond linearly to fields rather than via discrete photon absorption.
The amplified phase signal from each particle is present on each pass, albeit partially obscured by some amount of noise due to amplified vacuum fluctuations of the original radiation field and certain internal amplifier modes.

Therefore, quantum mechanical effects do not destroy cooling, but just introduce additional (additive) noise, approximately equivalent (prior to amplification) to about $O(\alpha\inv$) additional particles in a sample, setting an ultimate upper limit on cooling rates and on the gains in efficiency that can be expected from further beam dilution.

If sufficient care is given to the quantum mechanical aspects of the radiation and amplification, which in the end just
amounts to adding at least the minimum amount of amplifier noise that can physically be traced back to spontaneous emission in the gain medium, then optical stochastic cooling rates can be calculated using an essentially classical treatment with extra noise terms.  For ultra-fast OSC, where the correction consists of a relatively small number of relatively-large kicks compared to traditional RF-based schemes,  temporal smoothing or averaging is not really justified, but while individual particle behavior is subject to high uncertainty, properties the beam as a whole (mean per-particle energy and energy spread) can be reliably estimated because of averaging over many particles in the beam, rather than averaging over many time steps.

Using the cooling rates as derived above for the simplified model of a cooling section,  a very preliminary calculation
indicates that micro-second cooling for muons is possible under somewhat optimistic expectations for the precision of the
beam optics and an optimistic extrapolation for the power available in low-noise, high-bandwidth amplifiers.   Furthermore,
the bunch charge $N_{b} \sim O(10^{10})$ assumed in this study is smaller by an order of magnitude or two than what is
discussed in current proposals for a muon collider, although because of the additional stochastic cooling, the final luminosity can be just as high.

Now that concerns about potentially devastating quantum mechanical limitations have been assuaged, the possibility of ultra-fast cooling rates appears sufficiently promising to justify more careful investigation.  Attention can turn to of the more prosaic classical effects that might limit performance, in order to include important physics so far neglected.

A more realistic model of the classical wiggler radiation should be used, with the effects of off-axis particle injection and diffraction included.  A careful classical calculation will be somewhat complicated by the unusual parameter regime of OSC wigglers, involving high magnetic field strengths and a relatively small number of relatively long-wavelength poles.  The effects on particle orbits of angular deviations and  oscillating components of the longitudinal motion in a planar wiggler are not necessarily small, but have so far been neglected.  Most critically, sensitivity to the errors in the bypass lattice optics must be investigated, the effects of only partial mixing between passes or unwanted mixing within passes addressed, the Guoy phase included, and the effects of non-uniform gain and optical dispersion in the amplifier realistically treated.

Many technological challenges need to be addressed and assessed: the beam stretching needed to reversibly increase bunch length and decrease energy spread to manageable levels before cooling must be accomplished on microsecond time-scales; as must the re-compression after cooling, which is even more difficult since there will by design be less energy spread to be exploited in the high-dispersion lattice.  The bypass and kicker optics must be carefully engineered, sufficiently adjustable to calibrate particle orbits within a fraction of a micron tolerance, and presumably be dynamically stabilized and controlled through active feedback.  The optical amplifiers must be robust, stable, single-pass (non-regenerative), highly linear even at high power, variable gain but with high peak gain, low-noise, and probably actively-cooled.  Mixing (effective shifting longitudinal particle positions on the scale a radiation wavelength or more) must be very thorough between passes the cooling sections, but negligible between the pickup and cooler of any given cooling section.

Now, at least, it is confirmed that such cooling appears in principle completely consistent with quantum mechanical features of radiation and amplification.  It remains to be seen if it is practical in the face of these more mundane but important constraints. 

\section*{Acknowledgements}

This report is dedicated by one of us (A.E.C.) to the memory of his uncle, William Ross Charman, a man of uncommon curiosity and decency, who died while these ideas were being explored and this research completed.

We thank Max Zolotorev for bringing this problem to our attention, and for many subsequent conversations.  We also acknowledge Bill Fawley, Sasha Zholentz, Sam Heifets, Ray Chiao, and Ryan Lindberg for helpful advice, comments, and questions. This research was supported by the Advanced Accelerator Concepts (AAC) initiative of the High Energy Physics (HEP) Division of the US Department of Energy (DOE).


\bibliographystyle{unsrt}
\bibliography{osc_references}

\end{document}